\documentclass[prx, %preprint,
aps,
amsmath,
amssymb,
twocolumn,
nofootinbib,
superscriptaddress,
]{revtex4-2}
\usepackage{graphicx,dcolumn,bm}
\usepackage{amsfonts,amsmath,amssymb,amsthm,amscd}
\usepackage[english]{babel}
\usepackage{physics}
\usepackage{needspace}
\usepackage{xcolor}
\usepackage{bm}
\usepackage{slashed}
\usepackage[export]{adjustbox}
\usepackage{tabularx}

\definecolor{linkblue}{HTML}{2e3092}
\definecolor{textblue}{HTML}{0208D8}
\usepackage[colorlinks=true, allcolors = linkblue]{hyperref}
\renewcommand{\vec}[1]{\mathbf{#1}}

\newcommand{\tem}{t_{\mathrm{em}}}
\newcommand{\gem}{\gamma_{\mathrm{em}}}
\newcommand{\cem}{\chi_{\mathrm{em}}}
\AtBeginDocument{\renewcommand{\natexlab}[1]{#1}}
\newcommand{\Wrad}{W_{\mathrm{rad}}}
\newcommand{\WCS}{W_{\scalebox{.7}{\scriptsize $\mathrm{CS}$}}}
\newcommand{\Wcr}{W_{\mathrm{cr}}}
\newcommand{\WBW}{W_{\scalebox{.7}{\scriptsize $\mathrm{BW}$}}}
\newcommand{\Wradcr}{W_{\mathrm{rad,cr}}}
\newcommand{\WCSBW}{W_{\scalebox{.7}{\scriptsize $\mathrm{CS,BW}$}}}
\newcommand{\oeff}{\omega_{\mathrm{eff}}}
\newcommand{\omegadimless}{\omega^*}
\newcommand{\llgg}{\mathrel{\substack{\ll\\[-.05em]\gg}}}

\allowdisplaybreaks

\newcommand{\rmd}{d}

\renewcommand{\arraystretch}{2.2}

\newcommand*\widebar[1]{%  %%% Warning: weird stuff happens if I remove the seemingly-useless '%'
	\vbox{%
		\hrule height 0.5pt%    % Line above with certain width
		\kern0.15ex%          % Distance between line and content
		\hbox{%
			\kern-0.2em%   % Distance between content and left side of the box, negative values for lines shorter than content
			\ifmmode#1\else\ensuremath{#1}\fi%  % The content, typeset in dependence of mode
			\kern-0.1em%                        % Distance between content and left side of the box, negative values for lines shorter than content
		}% end of hbox
	}% end of vbox
}

\begin{document} 
	
%%% TITLE %%%%%%%%%%%%%%%%%%%%%%%%%%%%%%%%%%%%%%%%%%%%%%%%%%%%%%%%%%%%%%%%%%%%%%%%%%%

\title{Growth rate of self-sustained QED cascades induced by intense lasers}

\author{A. Mercuri-Baron}\thanks{These two authors contributed equally. \\ \url{anthony.mercuri@protonmail.com} \\ \url{mironov.hep@gmail.com}} 
\author{A.~A. Mironov}\thanks{These two authors contributed equally. \\ \url{anthony.mercuri@protonmail.com} \\ \url{mironov.hep@gmail.com}} 
\affiliation{LULI, Sorbonne Université, CNRS, CEA, École Polytechnique, Institut Polytechnique de Paris, F-75255 Paris, France}
\author{C. Riconda}
\affiliation{LULI, Sorbonne Université, CNRS, CEA, École Polytechnique, Institut Polytechnique de Paris, F-75255 Paris, France}
\author{A. Grassi}
\affiliation{LULI, Sorbonne Université, CNRS, CEA, École Polytechnique, Institut Polytechnique de Paris, F-75255 Paris, France}
\author{M. Grech}
\affiliation{LULI, CNRS, CEA, Sorbonne Universit\'{e}, École Polytechnique, Institut Polytechnique de Paris, F-91128 Palaiseau, France}

\begin{abstract}
	It was suggested [A. R. Bell \& J. G. Kirk, PRL 101, 200403 (2008)] that an avalanche of electron-positron pairs can be triggered in the laboratory by a standing wave generated by intense laser fields. Here, we present a general solution to the long-standing problem of the avalanche growth rate calculation.  We provide a simple formula that accounts for the damping of the growth rate due to pair and photon migration from the region of prolific generation. We apply our model to a variety of 3D field configurations including focused laser beams and show that i) the particle yield for the full range of intensity able to generate an avalanche can be predicted, ii) a critical intensity threshold due to migration is identified, iii) the effect of migration is negligible at a higher intensity and the local growth rate dominates. Excellent agreement with Monte Carlo and self-consistent PIC simulations is shown. The growth rate calculation allows us to predict when abundant pair production will induce a back-reaction on the generating field due to plasma collective effects and screening. Our model can be applied to study the generation of electron-positron pair avalanches in realistic fields to plan future experiments at ultra-high-intensity laser facilities.
\end{abstract}
\maketitle

\section{Introduction}

Relativistic electron-positron pair (or QED) plasma is a state of matter that is believed to be responsible for multiple striking and yet not fully explained astrophysical phenomena. It can be generated in the vicinity of compact objects such as black holes \cite{blandford1977electromagnetic, ford2018electron} or neutron stars \cite{michel1999electrodynamics} and pulsars \cite{goldreich1969pulsar, arons1979some, gueroult2019determining}. Interactions with such QED plasma can be the source of prominent hard cosmic radiation in Gamma Ray Bursts and in bright gamma flashes from relativistic jets \cite{piran2005physics, chang2008long, spitkovsky2008particle, kumar2015physics}. Abundant production of $e^-e^+$ pairs can take place in cascade processes developing in polar caps of a rotating compact star \cite{ford2018electron, timokhin2019maximum}. This mechanism opens a path to explain the nature of radio-pulsar emission \cite{cruz2021kinetic,cruz2022model} and the source for plasma populating the magnetosphere of a star and the magnetic reconnection layer \cite{cerutti2013simulations, philippov2014ab, hu2022axisymmetric}.

A laboratory study of dense relativistic electron-positron pair plasma would facilitate a breakthrough in our understanding of the impact of QED effects in astrophysical phenomena. However, the generation of such plasma appears to be exceptionally challenging \cite{chen2023perspectives, qu2023creating}. The first observation of a neutral $e^-e^+$ plasma state was reported only a few years ago \cite{sarri2015generation}, and the first investigation of collective behaviour was done very recently \cite{arrowsmith2023laboratory}. Still, in both cases, the density of electrons and positrons, generated via the Bethe-Heitler process, barely reached the value high enough to form a plasma. 

A prospective path to obtaining $e^-e^+$ plasma in the laboratory lies in using super-strong electromagnetic (EM) fields, for example, generated with ultra-high intensity lasers \cite{ridgers2013dense,narozhny2014creation, sarri2015overview, zhang2020relativistic, qu2021signature}. When interacting with elementary particles, such fields can induce a wide variety of nonlinear strong-field QED (SFQED) phenomena by sharing $N\ggg1$ soft photons $\gamma_L$ with $e^\pm$ \cite{fedotov2022high, gonoskov2022charged, popruzhenko2023dynamics, di2012extremely, ruffini2010electron}. The two leading order effects are the nonlinear inverse Compton scattering (for brevity, Compton emission) $e^\pm+N\gamma_L \rightarrow e^\pm+\gamma$ and the nonlinear Breit-Wheeler pair production\footnote{For brevity, we omit the word `nonlinear' throughout the paper when referring to both these processes.} 
$\gamma+N\gamma_L\rightarrow e^-+e^+$. The strong-field QED treatment of these processes is essential when the field experienced by a relativistic particle in its rest frame\footnote{For a photon of frequency $\omega_\gamma$ in the laboratory frame, the analogous frame can be defined as the one where $\hbar\omega'_\gamma=mc^2$.} is comparable to the critical field of QED $E_S=m^2c^3/(e\hbar)$, where $m$ and $-e$ are the electron mass and charge, respectively. Sequential Compton emissions and Breit-Wheeler pair productions can lead to cascades.

Cascades can be generated as \textit{shower-type} events by incident high-energy particles in a strong field \cite{gaisser2016cosmic, sokolov2010pair, bulanov2013electromagnetic}. According to the estimates \cite{blackburn2017scaling, qu2021signature, mercuri2021impact}, the shower particle yield is defined by the energy input from the incident bunch. The configuration envisioned in this scheme is similar to Bethe-Heitler process-based setups \cite{zerby1963studies}, with the only difference being that in the former the field is provided by high-Z nuclei. In the classical limit (namely, in weak fields or at low particle energy), the Compton emission describes classical radiation by a charge, while the Breit-Wheeler process is suppressed \cite{ritus1985}. This naturally limits the shower multiplicity.

\begin{figure}%%%
	\includegraphics[width=\linewidth]{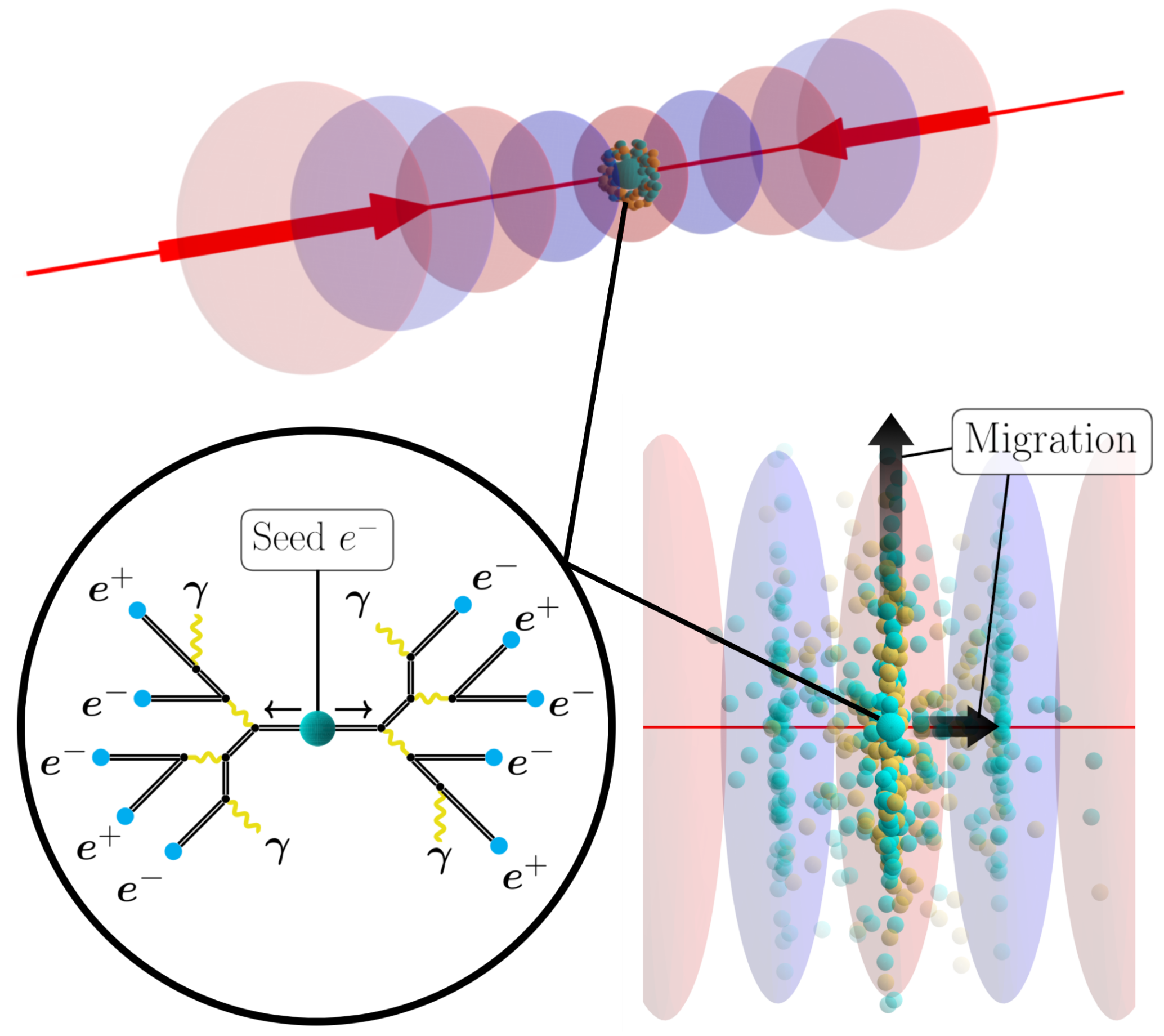}
	\caption{Schematic representation of an avalanche-type cascade developing at the mutual focus of two counterpropagating laser beams. Blue and yellow balls show $e^\pm$ and $\gamma$, respectively, while red and blue blobs depict the electric and magnetic antinodes of the standing-wave field. The initial electron is injected into the central $E$-antinode at the focus. The electric field accelerates the initial and secondary $e^\pm$ in the direction transverse to the optical axis. This acceleration renders the onset and further development of the cascade with prolific production of $e^+e^-$ pairs in the linked nonlinear Compton emission and Breit-Wheeler processes. As the avalanche develops, particles can migrate to the regions with a weaker electric field where the acceleration is less efficient, which reduces the cascade growth rate.} 
	\label{fig:avalanche}
\end{figure}

Strong EM fields can induce a different phenomenon originally predicted by Bell \& Kirk \cite{Kirk2008}: electron-seeded \textit{avalanche-type} (or self-sustained) QED cascades. They can be triggered by low-energy electrons injected into the strong field region (illustrated in Fig.~\ref{fig:avalanche}). The initial and secondary charged particles experience ongoing acceleration by the field, which restores their energy in between hard photon emissions, therefore sustaining the cascade. In an avalanche triggered even by few seed electrons, the number of produced $e^-e^+$ pairs can rapidly exponentiate with time and field strength \cite{Kirk2009, fedotov2010limitations}. The process is accompanied by a bright gamma-flash \cite{gonoskov2017ultrabright}. Such cascades induced by lasers can hence mimic the processes developing in the polar caps of rotating neutron stars \cite{ford2018electron, timokhin2019maximum, cruz2021kinetic}.

An avalanche-type cascade can onset in alternating electric fields that are strong enough \cite{Arseny2021}. The suitable configurations include electric antinodes of standing waves formed by two counterpropagating laser beams of an optical frequency and circularly \cite{Kirk2009, elkina2011qed, mironov2014collapse, grismayer2017seeded} or linearly \cite{bashmakov2014effect, Jirka2016} polarized; a multi-beam setup \cite{gelfer2015optimized, marklund2023towards} and a dipole wave as its limiting case \cite{gonoskov2017ultrabright}; a single ultra-intense beam focused in vacuum \cite{Arseny2021} or reflected from a plasma mirror \cite{vincenti2023plasma}; irradiation of a solid \cite{ridgers2012dense, jirka2017qed, Slade2019}, a plasma slab \cite{zhu2016dense, samsonov2021hydrodynamical}, or a gas target \cite{artemenko2017ionization, Tamburini_2017}, vortex laser pulses \cite{jirka2024effects} (for a more detailed review see Refs.~\cite{fedotov2022high, gonoskov2022charged, popruzhenko2023dynamics}). Simulations show that an electron-positron plasma can be generated in a single laser shot and reach a high density demonstrating collective effects \cite{narozhny2014creation,lobet2015shock,yu2018qed, luo2018qed} or even screening the field \cite{nerush2011analytical, ridgers2012dense, Grismayer2016, grismayer2017seeded, efimenko2018extreme}.

The estimated field needed to trigger an $e^-$-seeded avalanche-type cascade is at the scale of $\alpha E_S$ \cite{fedotov2010limitations}, where $\alpha=e^2/(4\pi\varepsilon_0 \hbar c)\approx 1/137$ is the fine structure constant. More detailed simulations for the mentioned setups show that the required intensity is of the order $10^{24}$ W/cm$^2$ for femtosecond optical lasers. Intensities approaching this value are anticipated to be within the reach of the new generation of multi-PW laser facilities, such as Apollon \cite{papadopoulos2019first}, ELI-Beamlines \cite{weber_mre2017},  CoReLS \cite{corels}, and more worldwide \cite{tanaka_mre2020, astragemini, gan_optexpr2017, pearl, danson_hplse2019}. The record intensity of $\approx 10^{23}$ W/cm$^2$ was recently reported  \cite{yoon2021realization}, and increasing efforts are invested to reach $10^{24}$--$10^{25}$ W/cm$^2$ in the near future \cite{bromage2019technology, nsf_opal, khazanov2023exawatt, shao2020broad}.

In this work, we address two general questions: (i)~What is the threshold and the scaling of the particle yield with the field strength in an avalanche-type cascade developing in a realistic field configuration? and (ii)~What are the requirements to reach plasma densities in such cascades? Assuming that initial electrons are already injected in the strong field, the cascade development splits into the onset and exponential phases, with the majority of particles produced in the latter. While the general onset conditions seem clear from the theoretical viewpoint \cite{Arseny2021}, a universal model for the exponential growth rate of particles is lacking. 

Most of the progresses in the avalanche-type cascade theory were made for a uniform rotating electric field (we will refer to it as `rotating $E$-field'). It corresponds to the electric antinode ($E$-antinode) of a circularly polarized (CP) standing wave formed by two counterpropagating laser beams. Cascades in this configuration were studied in the limiting cases of relatively low and high field strength \cite{fedotov2010limitations, elkina2011qed, Grismayer2016, bashmakov2014effect, grismayer2017seeded, kostyukov2018growth}, $E\llgg \alpha E_S$. When it comes to more realistic field configurations, the analytical predictions are lacking for the whole range of setup parameters. In the current work, we address this problem in general terms.
	
The exponential cut-off for the Breit-Wheeler process at low fields and photon energies render the cascade onset threshold. However, the threshold is not well-defined and specific to the field configuration. In particular, it depends on the dynamics of particles that can escape from the strong field region, i.e. the laser focus or the $E$-antinode in a standing wave  (as illustrated in Fig.~\ref{fig:avalanche}) \cite{Esirkepov2014, King2016classical, jirka2017qed, Tamburini_2017}. Particle migration can reduce the overall $e^-e^+$ pair yield \cite{Jirka2016}, and as we argue in this work, can induce a hard threshold for the cascade onset.

Under the semiclassic approximation, the kinetic approach provides a general basis for describing the particle distribution evolution in cascades \cite{elkina2011qed,nerush2011analytical, Niel2018, fauth2021collisional}. Indeed, common numerical approaches solve kinetic equations in the Monte Carlo scheme or particle-in-cell codes. The latter allows accounting for plasma effects by consistent treatment of the Vlasov and Maxwell equations \cite{gonoskov2015extended}. Notably, a kinetic description also allows to account for spin effects \cite{king2013photon, seipt2023kinetic, zhao2023cascade}. 

We adopt the kinetic approach for an ab initio analytical derivation of the particle growth rate expression in an avalanche-type cascade and rigorously define the approximations in use for a generic field configuration. It allows us to account for particle migration and highlight its effect on the growth rate. Then we investigate several field configurations through the optics of our approach: We first consider in detail cascades in an infinite CP standing wave, and then study more complex configurations combining 1, 2, and even 8 strongly focused laser beams.

We leave aside the nontrivial question of injecting the initial electrons \cite{sampath2018towards, jirka2021reaching} and assume that they are located in the strong field region. Let us mention that seeding can be implemented by using a high-energy electron bunch \cite{mironov2014collapse,mironov2016generation}, ionizing a jet of high-Z atoms \cite{artemenko2017ionization}, or solid targets \cite{ridgers2013dense, jirka2017qed}. At extreme fields, seed electrons can be created due to Schwinger $e^-e^+$ pair production, which can become feasible due to the large volume of the focal spot \cite{bulanov2010multiple, marklund2023towards}. As we aim at finding the field scaling of the avalanche-type cascade growth rate, for simplicity, we omit the Schwinger effect, as well as spin contribution in the current work.

The paper is organized as follows. At the beginning of Sec.~\ref{sec:ii} we introduce our notations and give an overview of the approximations in use. In Sec.~\ref{sec:ii_a} we present the probability rates of the basic processes. In Sec.~\ref{sec:ii_b}, following Ref.~\cite{Arseny2021}, we postulate the semiclassic short-time dynamics of a single electron in a general accelerating field. This is then used in Sec.~\ref{sec:ii_c} to calculate the characteristic time between Compton emissions for the electron and to find the key parameters defining the emission process. In Sec.~\ref{sec:ii_d} we discuss the related asymptotic limits. Section~\ref{sec:iii} is devoted to our major analytical results: In Sec.~\ref{sec:iii_a} we introduce the master equations for the particle numbers and the general formula for the avalanche-type cascade growth rate that takes into account particle migration; In Sec.~\ref{sec:iii_b} we derive the effective model for the growth rate. In Sec.~\ref{sec:iii_c}, we discuss its low-field limit and estimate the threshold field parameters required to trigger a cascade in a general field. The high-field limit is discussed in Sec.~\ref{sec:iii_d}. In Sec.~\ref{sec:iv}, we apply our model to study avalanches developing in a CP standing EM wave and test the model against simulations performed with the PIC-QED code SMILEI \cite{derouillat2018smilei}. Section~\ref{sec:iv_a} contains details about the numerical setup and data analysis. In Sec.~\ref{sec:iv_b}, we present the physical results and validate our model at relatively low- and mid-range fields. Section~\ref{sec:iv_c} continues this discussion for high fields. Section \ref{sec:v} is devoted to cascades developing in more intricate field models combining several focused laser beams: In Sec. \ref{sec:v_a} we study the impact of transverse migration in a standing wave formed by two Gaussian beams, in Sec. \ref{sec:v_b} calculate the particle yield in cascades triggered in a single strongly focused beam, and in Sec. \ref{sec:v_c} discuss cascades in a multi-beam setup combining 8 optimally polarized focused laser beams.
In Sec.~\ref{sec:vi} we apply our model to identify the field parameters at which the plasma state can be reached in avalanche-type cascades in the above-listed field configurations and test the model against PIC simulations. In Sec.~\ref{sec:conclusions} we summarise and discuss our results.

\section{Particle dynamics in a cascade}
\label{sec:ii}
The interaction of a relativistic particle (electron, positron, or photon) with a strong EM field can be characterized by two parameters. First, by the particle Lorentz factor $\gamma_e$ for $e^\pm$ or by the normalized energy $\gamma_\gamma=\hbar\omega_\gamma/mc^2$ for a photon with frequency $\omega_\gamma$. Second, by the invariant quantum dynamical parameter combining the electron\footnote{Hereinafter, we treat electrons and positrons identically, and for brevity refer to both as `electrons' unless mentioned explicitly.} or photon momentum $\vec{p}_{e,\gamma}$ and the electric and magnetic field components $\vec{E}$ and $\vec{B}$:
\begin{eqnarray}
	\label{eq:chi}
	\chi_{e,\gamma} = \dfrac{1}{E_S }\,\sqrt{ \left(\gamma_{e,\gamma}  \vec{E} + c \vec{u}_{e,\gamma} \times  \vec{B} \right)^2 - \left(\vec{u}_{e,\gamma}\cdot \vec{E}\right)^2 },
\end{eqnarray}
where $\vec{u}_{e,\gamma}=\vec{p}_{e,\gamma}/(mc)$. For an electron, $\chi_e$ equals the rest frame field strength in the units of the QED critical field $E_S = m^2c^3/(e \hbar)$. 

Parameter \eqref{eq:chi} allows to assess when the quantum effects in the interaction of a particle with an external EM field need to be considered. They become important at $\chi_{e,\gamma}\gtrsim 1$. As mentioned in the introduction, we consider two leading order field-induced effects: the nonlinear Compton emission and Breit-Wheeler processes. In this work, we adopt the probability rates for these processes within the locally constant field approximation (LCFA) \cite{nikishov1964quantum, nikishov1967pair, Baier1968, ritus1985}. It implies that the rates depend only on the field strength at the local position of the incoming relativistic particle, and the field is close to the constant crossed EM field in the particle reference frame  \cite{ritus1985, fedotov2022high}. Furthermore, in a quantum process, the produced and scattered particles propagate collinearly to the incoming one.

It should be noted, that the LCFA breaks if one of the particles in the process is nonrelativistic or the transverse field seen by the particle is relatively low \cite{gelfer2022nonlinear} (see more discussions in Refs.~\cite{DiPiazza2018_LCFA,di2019improved,ilderton2019extended,raicher2019semiclassical, heinzl2020locally, king2020uniform, di2021wkb, nielsen2022high}. However, the LCFA is reasonable for modelling avalanche-type cascades, as under optimal conditions for their onset, such events are rare and also do not contribute to the cascade particle growth rate \cite{gelfer2022nonlinear, Arseny2021}.

\subsection{Photon emission and pair creation probability rates}
\label{sec:ii_a}
The differential probability rate for an electron with the Lorentz factor $\gamma_e$ and quantum parameter $\chi_e$ to emit a photon with energy within the interval $mc^2\times(\gamma_\gamma, \gamma_\gamma+\rmd\gamma_\gamma)$ is given by 
\begin{widetext}
\begin{gather}
	\label{eq:rate_CS}
	\frac{\rmd \WCS (\gamma_{e}, \chi_{e}, \gamma_\gamma)}{\rmd \xi} = \frac{1}{\sqrt{3}\pi} \dfrac{\alpha}{ \tau_C \gamma_{e}}   \left[ \left(1-\xi+\frac{1}{1-\xi}\right) K_{2/3}(\mu)-\int_{\mu}^\infty \rmd s K_{1/3}(s)   \right] ,
\end{gather}
where $\tau_C=\hbar/(mc^2)$ is the Compton time, and we defined $\xi=\gamma_\gamma/\gamma_e\equiv \chi_\gamma/\chi_e$, $\mu = 2\xi/[3\chi_{e}(1 - \xi)]$. $K_{n}$ are the $n$-th order modified Bessel functions of the second kind. Note that the $\chi$-parameter is conserved, $\chi_e=\chi_\gamma+\chi_e'$, where $\chi_e'$ corresponds to the scattered electron. 
The differential probability rate for a photon with $\gamma_\gamma$ and $\chi_\gamma$ to produce a pair of $e^-e^+$, so that either $e^-$ or $e^+$ gets energy within the interval $mc^2\times(\gamma_e,\gamma_e+\rmd \gamma_e)$, reads:
\begin{equation}
	\label{eq:rate_BW}
	\frac{\rmd \WBW(\chi_{\gamma}, \gamma_{\gamma}, \gamma_e)}{\rmd\zeta}  = \frac{1}{\sqrt{3}\pi} \dfrac{\alpha}{ \tau_C \gamma_{\gamma}}  \left[\left( \frac{\zeta}{1-\zeta}+\frac{1-\zeta}{\zeta}\right)K_{2/3}(\mu') -\int_{\mu'}^\infty \rmd s K_{1/3}(s) \right], 
\end{equation}
where $\zeta=\gamma_e/\gamma_\gamma\equiv\chi_e/\chi_\gamma$, $\mu' = 2 \zeta/[3\chi_{\gamma} (1 - \zeta)]$, and $\chi_\gamma=\chi_e+\chi_e'$. Here, $\chi_e$ and $\chi_e'$ correspond to the electron and positron. 

\end{widetext}
The total probability rates $\WCS (\gamma_{e}, \chi_{e})$, $\WBW(\chi_{\gamma}, \gamma_{\gamma})$  are obtained by integrating out the final particle energies or, equivalently, $\xi$ and $\zeta$: 
\begin{gather}
	\label{eq:rate_CS_tot}
	\WCS(\gamma_{e}, \chi_{e})=\int_0^1 \frac{\rmd \WCS (\gamma_{e}, \chi_{e}, \gamma_\gamma)}{\rmd \xi}\rmd \xi,\\
	\label{eq:rate_BW_tot}
	\WBW(\chi_{\gamma}, \gamma_{\gamma})=\int_0^1 \frac{\rmd \WBW(\chi_{\gamma}, \gamma_{\gamma}, \gamma_e)}{\rmd\zeta} \rmd\zeta
\end{gather}
In the asymptotic case, they can be simplified:
\begin{gather}
	\renewcommand{\arraystretch}{1.5}
	\label{eq:rate_CS_asympt}
	\WCS(\gamma_{e}, \chi_{e}) \approx \frac{\alpha }{\tau_C\gamma_e}\times\left\lbrace
	\begin{array}{ll}
		1.44\,\chi_e, & \chi_e\ll 1, \\
		1.46\,\chi_e^{2/3}, & \chi_e\gg 1,
	\end{array}\right.\\
	\renewcommand{\arraystretch}{1.5}
	\label{eq:rate_BW_asympt}
	\WBW(\chi_{\gamma}, \gamma_{\gamma}) \approx \frac{\alpha}{\tau_C\gamma_\gamma} \times \left\lbrace
	\begin{array}{ll}
		0.23\,\chi_\gamma e^{-\frac{8}{3\chi_\gamma}}, & \chi_\gamma\ll 1, \\
		0.38\,\chi_\gamma^{2/3}, & \chi_\gamma\gg 1.
	\end{array}\right.
\end{gather}

The probability rates and the product spectrum width grow with the $\chi$-parameter of the incoming particle (if the Lorentz factor is fixed). On the other hand, the $\chi$-parameter is shared among the products in each quantum event, so its value decreases for subsequent particle generation. As a result, if a high-energy particle in a laser field triggers a cascade, at some point it will stop unless the $\chi$-parameter increases again during the particle propagation in between the quantum events  \cite{bulanov2013electromagnetic,mironov2014collapse}. For this reason, the multiplicity of shower-type cascades is defined by the energy of incident particles, as $\chi$ decreases for each generation of secondary particles.

\subsection{$\chi_e$ time dependence in accelerating fields}
\label{sec:ii_b}
The key idea of the avalanche-type cascade mechanism is that in between emission events electrons are re-accelerated by the field, which restores a high value of $\chi_e$. Then the cascade will be sustained as long as particles experience a strong field. The mean free path time of an electron or a photon in an external EM field can be estimated as the inverse total probability rate $\WCSBW^{-1}$. Let $\omega^{-1}$ be the time scale of the field variation. For a particle in a prolific cascade, we expect that $\omega \WCSBW^{-1}<1$ \cite{Kirk2008,fedotov2010limitations}, i.e. multiple quantum events take place during a field cycle. The cascade overall properties, such as the particle number growth rate and spectra, are defined by the single-particle $\chi_e$ evolution at the time scale $\omega t\ll 1$. 

Let us consider the motion of an electron injected in a generic strong field. Fields of electric type, $E>cB$, are favourable for the acceleration of charges and can restore $\gamma_e$ and $\chi_e$. Within the semiclassic approximation, the evolution of $\gamma_e$ and $\chi_e$ can be calculated from the particle trajectory given by the Lorentz equations. Following Ref. \cite{Arseny2021}, they can be solved perturbatively for a generic inhomogeneous time-dependent field of electric type at the time scale $\omega t\ll 1$ is the expansion parameter. In the leading order, the solution corresponds to a constant homogeneous background and is formulated in terms of the invariant field strength at the initial position of the electron:
\begin{equation}
    \label{eq:eps_definition1}
    \begin{split}
    &\epsilon = \sqrt{\sqrt{ \mathcal{F}^2 + \mathcal{G}^2 } + \mathcal{F}},\\
    &\eta = \sqrt{\sqrt{ \mathcal{F}^2 + \mathcal{G}^2 } - \mathcal{F}},
    \end{split} 
\end{equation}
where $\mathcal{F}=(\vec{E}^2 - c^2\vec{B}^2)/2$ and $\mathcal{G}=c\vec{E}\cdot\vec{B}$. The higher-order corrections in $\omega t$ can be calculated systematically. More details can be found in Appendix \ref{sec:appendix_weff} and Ref. \cite{Arseny2021}. Let us note here that Eq.~\eqref{eq:eps_definition1} defines the electric and magnetic fields, correspondingly, in the frame where the electric and magnetic components are parallel or one of them vanishes \cite{taub1948orbits, ritus1985}. 

For further analysis, it is enough to consider the leading terms for $\gamma_e(t)$ and $\chi_e(t)$. For the former we have:
\begin{equation}
    \label{eq:gamma_short_time}
    \gamma_{e}(t)  \simeq \dfrac{e\epsilon  t}{mc}.   
\end{equation}
The first nontrivial contribution to $\chi_e(t)$ results from the solution to the order $O((\omega t)^2)$ and reads:
\begin{equation}
    \chi_{e}(t) \simeq  \dfrac{\epsilon^2 \omega_{\rm eff}}{E_S^2\tau_C} t^2 \label{eq:chi_short_time},
\end{equation}
Here, the effective frequency $\omega_{\rm eff}\propto\omega$ represents the time and length scale of the field variation. The full expression for $\omega_{\rm eff}$ is presented in Appendix~\ref{sec:appendix_weff}. The effective frequency $\omega_{\rm eff}$ is defined via the local field derivatives and vanishes for a constant field. The combination $\oeff t^2$ is Lorentz-invariant.

Equations~\eqref{eq:gamma_short_time} and \eqref{eq:chi_short_time} are general for a wide class of EM fields, and are valid under the following assumptions:
\begin{enumerate}
	\item[(i)] \textit{The particle dynamics is semiclassic}, namely, photons propagate along straight lines and electron motion satisfies the Lorentz equations. This holds for subcritical fields, $E\ll E_S$, which is relevant for laser beams and astrophysical applications;
	\item[(ii)] The field components at the initial position of the electron satisfy the condition $\epsilon>0$, which applies to fields of \textit{electric type} $E>cB$, considered in this work. 
	\item[(iii)] By the time moment of emission $\WCS^{-1}\ll \omega^{-1}$, the \textit{energy gained by the electron from the field is high} compared to its initial energy (i.e. particles rapidly `forget' their initial conditions). 
\end{enumerate}
Then the electron trajectory can be approximated and substituted in Eq.~\eqref{eq:chi}, which results in Eqs.~\eqref{eq:chi_short_time}. We provide some details about the derivation in Appendix~\ref{sec:appendix_weff}, which follows Ref.~\cite{Arseny2021}.

One of the suitable configurations is the electric antinode of a standing wave formed by two counter-propagating circularly polarized (CP) laser beams \cite{Kirk2009,fedotov2010limitations,nerush2011analytical,elkina2011qed,bashmakov2014effect,grismayer2017seeded,jirka2017qed,Slade2019}, where the magnetic field is absent and the electric component is close to a uniform rotating electric field:
\begin{equation}
	\label{eq:rotating_E}
	\vec{E}=E_0\left[\cos(\omega t)\boldsymbol{\hat{x}}+\sin(\omega t)\boldsymbol{\hat{y}} \right],
\end{equation}
with $E_z=0$, $\vec{B}=0$. In this case, one has $\epsilon=E_0$, $\omega_{\rm eff}=\omega/2$, and Eqs.~\eqref{eq:gamma_short_time}-\eqref{eq:chi_short_time} read \cite{fedotov2010limitations, elkina2011qed, nerush2011analytical}:
\begin{equation}
	\label{eq:gamma_chi_tem_rotating}
	\gamma_{e}(t)  \simeq \frac{e E_0 t}{mc}, \quad
	\chi_{e}(t) \simeq   \frac{E_0^2 \omega t^2}{2 E_S^2 \tau_C}.
\end{equation}

\subsection{Characteristic $\gamma_e$ and $\chi_e$ at emission}
\label{sec:ii_c}
Once the evolution of $\chi_e$ for the accelerated electron is determined, it can be used to calculate the time instant of photon emission and the corresponding values of $\gamma_e$ and $\chi_e$ at this moment. These quantities identify whether the electron-seeded cascade can onset \cite{Arseny2021} and the particle growth rate (as we will show in Sec.~\ref{sec:iii}).

In a prolific avalanche-type cascade, the number of particles rapidly exponentiates with time and therefore is sensitive to small variations of the parameters that enter in the growth rate. For this reason, in our model, we go beyond the simple estimate for the electron mean free path time as the inverse emission rate, $\WCS^{-1}$. Let us introduce the characteristic time of electron propagation between photon emissions $t_{\mathrm{em}}$:
\begin{equation}\label{eq:tem_def}
	\int_0^{t_{\mathrm{em}}}  \WCS(t') \rmd t' := 1,
\end{equation}
where $\WCS(t)=\WCS(\gamma_e(t),\chi_e(t))$, and $\gamma_e(t)$ and $\chi_e(t)$ are given by Eqs.~\eqref{eq:gamma_short_time}-\eqref{eq:chi_short_time}. It is convenient to introduce the corresponding characteristic values:
\begin{equation}
	\label{eq:gamma_chi_em}
	\gamma_\mathrm{em}=\gamma_e(\tem),\quad \chi_\mathrm{em}=\chi_e(\tem),
\end{equation}
obtained by substituting $\tem$ into Eqs.~\eqref{eq:gamma_short_time}-\eqref{eq:chi_short_time}. As we will show, the cascading regime is characterised by $\cem$. Let us stress here, that $\tem$, $\gem$, and $\cem$ depend only on the field parameters via $\epsilon$ and $\oeff$:
\begin{equation}
	\label{eq:mapping_eps_cem}
	\epsilon,\,\oeff \mapsto \{\tem,\, \gem,\, \cem\}. 
\end{equation}

In view of Eqs.~\eqref{eq:gamma_short_time}-\eqref{eq:chi_short_time}, $\chi_\mathrm{em}$ can be chosen as an independent variable instead of $\tem$ (assuming that $\epsilon$ and $\omega_\mathrm{eff}$ are nonzero). In particular, by inverting Eq.~\eqref{eq:chi_short_time} and passing to the corresponding variable in Eq.~\eqref{eq:tem_def}, we can rewrite the latter in the Lorentz-invariant form:
\begin{equation}
	\label{eq:chi_em_invariant_def}
	\int_0^{\chi_{\mathrm{em}}}d\chi\, \frac{\tau_C \WCS(1,\chi)}{2\alpha \chi}=\frac{\epsilon}{\alpha E_S},
\end{equation}
where we moved the physical parameters to the RHS, and the integrand in the LHS is a dimensionless special function of $\chi$ (note that the dependence on $\alpha$ cancels in the LHS as $\WCS\propto\alpha$). This expression defines $\chi_\mathrm{em}$ (that is Lorentz invariant) in the electron reference frame. As one may notice, $\chi_\mathrm{em}$ is independent of $\omega_{\rm eff}$.

\subsection{Weak and high field regimes of photon emission by accelerated electrons}
\label{sec:ii_d}
Formula \eqref{eq:chi_em_invariant_def} establishes the general functional dependence $\cem=\cem(\epsilon/\alpha E_S)$. At $\cem\ll 1$ or $\gg 1$ Eq.~\eqref{eq:chi_em_invariant_def} simplifies, as we can use the asymptotic expressions \eqref{eq:rate_CS_asympt} for $\WCS$ to approximate the integral. The result sets the direct correspondence 
\begin{equation}
\label{eq:low_high_chi_field_correspondence}
\cem \llgg 1 \, \Longleftrightarrow \, \epsilon \llgg \alpha E_S,
\end{equation}
so that
\begin{equation}
	\label{eq:chi_em_asympt}
	\cem \simeq \left\lbrace
	\begin{array}{ll}
		1.39 \dfrac{\epsilon }{\alpha E_S}, &  \epsilon\ll \alpha E_S ,\\
		0.87  \left(\dfrac{\epsilon}{\alpha E_S}\right)^{3/2}, &  \epsilon\gg \alpha E_S.\\
	\end{array}
	\right.
\end{equation}
Relation~\eqref{eq:low_high_chi_field_correspondence} defines the weak and strong field regime of an avalanche-type cascade, with $\alpha E_S$ being the threshold. 

Analogously to Eq.~\eqref{eq:chi_em_asympt}, we can also calculate the asymptotic behavior of $\tem$ and $\gem$. The resulting expressions with the full numerical coefficients are presented in Appendix~\ref{sec:appendix_asymot}. 

\begin{figure}%%%
	\includegraphics[width=\linewidth]{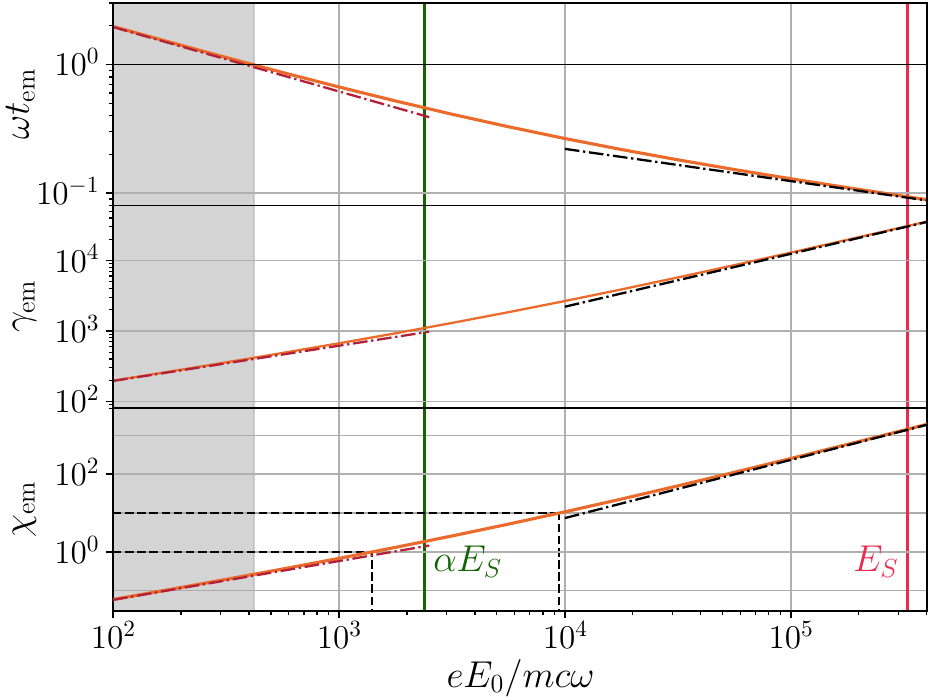}
	\caption{The emission characteristic time $\tem$ and the corresponding values $\gem$ and $\cem$ [see Eqs.~\eqref{eq:tem_def}-\eqref{eq:gamma_chi_em}] as functions of $E_0$ for an electron-seeded cascade in a uniform rotating electric field (solid lines). The dot-dashed lines show the low- and high-field asymptotic behavior. The short-time dynamics model given in Eqs.~\eqref{eq:gamma_short_time}-\eqref{eq:chi_short_time} is valid in the unshaded area, where $\omega \tem <1$. Thin dashed lines in the bottom panel show the field strength at which $\cem=1$ and $10$.  The vertical green line $E_0=\alpha E_S$ corresponds to $eE_0/(mc\omega)\approx2400$ and the vertical red line $E_0=E_S$ --- to $eE_0/(mc\omega)\approx 3.3\times 10^5$. The field rotation frequency $\omega$ corresponds to the wavelength $\lambda=0.8\mu$m.} 
	\label{fig:tem}
\end{figure}

For illustration, in Fig.~\ref{fig:tem}, we show the dependence of $\tem$, $\gem$, and $\cem$ on the field strength for an electron accelerated by a rotating $E$-field [Eq.~\eqref{eq:rotating_E}]. As $\tem$ and $\cem$ are defined via the full expression \eqref{eq:rate_CS_tot} for $\WCS$ [see Eqs.~\eqref{eq:tem_def}, \eqref{eq:chi_em_invariant_def}], we calculate them by numerical integration. We confirm that our initial assumption $\omega \tem < 1$, required in Eqs.~\eqref{eq:gamma_short_time}-\eqref{eq:chi_short_time}, is met in a wide range of $E_0$ and improves with a growing field strength. We also plot the asymptotic expressions for $\cem$ [given in Eq.~\eqref{eq:chi_em_asympt}], as well as for $\tem$ and $\gem$ (see Appendix~\ref{sec:appendix_asymot}). As expected, the low and high field asymptotics set in at $E_0\ll \alpha E_S$ and $\gg \alpha E_S$, respectively. Finally, let us note that in the transition regime $E_0\sim\alpha E_S$, $\cem$ takes the value $1\lesssim \cem < 10$. This interval will also correspond to the transition between the models for the cascade growth rate, which we present in the following Section.

\section{Steady-state particle growth rate of a cascade in a general field}
\label{sec:iii}
Let us now address the evolution of the particle number in an avalanche-type cascade and its scaling with the field strength, which is the central question of this work. We consider a field that meets the validity conditions for Eqs.~\eqref{eq:gamma_short_time}, \eqref{eq:chi_short_time} in a localized region. In the context of a standing EM wave, this can be an electric antinode, with the limitation that particles can be expelled (or `migrate') from the region of a strong electric field. In this section, we discuss the master equations for the particle number evolution and take this effect into account.

We show explicitly by an ab initio calculation that the master equations, proposed phenomenologically in the absence of migration in  Ref.~\cite{bashmakov2014effect} and later used in Refs.~\cite{grismayer2017seeded, luo2018qed}, are exact if the photon emission and pair creation rates entering in the equations are calculated by averaging over the particle distributions. 
However in general the analytic expression for the distribution functions and hence the rates are unknown. Here, we propose an approximation for the latter based on the small-time electron dynamics discussed in the previous Section. As a result, this allows us to solve the master equations and find the explicit scaling of the cascade growth rate with the field strength.

\subsection{Cascade growth rate and onset threshold in a steady state}
\label{sec:iii_a}
The onset stage of the avalanche-type cascade is highly nonstationary. The seed particle distribution rapidly evolves as new $e^-e^+$ pairs and photons are produced. Provided that the particle yield is high at the scale of field duration, the particle energy quickly averages out as the cascade develops. The cascade can enter the so-called (quasi-)steady state, in which the energy distribution relaxes to a stationary function \cite{elkina2011qed,grismayer2017seeded}.  We assume that the system quickly reaches this state (which we further confirm with numerical simulations), where the particle number grows exponentially at a constant rate. However, depending on the field strength, the relaxation can take up to multiple field periods. Hence, depending on the field configuration, electrons and photons can migrate from the strong-field area, which competes with the pair creation process and can reduce the overall particle production rate.

The steady-state master equations for the number of pairs $N_p$ and photons $N_\gamma$ generated in a cascade can be derived \textit{exactly} by using the kinetic approach, as we show in Appendix~\ref{sec:appendix_kinetic}:

\begin{eqnarray}
	\label{eq:const_rate_eq1}
	\dfrac{\rmd N_{p}}{\rmd t} &=& W_{\mathrm{cr}}N_{\gamma} - \nu_e N_{p},\\
	\label{eq:const_rate_eq2}
	\dfrac{\rmd N_{\gamma}}{\rmd t} &=& - W_{\mathrm{cr}} N_{\gamma} + 2 W_{\mathrm{rad}} N_{p} - \nu_\gamma N_\gamma,
\end{eqnarray}
where $\Wradcr$ are the effective constant rates of photon emission and pair creation, respectively, and the migration rates $\nu_{e,\gamma}$ represent the average stationary flow of electrons and photons leaving the cascading region. $\Wradcr$ result from averaging $\WCSBW(\gamma,\chi)$ over the steady-state particle distributions $f_{p,\gamma}(\gamma,\chi)$:
\begin{equation}
\label{eq:rates_average}
\Wradcr = \langle \WCSBW(\gamma,\chi )\rangle.
\end{equation}
In the derivation of Eqs.~\eqref{eq:const_rate_eq1}-\eqref{eq:rates_average}, we rely only on the semiclassic approximation and the LCFA. These equations are valid for a general field that allows the formation of a steady state. If the strong-field region is finite in space (e.g. laser field focus), $N_{p,\gamma}$ should be interpreted as local particle numbers in this region. In the presence of migration, depending on the setup parameters, $N_{p,\gamma}$ can either increase or decrease with time.

Eqs.~\eqref{eq:const_rate_eq1}-\eqref{eq:const_rate_eq2} can be solved by the substitution $e^{\lambda t}$, resulting in two eigenvalues $\lambda$. The largest of them, denoted by $\Gamma$, can take positive values. Hence, the particle number can increase exponentially with time, $N_{p,\gamma}(t) \sim e^{\Gamma t}$. In effect, the cascade growth rate reads:
\begin{equation}\label{eq:growth_rate_formula}
	\begin{split}
	&\Gamma [\Wrad, \,  \Wcr,\,
	\nu_e,\,\nu_\gamma] = \dfrac{W_{\mathrm{cr}} + \nu_e + \nu_\gamma}{2} \\
	 &\quad\quad\times \left[-1+\sqrt{1 + 4\frac{W_{\mathrm{cr}}(2W_{\mathrm{rad}} - \nu_e) -\nu_e\nu_\gamma}{(W_{\mathrm{cr}} + \nu_e + \nu_\gamma)^2}} \right].
	\end{split}
\end{equation}
For a given EM field configuration, $\Wradcr$ and $\nu_{e,\gamma}$ depend on the field parameters, and hence $\Gamma$ too.

A calculation of the effective rates $\Wradcr$ and $\nu_{e,\gamma}$ requires averaging over the particle distribution functions $f_{p,\gamma}$, which are generally unknown. However, they can be extracted from simulations or estimated under additional approximations.
The rate of migration from a region of size $\sim 2\pi c/\omega$ can be estimated as $\nu_{e,\gamma}\sim\omega/(2\pi)$. For higher precision calculations we use the numerical data. Anticipating further discussion in Section~\ref{sec:iv}, let us note here that $\nu_{e,\gamma}$ scale weakly with the field strength, and in many cases it is sufficient to use an order of magnitude estimate. In contrast to this, $\Wradcr$ strongly depend on the field parameters. In what follows, we propose a refined analytical model for $\Wradcr$, which is valid in a wide class of EM field models.

If $\nu_e$ and $\nu_\gamma$ take relatively high values, the square root term in Eq.~\eqref{eq:growth_rate_formula} can become smaller than 1 and, hence, $\Gamma\leq 0$. A negative growth rate corresponds to a situation when the particle average flow from the finite-sized strong-field region exceeds the overall pair creation rate in this area. The condition $\Gamma=0$ determines a well-defined \textit{critical}-like threshold in the parameter space for the cascade onset. In particular, if $\Gamma$ can be explicitly written as a function of the invariant field strength $\epsilon$, we can define the critical value:
\begin{equation}
	\label{eq:eps_c_def}
	\epsilon^*:\,\, \Gamma(\epsilon^*)=0.
\end{equation}

Notice that if one or both migration rates $\nu_{e,\gamma}$ are negligible, electrons or/and photons can stay in a strong-field region for a long time. In this case, as follows from Eq.~\eqref{eq:growth_rate_formula}, $\Gamma$ is always positive [we take into account that $\Wrad>\omega/(2\pi)\gtrsim \nu_e$, see Eq.~\eqref{eq:rate_CS_asympt}]. Then a cascade can onset even in a weak field, although this can require an exponentially large time. This type of threshold can be viewed as a smooth \textit{crossover}.

\subsection{Effective emission and pair creation}
\label{sec:iii_b}
In this subsection, we propose an approximation for the rates $\Wradcr$, that will lead to an explicit formula for the growth rate $\Gamma$ in terms of the relevant parameters. We adapt the general idea of Ref.~\cite{fedotov2010limitations}, where it was proposed to estimate $\Wrad$ at the characteristic $\gamma_e$ and $\chi_e$ gained by $e^-$ in a uniform rotating field before the instant of emission. We go beyond this simple estimate as we (i) consider a general accelerating field [in the context of Eqs.~\eqref{eq:gamma_short_time}-\eqref{eq:chi_short_time} validity], (ii) use the refined expressions for $\gem$ and $\cem$ to describe the radiating $e^-$, which we propose in Section~\ref{sec:ii_c}, and (iii) show that depending on the field strength, different parts of the photon distribution need to be considered.

To evaluate the effective emission rate $W_{\mathrm{rad}}$ by an electron contributing to an avalanche-type cascade, we plug the characteristic values for $\gamma_e\sim\gamma_{\mathrm{em}}$ and $\chi_e\sim\chi_{\mathrm{em}}$ at the time of emission [see Eqs.~\eqref{eq:tem_def}-\eqref{eq:gamma_chi_em}] into the full probability rate for the Compton emission [Eq.~\eqref{eq:rate_CS_tot}]:
\begin{equation}
	\label{eq:Wrad_at_tem}
	W_{\mathrm{rad}} \simeq  \WCS(\gamma_{\mathrm{em}}, \chi_{\mathrm{em}}).
\end{equation}
Note that in a strong field such that $\cem\gg 1$, the field dependence of the emission probability rate can be written explicitly [see also Eq.~\eqref{eq:rate_CS_tem} in Appendix~\ref{sec:appendix_asymot}]:
\begin{equation}
\label{eq:rate_CS_tem_text}
\WCS(\chi_{\mathrm{em}}\gg 1)\approx  1.43\, \alpha \sqrt{\frac{\oeff}{\tau_C}}\left(\dfrac{\epsilon}{\alpha E_S}\right)^{1/4}.
\end{equation}

The estimate for the effective pair creation rate $W_{\mathrm{cr}}$ is less straightforward. Unlike for electrons, $\chi_\gamma$ can vary only at the scale $\sim\omega^{-1}$ as the photon traverses through the field. In a cascade, the emission and pair creation processes happen successively. To select the photons contributing the most, we should consider the interplay between two factors: (i) the pair creation is exponentially suppressed for $\chi_\gamma<1$ but contributes at $\chi_\gamma\gtrsim1$ [see Eq.~\eqref{eq:rate_BW_asympt}], and (ii) in the emission spectrum, softer photons with $\chi_\gamma\ll \chi_e$ dominate over the hardest ones with $\chi_\gamma\sim\chi_e$. Furthermore, one should note that $\gamma_\gamma$ and $\chi_\gamma$ are related, since $\gamma_e$ and $\chi_e$ of the emitting electron are mutually dependent in view of Eqs.~\eqref{eq:gamma_short_time}-\eqref{eq:chi_short_time}. 

\begin{figure}%%%
        \begin{center}
        \includegraphics[width=\linewidth]{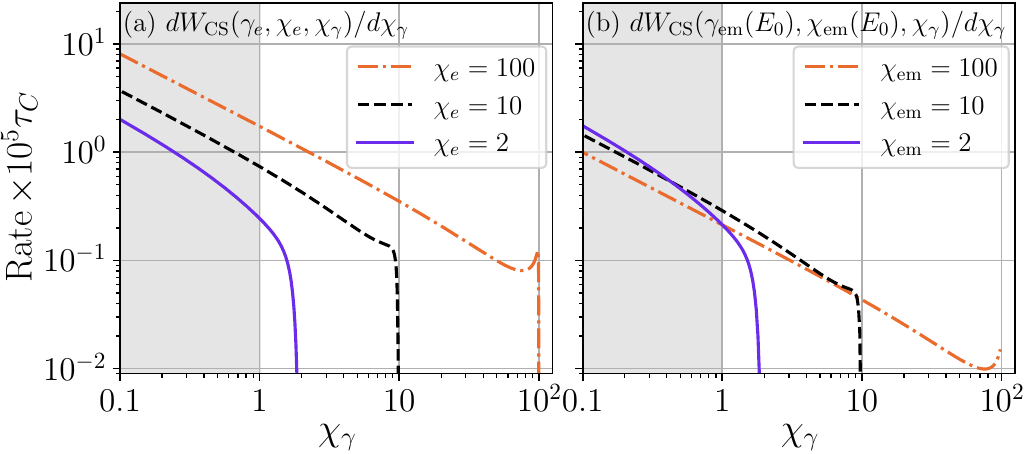}\\
        \includegraphics[width=0.58\linewidth]{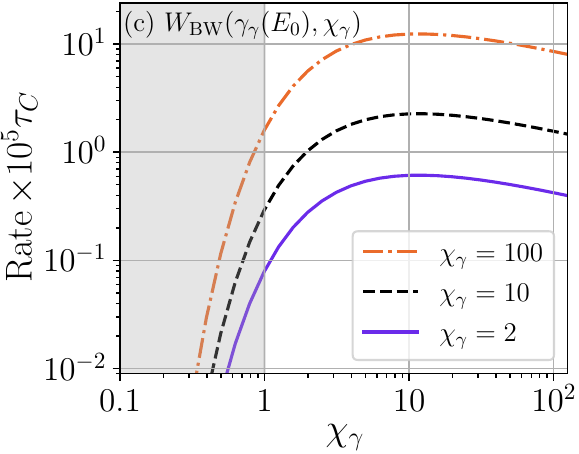}            
        \end{center}
	\caption{The spectrum of photons emitted by an electron [(a) and (b)] [see Eq.~\eqref{eq:rate_CS}]  and the total probability rate of $e^-e^+$ pair creation by a photon (c) [see Eq.~\eqref{eq:rate_BW_tot}] as a function of $\chi_\gamma$. Panel (a) corresponds to emission by an electron with $\gamma_e=10^3$ and different values of $\chi_e$; panel (b) --- to emission by an electron accelerated by a rotating $E$-field, so that $\gamma_e=\gem$ and $\chi_e=\cem$ [see Eqs.~\eqref{eq:gamma_chi_tem_rotating}, \eqref{eq:tem_def}]. Panel (c) shows $\WBW$ for a photon emitted in a rotating $E$-field, so that $\gamma_\gamma$ and $\chi_\gamma$ are consistent, $\gamma_\gamma\sim \chi_\gamma E_S/E_0$. In panels (b) and (c) the following values were taken for $eE_0/(mc\omega)$: 2550 (solid), 9436 (dashed), and 51581 (dot-dashed), which gives $\cem = 2$, $10$, and $100$, respectively, for the field wavelength $\lambda=2\pi c/\omega=0.8\mu$m. In the grey-shaded area $\chi_\gamma<1$, the pair creation probability is suppressed [see Eq.~\eqref{eq:rate_BW_asympt}]. }
	\label{fig:CS_spectra}
\end{figure}

In Fig.~\ref{fig:CS_spectra}, we show examples of the emission spectra for two cases: for electrons with fixed $\gamma_e$ and $\chi_e$, and for electrons accelerated by a rotating $E$-field so that $\gem$ and $\cem$ are consistent with Eqs.~\eqref{eq:gamma_chi_tem_rotating} and \eqref{eq:tem_def}. We compare these curves to the probability rate of $e^-e^+$ pair creation by a photon in a rotating electric field. For this, we estimated $\chi_\gamma\sim \gamma_\gamma \epsilon/E_S$. The result suggests that at relatively low and high fields the dominating contribution to pair production is rendered by {\it{different parts of the emission spectrum}}. Let us discuss these cases one by one.

At a relatively low field, such that $\cem$ barely exceeds 1, only the rightmost edge of the photon $\chi_\gamma$ distribution can contribute to the cascade (see the $\chi_e=2$ line in Fig.~\ref{fig:CS_spectra}). Then the pair creation rate can be estimated as
\begin{equation}
	\label{eq:Wcr_model_low_cem}
	\left .W_{\mathrm{cr}}\right|_{\cem<\mathcal{X}} \sim \WBW(\gamma_\gamma=\gem,\, \chi_\gamma=\cem),
\end{equation}
where $1\lesssim\mathcal{X}<10$ sets the effective applicability range of the formula. We discuss it at the end of this subsection. Note that this is consistent with the (rough) cascade threshold condition proposed in  Refs.~\cite{fedotov2010limitations,Arseny2021}. Namely, $\cem\gtrsim 1$ is the necessary requirement ensuring that the emitted photon can have $\chi_\gamma\sim 1$ and, hence, create a pair in a successive quantum event.

In a high field, electrons emit at $\cem\gg 1$, as we discussed in Section~\ref{sec:ii_d}. All of the photons with $1\lesssim \chi_\gamma\leq\cem$ can contribute to pair production, which represents a wide part of the distribution (see the orange lines in Fig.~\ref{fig:CS_spectra} corresponding to $\cem=100$). Furthermore, the fraction of photons with $\chi_\gamma\sim 1$ significantly exceeds that of $\chi_\gamma\sim \cem$, which is not compensated for by the (slower) increase of the pair creation rate. Therefore, we evaluate the pair creation rate as
\begin{equation}
	\label{eq:Wcr_model_high_cem}
	\begin{split}
		\left.W_{\mathrm{cr}}\right|_{\cem>\mathcal{X}} 
		\sim\, & \WBW(\gamma_\gamma=\frac{E_S}{\beta\epsilon},\chi_\gamma=1)\\
		&=\beta\dfrac{\epsilon}{E_S}\WBW(1,1).
	\end{split}
\end{equation}
Here, we estimated $\gamma_\gamma$ from $\chi_\gamma\sim \beta\gamma_\gamma \epsilon/E_S:=1$. In the second line, we used that $\WBW$ depends on $\gamma_\gamma$ only in the prefactor [see Eq.~\eqref{eq:rate_BW}]. The geometric coefficient $\beta$ is determined by the characteristic angle between the photon momentum and the field at the instance of pair creation. We will assume that $\beta\sim 1$ for the sake of an estimate unless noted explicitly. 

Summarising, $\Gamma$ depends on the (invariant) field strength $\epsilon$ entering the effective rates parametrically via $\gem(\epsilon)$, $\cem(\epsilon)$. This dependence can be written for a general field by plugging the effective rates into Eq.~\eqref{eq:growth_rate_formula}:
\begin{align}
	\label{eq:growth_rate_low_cem}
	& \left.\Gamma\right|_{\cem<\mathcal{X}}(\epsilon) \simeq \Gamma[\WCS(\gem,\cem),\WBW(\gem,\cem)],\\
	\label{eq:growth_rate_high_cem}
	& \left.\Gamma\right|_{\cem>\mathcal{X}}(\epsilon) \simeq \Gamma[\WCS(\gem,\cem),\beta\frac{\epsilon}{E_S}\WBW(1,1)].
\end{align}
In these expressions, the rates $\WCS$ and $\WBW$ can be calculated numerically by using Eqs.~\eqref{eq:rate_CS}-\eqref{eq:rate_BW_tot}, and $\nu_e$ can be extracted from simulations. We require that Eqs.~\eqref{eq:growth_rate_low_cem} and \eqref{eq:growth_rate_high_cem} match at $\cem\sim\mathcal{X}$. This condition can be reexpressed in terms of the field strength in view of Eq.~\eqref{eq:chi_em_invariant_def}. 

It is important to note that the crossover region for Eqs.~\eqref{eq:growth_rate_low_cem}-\eqref{eq:growth_rate_high_cem} is finite, which is due to the following reason. The transition between the approximations for $\Wcr$ in Eqs.~\eqref{eq:Wcr_model_low_cem} and \eqref{eq:Wcr_model_high_cem} is smeared, as for the electrons radiating with $1\lesssim \cem <10$ the relative decrease of the number of emitted photons with $\chi_\gamma\sim 1$ and $\chi_\gamma\sim\cem$ can be comparable [see the black curves in Fig.~\ref{fig:CS_spectra}, corresponding to $\cem=10$]. The photons with $\chi_\gamma\sim 1$ start to dominate in the spectrum at relatively high $\cem\gtrsim 10$. 
%As a result, we can estimate the transition point $\mathcal{X}$ by picking a value from the interval $1\lesssim \mathcal{X}< 10$. Alternatively, it can be evaluated with higher precision by extracting  Eqs.~\eqref{eq:growth_rate_low_cem} and \eqref{eq:growth_rate_high_cem} from simulations and matching them.

To summarize, in order to obtain the particle number growth rate $\Gamma$ in a cascade, one should (i) calculate the local values for $\epsilon$ and $\oeff$ [see Eqs.~\eqref{eq:eps_definition1}, \eqref{eq:omega_eff}], (ii) evaluate $\gem$ and $\cem$ by using Eqs.~\eqref{eq:tem_def}, \eqref{eq:gamma_short_time}, and \eqref{eq:chi_short_time}, (iii) calculate the probability rates $\Wradcr$, (iv) evaluate or estimate the migration rates $\nu_{e,\gamma}$, and, finally, (v) use Eq.~\eqref{eq:growth_rate_formula} to obtain $\Gamma$ with appropriate values [leading to Eq. \eqref{eq:growth_rate_low_cem} or \eqref{eq:growth_rate_high_cem}]. The integration in steps (ii) and (iii) can be done numerically. The analytic derivation of $\oeff$ is possible even for realistic field models, however, the calculation can be cumbersome. Moreover, some field models may be accessible only numerically. We developed a dedicated computational package that allows to calculate all the listed quantities numerically (including $\oeff$) based on a given numerical data for the field distribution in time and space. The package was developed in Julia \cite{bezanson2017julia} and is available online \cite{PairAvalanchesQED}.

\subsection{$\Gamma$ at low field, the cascade threshold}
\label{sec:iii_c}
Let us discuss the behavior of the growth rate in the limiting cases of a low and high field. We define a low/high field by the correspondence given in Eq.~\eqref{eq:low_high_chi_field_correspondence}. In a weak field $\epsilon\ll \alpha E_S$, i.e. at $\cem\ll 1$, the emitted photons are soft, $\chi_\gamma\ll 1$, and the pair creation rate is exponentially small, see Eq.~\eqref{eq:rate_BW_asympt}. At the same time, the probability for an electron to emit a soft photon stays finite. 
Assuming that $\nu_{e,\gamma}\sim\omega/(2\pi)$, let us expand Eq.~\eqref{eq:growth_rate_low_cem} at small $\Wcr\ll \nu_{e,\gamma},\, |\nu_e-\nu_\gamma|,\,  \Wrad$ to the leading order:
\begin{equation}
	\label{eq:growth_rate_low_eps_nu}
	\begin{split}
		&\left.\Gamma\right|_{\nu_{e}\neq\nu_\gamma}  (\epsilon\ll \alpha E_S)  \simeq -\frac{\nu_e+\nu_\gamma -|\Delta\nu|}{2}\\
		&\quad + \frac{\WBW(\gem,\cem)}{2}\\
		&\quad\quad \times \left[-1+ \frac{\left(4\WCS(\gem,\cem) - \Delta\nu \right) }{ |\Delta\nu| }\right].
	\end{split}
\end{equation}
where $\Delta\nu=\nu_e-\nu_\gamma$. This expansion is also valid if one (and only one) of the migration rates is zero. If $\nu_e=\nu_\gamma=\nu$, the expansion reads:
\begin{equation}
\label{eq:growth_rate_low_eps_nu_zero}
\begin{split}
 \left.\Gamma\right|_{\nu_e=\nu_\gamma} & (\epsilon \ll \alpha E_S) \simeq\\
& -\nu+\sqrt{2\WCS(\gem,\cem)\WBW(\gem,\cem)}.
\end{split}
\end{equation}
Note that, as mentioned above, at $\nu=0$ the growth rate is positive, $\Gamma \propto e^{-4/(3\cem)}$ (c.f. Eq. (8) in \cite{grismayer2017seeded}). Particle migration leads to significant suppression of the cascade growth rate in the low-field regime. The field dependence at low $\epsilon$ can be found explicitly by using the low-$\chi$ asymptotic of Eqs.~\eqref{eq:rate_CS_asympt}, \eqref{eq:rate_BW_asympt} and the low-$\epsilon$ expansion for $\gem(\epsilon)$ and $\cem(\epsilon)$ [see also Eqs.~\eqref{eq:gamma_tem_asympt_exact}, \eqref{eq:chi_tem_asympt_exact}]:
\begin{equation}
		\label{eq:growth_rate_low_eps_field_dep}
		\begin{split}
			&\Gamma(\epsilon\ll \alpha E_S)  \approx -\dfrac{\nu_e+\nu_\gamma-|\Delta\nu|}{2} \\
			& \quad\quad + \frac{\alpha}{\tau_C} 
			\left\lbrace
			\begin{array}{ll}
				1.52 \dfrac{\epsilon\oeff }{ E_S |\Delta\nu| } e^{-1.92 \frac{\alpha E_S}{\epsilon}}, & |\Delta \nu|>0,\\
				1.23\sqrt{\dfrac{\epsilon\oeff \tau_C }{\alpha E_S}}\, e^{-0.96 \frac{\alpha E_S}{\epsilon}}, & \nu_e=\nu_\gamma.
			\end{array}
			\right.
		\end{split}
\end{equation}
The second line is also valid for $\nu_e=\nu_\gamma=0$. The asymptotic expressions with full numerical coefficients are given in Appendix~\ref{sec:appendix_asymot}, see Eq.~\eqref{eq:gamma_all_asympt}.

Let us estimate the threshold field strength defined by Eq.~\eqref{eq:eps_c_def} from Eq.~\eqref{eq:growth_rate_low_eps_field_dep}. The equation $\Gamma(\epsilon^*)=0$ can be solved explicitly in terms of the Lambert function $W_0$:
\begin{eqnarray}
	\label{eq:eps_c_estimate}
	\epsilon^*\simeq 1.92 \alpha E_S\left[W_0\left( 5.84\dfrac{\alpha^2 \oeff}{\tau_C \tilde{\nu}^2} \right)\right]^{-1},
\end{eqnarray}
where we introduced
\begin{eqnarray}
	\nonumber
	\tilde{\nu}^2=\left\lbrace\begin{array}{ll}
		|\Delta\nu| (\nu_e+\nu_\gamma-|\Delta\nu|), & |\Delta\nu|>0,\\
		2\nu^2, & \nu_e=\nu_\gamma=\nu.
	\end{array}
	\right.
\end{eqnarray}
For example, for a CP standing wave obtained via two optical ($\lambda=0.8\mu$m) counter-propagating laser beams focused to a diffraction limit, such that $\nu\approx \omega/(2\pi)$ and $\oeff=\omega/2$, $\epsilon^*\equiv E^*$, the critical field strength reads $E^*\approx 0.36\alpha E_S$ or $eE^*/(mc\omega)\approx 880$ in dimensionless units. We will discuss the cascade threshold for this setup in more detail in the following. For field configurations such that $\nu_{e,\gamma}\gg\oeff$,
%(e.g. a single weakly focused laser beam, see further discussion and Ref.~\cite{Arseny2021})
migration suppresses the cascade formation and the critical field is proportional to $\epsilon^*\propto \nu_{e,\gamma}^2/\oeff$. As expected, at vanishing migration rates, $\nu_{e,\gamma}\rightarrow0$, the critical field strength also tends to zero as $\epsilon^*\propto 1/|\ln\nu_{e,\gamma}|$.

\subsection{$\Gamma$ at high fields}
\label{sec:iii_d}
At high field $\epsilon>\alpha E_S$, namely, in the regime when Eq.~\eqref{eq:growth_rate_high_cem} for $\Gamma$ applies, the probability rates increase as $\Wrad \propto \epsilon^{1/4}$ and $\Wcr \propto \epsilon$ [see Eqs.~\eqref{eq:rate_CS_tem_text} and \eqref{eq:Wcr_model_high_cem}]. The migration rates are bounded, $\nu_{e,\gamma}\lesssim\omega/(2\pi)$, hence, as the field grows, we can neglect them compared to $\Wrad$ and $\Wcr$ in Eq.~\eqref{eq:growth_rate_high_cem}: 
\begin{equation}
	\label{eq:growth_rate_formula_high_field}
	\begin{split}
	\Gamma & (\epsilon> \alpha E_S) \simeq \\
	& \dfrac{\epsilon\WBW(1,1)}{2E_S}\left[\sqrt{1 + \frac{8E_S\WCS(\gem,\cem)}{\epsilon\WBW(1,1)}} -1 \right]. 
	\end{split}
\end{equation}
It means that at high $\epsilon$, regardless of the global field structure in space, the cascade will be effectively confined to the strong field region, where the particle production is the most efficient. Put differently, only the local structure of the field, where the conditions are optimal, affects the cascade growth rate. This effect is universal for the field configurations, in which avalanche cascades can onset and reach a steady state. 

By substituting in Eq.~\eqref{eq:growth_rate_formula_high_field} the asymptotic expressions for the rates given by Eqs.~\eqref{eq:rate_CS_tem_text}, \eqref{eq:Wcr_model_high_cem}, we obtain the explicit field dependence of the growth rate in the high-field regime:
\begin{equation}
\label{eq:growth_rate_formula_high_field_explicit}
\begin{split}
\Gamma(\epsilon> & \alpha E_S) \simeq  \\
 &\dfrac{ c_1\epsilon}{\tau_C E_S} \left[ \sqrt{1+ c_2\sqrt{\oeff \tau_C}\left(\dfrac{\alpha E_S}{\epsilon} \right)^{3/4}} -1\right],
\end{split}
\end{equation}
where the numerical coefficients $c_1\approx 0.52\times 10^{-4}$, $c_2\approx1.1\times 10^5$ are defined in Eqs.~\eqref{eq:c1}, \eqref{eq:c2} in Appendix~\ref{sec:appendix_asymot}. Here, we note that their magnitude is determined by $\tau_C\WBW(1,1)\approx 1.03\times 10^{-4}$.

Since $\Wrad\propto\epsilon^{1/4}$  grows slower with $\epsilon$ as compared to $\Wcr\propto\epsilon$, at asymptotically high fields (comparable to $E_S$) the latter will become so large that $8\Wrad/\Wcr\ll 1$, and Eqs.~\eqref{eq:growth_rate_formula_high_field}, \eqref{eq:growth_rate_formula_high_field_explicit} simplify even further:
\begin{equation}
	\begin{split}
	\Gamma(\epsilon\gg \alpha E_S) \simeq &\, 2\WCS(\gem,\cem) \\ 
	&\propto \alpha\sqrt{\frac{\oeff}{\tau_C}}\left(\frac{\epsilon}{\alpha E_S}\right)^{1/4}.
	\end{split}
\end{equation}
For the rotating electric field configuration [see Eq.~\eqref{eq:rotating_E}] this corresponds to the scaling proposed by Fedotov et. al. \cite{fedotov2010limitations} (recall that for this case $\epsilon=E_0$, $\oeff=\omega/2$):
\begin{equation}
	\label{eq:growth_rate_Fedotov}
	\Gamma_{\text{rot.-$E$}}(E_0\gg \alpha E_S) \sim \alpha\sqrt{\frac{\omega}{\tau_C}}\left(\frac{E_0}{\alpha E_S}\right)^{1/4}.
\end{equation}
The condition $\Wcr> 8\Wrad$ for this asymptotic to set in corresponds to extreme values of the field. By plugging in the explicit high-field expressions \eqref{eq:rate_CS_tem_text} and \eqref{eq:Wcr_model_high_cem} for the rates, we arrive at the condition $E_0> 3.35\times 10^6 \alpha (\omega \tau_C)^{2/3}$. For the optical wavelength $\lambda=0.8\mu$m, this corresponds to the field strength $E_0 > 5 E_S$, which is beyond the applicability range of the semiclassic approach. Notably, the scaling \eqref{eq:growth_rate_Fedotov} was never confirmed by simulations in past works \cite{elkina2011qed,bashmakov2014effect,grismayer2017seeded}, as this range of field strengths was not studied.

\section{Cascades in a CP standing wave}
\label{sec:iv}
Let us now apply our general framework for studying the development of $e^-$-seeded cascades in various field configurations. We will gradually increase the complexity of the field model, test the growth rate formulae \eqref{eq:growth_rate_low_cem}, \eqref{eq:growth_rate_high_cem} against simulations, and identify the cascade threshold and regimes in realistic configurations.

\begin{figure}%%%
	\includegraphics[width=\linewidth]{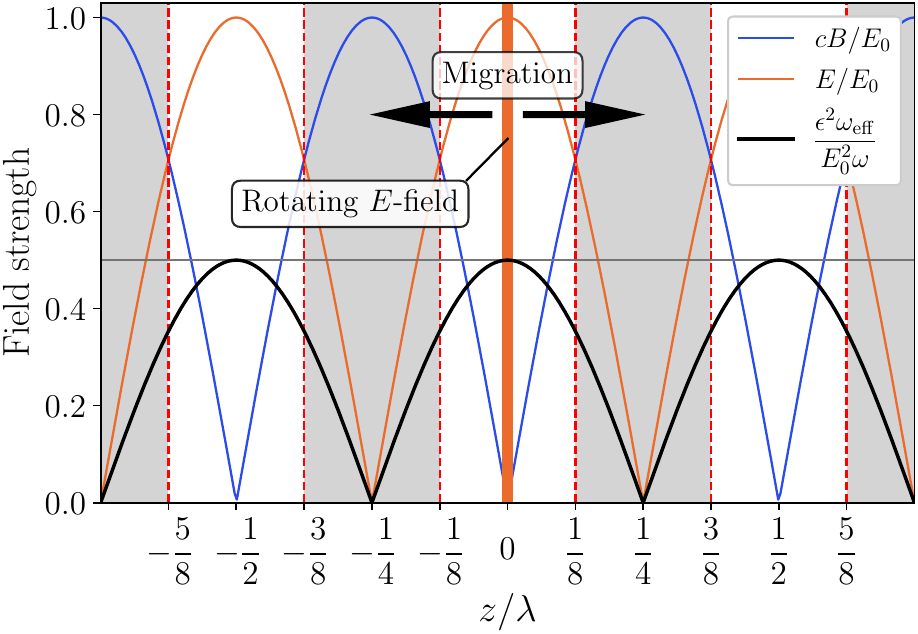}
	\caption{The longitudinal coordinate dependence for the field amplitude and $\epsilon^2 \omega_{\rm eff}$ in a CP standing wave [see Eqs. \eqref{eq:standing_wave_E_field}-\eqref{eq:standing_wave_B_field}]. The conditions for the cascade development are optimal near the $E$-antinode center (shown by the thick vertical line), where $B=0$ and the $E$-field has a constant amplitude and rotates in the transverse plane [as described in Eq.~\eqref{eq:rotating_E}], so that $\epsilon^2 \omega_{\rm eff}$ is maximal [see also Eq.~\eqref{eq:chi_short_time}]. As the cascade develops, $e^\pm$ migrate from the $E$-antinode to the grey-shaded areas at characteristic time $\nu_e^{-1}\sim 2\pi/\omega$ (shown by arrows) \cite{lehman2012,Esirkepov2014,Gong2016_attractors,Kirk2016,King2016classical}.}
	\label{fig:fields_invariant_1D_standing}
\end{figure}

First, let us consider the rotating $E$-field and CP EM standing wave configurations. Let us stress that the rotating $E$-field is not just an idealized model but the limiting solution for different configurations at high fields. The latter can be formed by two counterstreaming CP EM plane waves (with the opposite sign of polarization) so that the total electric and magnetic field components read:
\begin{align}
	&\vec{E} = E_0 \left[ \cos (kz) \sin (\omega t) \boldsymbol{\hat{x}} + \cos(kz) \cos(\omega t) \boldsymbol{\hat{y}}   \right], \label{eq:standing_wave_E_field} \\
	&\vec{B} = -\dfrac{E_0}{c}\left[ \sin (kz) \sin (\omega t) \boldsymbol{\hat{x}} +  \sin (kz) \cos (\omega t) \boldsymbol{\hat{y}}    \right] \, , \label{eq:standing_wave_B_field}
\end{align}
where $k = \omega/c$, $z$ is the axis of propagation, and $E_{0}$ is the electric field amplitude. The distribution of the electric and magnetic field strength along the longitudinal axis is illustrated in Fig. \ref{fig:fields_invariant_1D_standing}. At the electric antinode center, the magnetic field vanishes, and the electric field corresponds to the rotating $E$-field given in Eq.~\eqref{eq:rotating_E}. The values of $\epsilon$ and $\oeff$ depend on the longitudinal coordinate. At the $E$-antinode centre ($z=0$ in Fig.~\ref{fig:fields_invariant_1D_standing}), $\epsilon=E_0$ and $\oeff=\omega/2$ as in a rotating $E$-field, and the combination $\epsilon^2 \omega_{\rm eff}$ is maximal. In the $B$-antinodes (e.g. $\lambda/8<z<3\lambda/8$), $\epsilon^2 \omega_{\rm eff}$ is smaller and null at the center, so that electrons do not gain $\chi_e$ efficiently [see Eq.~\eqref{eq:chi_short_time}] and the cascade is not sustained in these regions.

As the cascade onset conditions are optimal near the $E$-antinode centers, the rotating $E$-field can serve as a reference field model. Formally this corresponds to assuming that particle migration in the cascade is zero, $\nu_e=\nu_\gamma=0$. However, it is known that $e^\pm$ injected into the $E$-antinode of a standing wave will migrate longitudinally to the $B$-antinodes at the rate $\nu_e\sim\omega/(2\pi)$, as they are the spiralling attractors for charges \cite{lehman2012,Esirkepov2014,Gong2016_attractors,Kirk2016,King2016classical}, and where the conditions are not favourable for cascade development. A comparison of the cascade growth rates calculated for both field models (rotating $E$-field or CP EM standing wave) allows to quantify the effect of $e^\pm$ migration. Notice that in both cases we will still assume that $\nu_\gamma=0$, as photons are emitted predominantly in the transverse direction. We will discuss the effect of transverse migration in the context of focused fields in the next section. 

\subsection{PIC simulations}
\label{sec:iv_a}
To study the development of electron-seeded cascades numerically, we perform simulations with the PIC-QED code SMILEI \cite{derouillat2018smilei}. For the rotating $E$-field and CP standing wave configurations, we use 1D3V geometry. For comparison, in this section, we will also consider the results of full 3D PIC-QED simulations for a realistic setup where the standing wave is formed by two counterpropagating Gaussian beams with relatively large waist $w_0=3\lambda$.

The numerical setup is detailed in Appendix~\ref{sec:appendix_sim_setup}, though, let us mention some important points. The seed electrons are injected at time $t=0$ into the electric antinode center $z=0$ in the simulations with a standing wave, and distributed homogeneously in the simulation box for the rotating $E$-field configuration. We fix the field frequency so that it corresponds to the optical laser wavelength $\lambda=0.8\mu$m, and we study the development of avalanches with time at varying $E_0$. The particle density is kept low during the whole simulation so that collective effects are insignificant.
 
To compare $\Gamma$ [given in Eq.~\eqref{eq:growth_rate_formula}] in the standing wave configuration with the PIC results, we use in the formula the migration rate field dependence $\nu_e(E_0)$ calculated from numerical simulations as follows. As the cascade develops, the initial and secondary electrons will migrate longitudinally to the $B$-antinodes. Let us consider a particle as dropped out of the cascade if it reaches a $B$-antinode, in particular a spatial region at $|z|>\lambda/8$. In light of this, we define $\nu_e$ as the average inverse time at which the initial particles propagate at distance $\lambda/8$ from the origin. The rate $\nu_e$ can be extracted from simulations of single particle motion (for more details, refer to Appendix ~\ref{sec:appendix_sim_setup}). We anticipate that as the $E_0$ grows $\nu_e$ varies slowly from a maximum value of $2\pi/\omega$ to a fraction of this value and that $\Gamma$ is not very sensitive to small variations of this parameter. Thus a simple estimate can be obtained by considering $\nu_e=2\pi /\omega$.

In each simulation, we ensure that the cascade reaches a steady state so that it exhibits exponential growth of the particle number. We extract the growth rate by using the two following methods. First, we directly calculate $\Gamma$ by fitting the $e^-e^+$ pair number time dependence $N_{p}(t)$ with the exponential function $e^{\Gamma t}$ (the full procedure is described in Appendix~\ref{sec:appendix_sim_setup}). Second, as discussed in Section~\ref{sec:iii_a}, we calculate the exact growth rate given in Eq.~\eqref{eq:growth_rate_formula} with the effective rates $\Wradcr$ evaluated as prescribed by Eq.~\eqref{eq:rates_average}. We extract the steady-state distribution functions $f_{e,\gamma}(\gamma_{e,\gamma},\chi_{e,\gamma})$ from the numerical data and calculate the corresponding values $\langle \WCSBW(\gamma,\chi )\rangle$ [see also Eqs.~\eqref{eq:Wrad_stationary}-\eqref{eq:Wcr_stationary} in Appendix~\ref{sec:appendix_kinetic}]. Then we plug this result and the migration rate $\nu_e(E_0)$ (calculated as explained above) in Eq.~\eqref{eq:growth_rate_formula} to obtain $\Gamma$.

\begin{figure}\centering%%%
	\includegraphics[width=\linewidth]{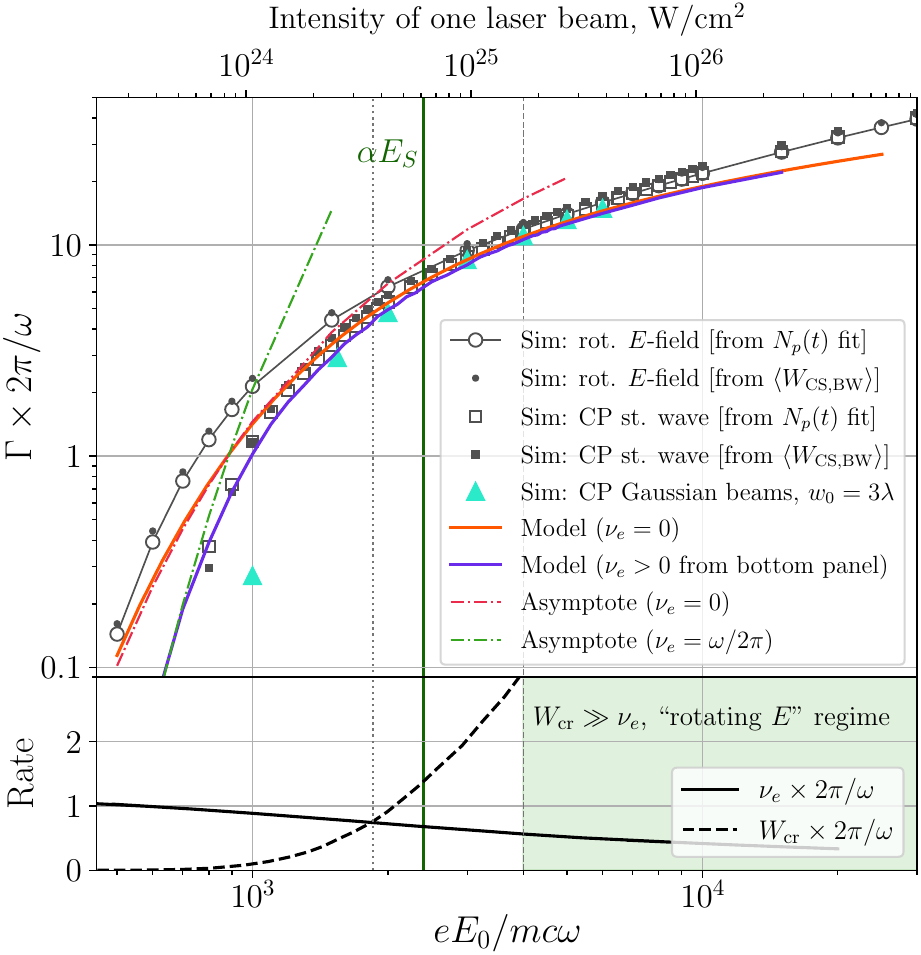}
	\caption{Top panel: the growth rate of an avalanche-type cascade as a function of the field amplitude $E_0$ in different field configurations. We plot the results of simulation obtained with SMILEI PIC for: CP standing wave (1D3V), a uniform rotating electric field (1D3V), CP standing wave formed by two Gaussian beams with $w_0=3\lambda$ (full 3D simulation). Empty circles and squares correspond to $\Gamma$ extracted from the exponential fit of the pair number time dependence $N_p(t)$ [see Appendix~\ref{sec:appendix_sim_setup}]. Small filled circles and squares correspond to Eq.~\eqref{eq:growth_rate_formula} with $\Wradcr=\langle \WCSBW(\gamma,\chi)\rangle$ averaged over the particle distributions reached at the end of each simulation. The numerical data is compared to our model given in Eq.~\eqref{eq:growth_rate_low_cem} and the asymptotic formula \eqref{eq:growth_rate_low_eps_field_dep} (here, $\nu_\gamma=0$). Bottom panel: migration rate $\nu_e$ in a CP standing wave extracted from 1D3V PIC simulations, and estimated pair production rate [see Eq.~\eqref{eq:Wcr_model_low_cem}]. For reference purposes, we put a secondary horizontal axis at the top in the intensity units of a single laser beam of the amplitude $E_0/2$, implying that the standing wave is formed by two of such laser beams [c.f. Eqs.~\eqref{eq:standing_wave_E_field}-\eqref{eq:standing_wave_B_field}]. %For all the curves, the field wavelength is set to $\lambda=0.8\mu$m.
	}
	\label{fig:model_simus_migration_rate}
\end{figure}

The simulation results and the comparison with the theoretical models for the described field configurations are presented in Fig.~\ref{fig:model_simus_migration_rate}. We plot the extracted cascade growth rate for a large range of $E_0$, going from a (relatively) low value, when the cascade multiplicity is exponentially small, to strong fields with high particle yield. The migration rate field dependence $\nu_e(E_0)$ is also shown in the bottom panel.

As visible in Fig.~\ref{fig:model_simus_migration_rate},  we validate the full consistency of the numerical data for $\Gamma$ obtained via the two methods described above: directly from $N_{p}(t)$ (shown in Fig.~\ref{fig:model_simus_migration_rate} with empty circles and squares for the rotating $E$-field and standing wave configurations, respectively) with formulas \eqref{eq:rates_average}-\eqref{eq:growth_rate_formula} (depicted with small filled round and square markers). This result substantiates our argumentation from Section~\ref{sec:iii_a} and Appendix~\ref{sec:appendix_kinetic} that the master equations and their solution, given in Eqs.~\eqref{eq:const_rate_eq1}-\eqref{eq:growth_rate_formula}, are exact. It also validates our method of accounting for particle migration by using the rate $\nu_e(E_0)$. A small discrepancy between the two methods in the low field regime for $\Gamma$ extraction is due to the limitations of our numerical approach in the determination of the steady state at a very low field. 
We now discuss in detail the comparison of the numerical results with the model, Eq.~\eqref{eq:growth_rate_low_cem} and Eq.~\eqref{eq:growth_rate_high_cem}. 

\subsection{Growth rates at low and moderate fields ($\cem<\mathcal{X}$)}
\label{sec:iv_b} 
Let us consider relatively moderate fields, namely, such that the characteristic value of the $\chi$-parameter for emitting electrons is $\cem<\mathcal{X}$. We plot the corresponding model for $\Gamma$ given by Eq.~\eqref{eq:growth_rate_low_cem}, see solid lines in Fig.~\ref{fig:model_simus_migration_rate}. Here, to find $\Gamma$, we evaluate $\gem$ and $\cem$ by combining Eqs.~\eqref{eq:gamma_chi_tem_rotating} and numerically integrated Eq.~\eqref{eq:tem_def} and plug the result into $\WCSBW(\gem,\,\cem)$.\footnote{We assume that the main contribution to the cascade growth rate comes from the particles the $E$-antinode center of a standing wave, therefore, when calculating $\gem$ and $\cem$, we set $\epsilon=E_0$ and $\oeff=\omega/2$ as in a rotating $E$-field.} Up to the high field region (discussed in the next subsection), the model shows excellent agreement with the simulation results for both configurations: the standing wave (compare the purple curve and the empty squares in Fig.~\ref{fig:model_simus_migration_rate}) and rotating $E$-field (compare the orange curve and the empty circles ibid). In the former, we take into account the particle migration by using the numerical data for $\nu_e(E_0)$. However, as anticipated, we note that by using a simple estimate $\nu_e=\omega/(2\pi)$ we obtain a very close result. 

Among the cases we studied, the uniform rotating $E$-field provides the highest pair yield. As compared to it, $\Gamma$ for a cascade in a standing wave is substantially smaller at low $E_0$, which is due to $e^-e^+$ migration. The exponential decay of the rate for the asymptotically low fields is faster for the standing wave configuration, which is consistent with the prediction by our model, see Eqs.~\eqref{eq:growth_rate_low_eps_nu}--\eqref{eq:growth_rate_low_eps_field_dep} (also plotted in Fig.~\ref{fig:model_simus_migration_rate} with thin dot-dashed lines, note that here $\Delta\nu=\nu_e$). For comparison we also show the growth rate in the field of two Gaussian beams of waist $w_0=3 \lambda$, shown by triangles in Fig.~\ref{fig:model_simus_migration_rate}.
In the low-field regime, the effect of migration is stronger and the growth rate is reduced when compared to the case of the standing wave. A more detailed discussion of migration in Gaussian beams and the growth rate dependence on $w_0$ will be provided in the following Section.  

At low $E_0$, the growth rate can be small as compared to the inverse field period, $\Gamma\times 2\pi/\omega\ll 1$, meaning that the setting time for the steady state can be also much larger than $2\pi/\omega$. In this case, the effect of finite pulse duration can be important, however, we leave its consideration for future study.

With growing $E_0$, the simulation results and the model predictions for $\Gamma$ in the standing wave configuration approach the optimal case of a rotating $E$-field. Moreover, the growth rate of a cascade in the field of Gaussian beams also shows the same behavior. First, this is simply because the cascade reaches the exponential phase at a sub-cycle time before particles can escape, as $\Wcr \gg \omega/(2\pi)$. Second, $\nu_e$ decreases with growing $E_0$ (as shown in the bottom panel of Fig.~\ref{fig:model_simus_migration_rate}), meaning that the particles are trapped in the antinode center, where the electric field is maximal. In effect, the cascade develops at the $E$-antinode center. The model of a uniform rotating $E$-field becomes {\it{universal}} for all CP standing wave-like configurations. Notably, the orbits of individual particles become 2-dimensional. This illustrates the discussion in Section~\ref{sec:iii_c}. 

Note that the $\nu_e(E_0)$ behaviour resembles the anomalous radiative trapping effect found for electrons moving in a linearly polarized standing wave \cite{gonoskov2014anomalous} and in dipole waves \cite{gonoskov2017ultrabright, bashinov2022particle}. This phenomenon naturally arises in numerical simulations, as recoil is taken into account exactly in electron dynamics. While building a fully analytical model for $\nu_e(E_0)$ is challenging due to the chaotic nature of the electron motion in a standing wave, we were able to account for this effect in the model Eq.~\eqref{eq:growth_rate_formula} by using the numerical data for average $\nu_e(E_0)$.

The transition to the reduced dynamics takes place at $E_0\sim \alpha E_S$, therefore, at higher fields it is enough to consider only the rotating electric field model for calculating the particle growth rate. Moreover, as the field increases the growth rate given by  Eq.~\eqref{eq:growth_rate_low_cem}, as expected, deviates from the numerical results (see Fig.~\ref{fig:model_simus_migration_rate}) and has to be replaced by Eq. \eqref{eq:growth_rate_high_cem}. This will be addressed in the following part.

\subsection{Growth rates at a high field ($\cem>\mathcal{X}$)}
\label{sec:iv_c}
\label{sec:v_a}
\begin{figure}%%%
	\includegraphics[width=\linewidth]{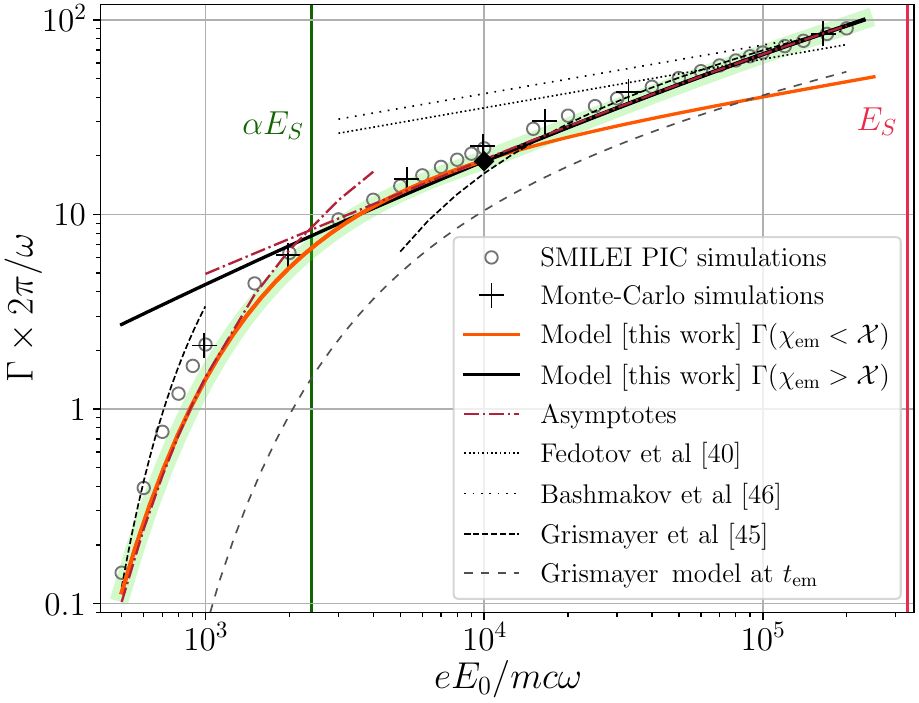}
	\caption{Cascade growth rate dependence on $E_0$ in a uniform rotating $E$-field obtained in simulations [circles (as in Fig.~\ref{fig:model_simus_migration_rate}) and cross markers]. The solid curves represent the model proposed in this work. The full model (highlighted in green) is obtained by switching from Eq.~\eqref{eq:growth_rate_low_cem} [in orange] to Eq.~\eqref{eq:growth_rate_high_cem} [in black] when they overlap (marked by the black diamond).}
	\label{fig:growth_rate_rotating}
\end{figure}

Let us discuss the cascade growth rate in a CP standing-wave-like field at very high fields. As shown above, in this case, particle migration becomes negligible, $\nu_e=\nu_\gamma=0$. We extend our simulations with SMILEI to higher $E_0$ and, in addition, perform independent 3D simulations with a semiclassic Monte Carlo code \cite{mironov2014collapse}. In the latter case, the external field is described by Eq.~\eqref{eq:rotating_E}, and the results are averaged over $\sim 10^2$ runs, where each run is initialized with one seed electron at rest. The results are presented in Fig.~\ref{fig:growth_rate_rotating}. The data obtained with both codes matches with high precision.

In Fig.~\ref{fig:growth_rate_rotating}, we plot  Eq.~\eqref{eq:growth_rate_low_cem} as well as the growth rate model developed for high fields, namely, Eq.~\eqref{eq:growth_rate_high_cem}. The latter expression matches the simulation data points in the high field region, which is not covered by Eq.~\eqref{eq:growth_rate_low_cem}.
The curves corresponding to Eqs.~\eqref{eq:growth_rate_low_cem}-\eqref{eq:growth_rate_high_cem} overlap in an extended interval of $E_0$ at $E_0\gtrsim \alpha E_S$. The full model is rendered by linking both expressions (highlighted with green in Fig.~\ref{fig:growth_rate_rotating}). At the overlap, $\cem$ varies from $\approx 3$ to 10 (also shown in Fig.~\ref{fig:tem}). This corresponds to the matching condition for Eqs.~\eqref{eq:growth_rate_low_cem} and \eqref{eq:growth_rate_high_cem} discussed in Section~\ref{sec:iii_b}.

In Fig.~\ref{fig:growth_rate_rotating}, we also plot the asymptotic expressions for $\Gamma$ at low and high fields, given in Eqs.~\eqref{eq:growth_rate_low_eps_field_dep} and \eqref{eq:growth_rate_formula_high_field_explicit}, respectively. The latter fits the black curve at $E_0>\alpha E_S$ with high precision, namely, almost in the whole region of Eq.~\eqref{eq:growth_rate_high_cem} applicability. The two asymptotic expressions cross at $E_0\sim\alpha E_S$. As a result, when combined, they provide a simple yet robust explicit formula for the field dependence of the avalanche growth rate.

\begin{figure}%%%
	\includegraphics[width=\linewidth]{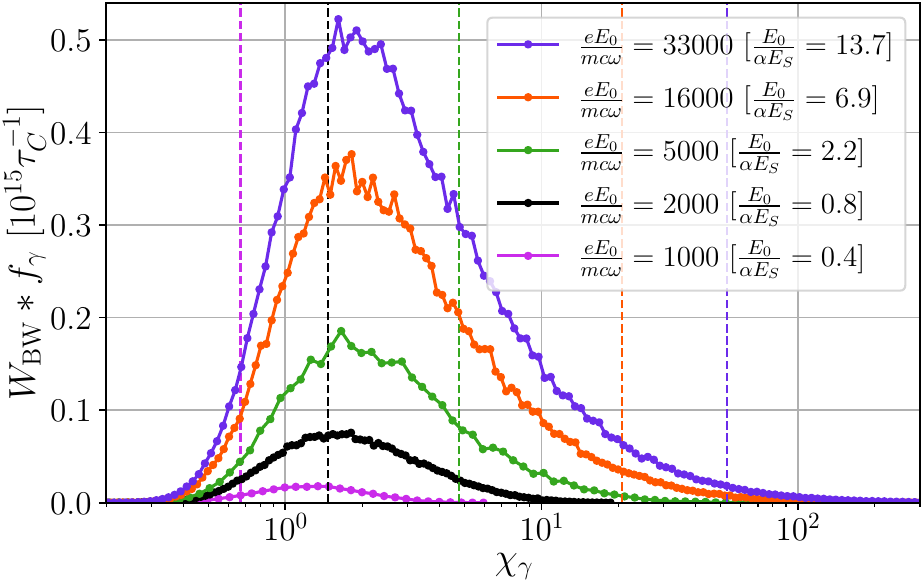}
	\caption{Distribution of photons $f_\gamma(\chi_\gamma)$ in a cascade in the steady state weighted with the probability rate of pair creation $\WBW$ for different values of field strength [for the definition, see Eq.~\eqref{eq:photon_distribution_weighted}]. The data is extracted from Monte Carlo simulations. The vertical dashed lines represent the characteristic value of $\cem$ for electrons at the emission event [as defined in Eq.~\eqref{eq:chi_em_invariant_def}]. For each of the dashed lines, $\cem$ is calculated at $E_0$ respective to a $\Wcr f_\gamma$ graph of the same color.}
	\label{fig:photon_distrib_weigted}
\end{figure}

Recall that our model in \eqref{eq:growth_rate_high_cem} for high fields $E_0>\alpha E_S$ incorporates the pair creation rate by photons at $\chi_\gamma\sim 1$, as we expect them to dominate in the emission spectrum. We confirm this assertion with simulations. We extract the steady-state photon distribution function $f_\gamma(\gamma_\gamma,\chi_\gamma)$, which is normalized by the condition $\int d\gamma_\gamma\, d\chi_\gamma f_\gamma(\gamma_\gamma,\chi_\gamma)=1$. We then weight it with the rate $\WBW$ to identify the photons that contribute to the process of pair creation. The weighted distribution in $\chi_\gamma$ is defined by
\begin{equation}
\label{eq:photon_distribution_weighted}
\left[ \WBW \ast f_\gamma  \right](\chi_\gamma)=\int d\gamma_\gamma\,\WBW(\gamma_\gamma,\chi_\gamma) f_\gamma(\gamma_\gamma,\chi_\gamma).
\end{equation}
We plot this expression for different $E_0$ in Fig.~\ref{fig:photon_distrib_weigted}. As we anticipated in our model, the weighted distribution peaks at $\chi_\gamma\approx 2$ for $E_0>\alpha E_S$. At the same time, $\cem$ of the emitting electrons increases with the field (see also Fig.~\ref{fig:tem}). In the example of a lower field, $E_0\approx0.4\alpha E_S$, the curve maximum is shifted to the left as compared to the others, and the corresponding $\cem$ is shifted leftwards even further. Thus only the exponential tail of the photon distribution can contribute to pair creation, so the avalanche development is suppressed (see the corresponding growth rate in Fig.~\ref{fig:growth_rate_rotating}). At $E_0\approx 0.8 \alpha E_S$, the value of $\cem\approx 1.4$ is close to the peak of the corresponding weighted distribution. Hence, both the emission and pair creation rates are far from their asymptotic given in Eqs.~\eqref{eq:rate_CS_asympt}, \eqref{eq:rate_BW_asympt}, and we can associate the region $E_0\sim\alpha E_S$ to the transition between the low and high field regimes. 

In Fig.~\ref{fig:growth_rate_rotating}, we compare our results to the models for $\Gamma$ in a rotating $E$-field existing in the literature. The well known scaling $\Gamma \propto E_0^{1/4}$ at $E_0\gg\alpha E_S$ proposed by Fedotov et al \cite{fedotov2010limitations,elkina2011qed} and later refined by Bashmakov et al \cite{bashmakov2014effect} does not set in at $E_0<E_S$, which is consistent with the discussion following Eqs.~\eqref{eq:growth_rate_Fedotov}. We also plot the low- and high-field asymptotic models proposed by Grismayer et al \cite{Grismayer2016,grismayer2017seeded}, which are based on the kinetic approach and some phenomenological argumentation. As one can see from the plot, our new model for the high-field asymptotic [see Eq.~\eqref{eq:growth_rate_formula_high_field_explicit}] not only corrects these estimates at very high fields but also provides a reliable scaling at intermediate field strength as the covered range corresponds to $E_0\gtrsim\alpha E_S$. A detailed discussion of this comparison and the discrepancies with other existing models is performed in Appendix~\ref{sec:appendix_comparison}, and  Table~\ref{tab:rotating_E_models}.

\section{Cascades in focused laser beams}
\label{sec:v}
The cascade onset requires high fields, and one of the ways to achieve it with laser fields is the tight focusing of the beams. For the cascade dynamics, the main effect of focusing is the appearance of particle migration in the direction transverse to laser beam propagation. In what follows, we apply our model and simulations to evaluate the cascade growth rates in several realistic field configurations and identify the effect of particle migration. As we anticipated in the discussion in Sec.~\ref{sec:iii_a}, if both photons and electrons can escape the cascading region, the cascade onset threshold becomes critical-like.

\subsection{Counter-propagating CP Gaussian beams}
\label{sec:v_a}
\begin{figure}
	\includegraphics[width=\linewidth]{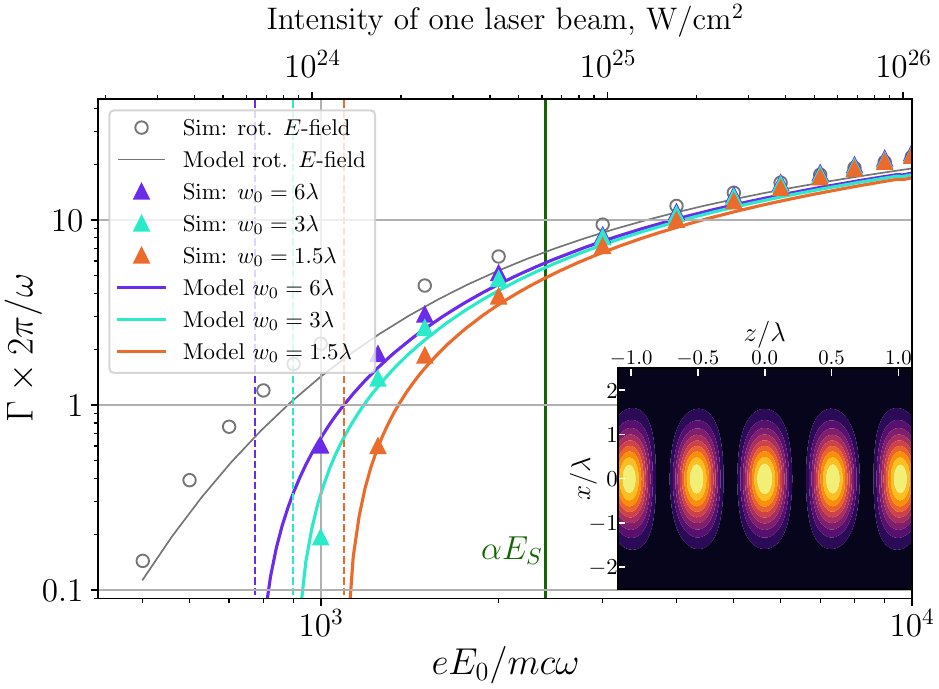}
	\caption{Particle number growth rate dependence on the peak field strength $E_0$ for a cascade developing in a CP standing wave formed by two counter-propagating Gaussian beams. The 3D PIC simulation results for 3 different beam waists $w_0$ are shown with colored triangle markers. The solid lines show our model given by Eq.~\eqref{eq:growth_rate_low_cem} with $\nu_{e}=2c/w_0+\omega/(2\pi)$ and $\nu_\gamma=2c/w_0$. The vertical dashed lines correspond to the threshold field $E_0^*:$ $\Gamma(E_0^*)=0$ [see Eq.~\eqref{eq:eps_c_def}] for each value of $w_0$ (with the respective color). The inset shows a snapshot of the electric field distribution $|\vec{E}(x,z)|$ (lighter color corresponds to a higher field value in a.u.; here, $w_0=2\lambda$), seed electrons are injected near $\vec{r}=0$. As in Fig.~\ref{fig:model_simus_migration_rate}, the axis at the top shows the intensity for a single laser beam with the peak field $E_0/2$.}
	\label{fig:GR_Gauss}
\end{figure}

Let us consider cascade seeded in the $E$-antinode of a CP standing wave formed by two counter-propagating Gaussian beams with waist $w_0$. The setup is similar to the one studied in the previous section, but the interaction region is finite in the transverse direction.

As in Sec.~\ref{sec:iv_b}, we assume that electrons can migrate along the optical axis $z$ at the rate $\sim \omega/(2\pi)$. At the same time, both electrons and photons can escape the strong-field region in the transverse direction. We estimate its characteristic size by $\sim w_0$. Then the migration rates $\nu_{e,\gamma}$ are defined by the particle flux trough the surface defined by a cylinder $\mathcal{C}$ of the radius $w_0/2$ and height $\lambda/4$, which is aligned with the $E$-antinode center, $-\lambda/8\leq z\leq \lambda/8$ [recall Eq.~\eqref{eq:nu_def}]. We estimate them as $\nu_{e}\sim 2c/w_0+\omega/(2\pi)$ and $\nu_\gamma\sim2c/w_0$. Then we can plug $\nu_{e,\gamma}$ into Eq.~\eqref{eq:growth_rate_low_cem} to obtain the cascade growth rate in the low- and moderate-field regimes.

We performed 3D PIC simulations for different values of $w_0$ starting from $E_0$ near the cascade threshold. For a meaningful comparison of the results with our model, the procedure for the growth rate extraction from the simulation data was modified: We track the number of particles inside cylinder $\mathcal{C}$ (instead of the full simulation box). Otherwise, we proceed as described in Appendix~\ref{sec:appendix_sim_setup}. 

The simulation results and the model predictions (plotted in Fig.~\ref{fig:GR_Gauss}) are in good agreement. We find a well-defined threshold for the cascade onset, which is in accordance with the model. Notice that for simulation points below the threshold at $eE_0/(mc\omega)=750$ for $w_0=6\lambda$ and $3\lambda$, and $eE_0/(mc\omega)=1000$ for $w_0=1.5\lambda$, the growth rates obtained from simulations are negative, meaning that the particle outflow from $\mathcal{C}$ exceeds the gain from pair creation.

Our model provides a lower boundary for the cascade threshold. The corresponding values for $E_0^*$ and laser intensity are summarized in Table~\ref{tab:tab_gauss_threshold} (also shown by the vertical lines in Fig.~\ref{fig:GR_Gauss}). The model estimate for $w_0=\lambda$ is also provided for comparison. As $w_0$ increases, the threshold field goes down, however, in terms of laser power optimization, it is beneficial to use tight focusing. Note that the effects of finite pulse duration are not included, but could be also important near the threshold.

\begin{table}
	\caption{Threshold field strength required for the $e^-$-seeded cascade onset in the $E$-antinode of two CP counter-propagating Gaussian beams as predicted by Eqs.~\eqref{eq:growth_rate_low_cem} and \eqref{eq:eps_c_def} for different waists $w_0$. $E_0^*$ corresponds to the total peak field and $P^*=2\times(I^*\pi w_0^2/2)$ to the total power, where  $I^*$ is the intensity of each beam (corresponding to $E_0^*/2$).}
	\label{tab:tab_gauss_threshold}
	\setlength{\tabcolsep}{11pt}
	\begin{tabular}{lcccc}
		\hline\hline
		$w_0$& $\lambda$ & $1.5\lambda$ & $3\lambda$ & $6\lambda$\\
		\hline
		$eE_0^*/(mc\omega)$ & 1269 & 1094 & 898 & 774\\
		$I^*$, $10^{23}$W/cm$^2$ & $17.2$ & $12.8$ & $8.6$ &  $6.4$\\
		% $P^*$, PW &  $17$ & $29$ & $78$ & $232$\\
            $P^*$, PW &  $34$ & $58$ & $156$ & $464$\\
		\hline\hline
	\end{tabular}
\end{table}

At higher fields, the growth rate becomes independent of $w_0$. The simulation results and the model curves for different $w_0$ converge to the prediction of the rotating $E$-field configuration. Namely, at $E_0>\alpha E_S$, the pair creation process at the $E$-antinode center dominates in the cascade dynamics, and the rotating $E$-field model is reliable.

\subsection{Single strongly focused laser beam}
\label{sec:v_b}
As a second example, let us consider cascades triggered in the focus of a single laser beam. For this setup, we extend the findings of Ref.~\cite{Arseny2021}. The key difference from the cases studied above is that the electric and magnetic components are both present and are of the same order, $E\sim cB$. The paraxial approximation is not applicable at very strong focusing: Unlike in a plane wave, the longitudinal field component and the field invariants $\mathcal{F}$ and $\mathcal{G}$ can be non-zero. In particular, the field in the focus can be of the electric type ($\mathcal{F}>0$), so that the invariant field $\epsilon>0$. In this case, injected electrons can be accelerated efficiently and gain $\chi_e$ as given in Eq.~\eqref{eq:chi_short_time}.

\begin{figure}%%%
	\includegraphics[width=\linewidth, trim={0 0 0 1}, clip]{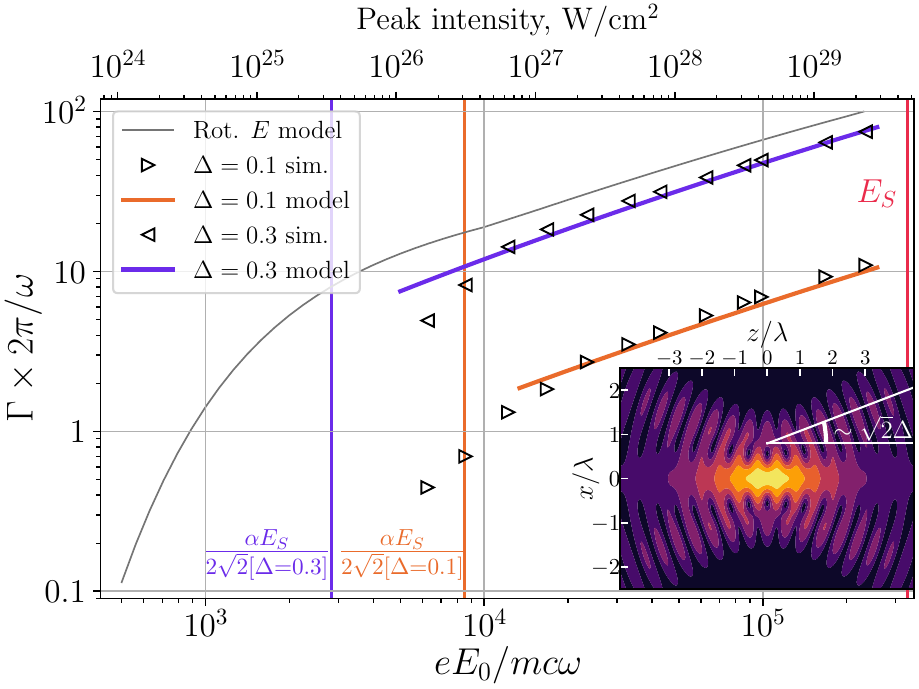}
	\caption{Particle number growth rate dependence on the peak field strength $E_0$ for a cascade developing in a single strongly focused laser beam at different values of the focusing parameter $\Delta$. The 3D Monte Carlo simulation results are shown with triangle markers. The solid lines show our high-field model given by Eq.~\eqref{eq:growth_rate_high_cem}. $\Gamma(E_0)$ for the rotating $E$-field configuration is shown for reference, however, note that the top axis in the laser intensity is offset as compared to other figures presented above. The vertical lines correspond to the condition $\epsilon=2\sqrt{2}\Delta E_0=\alpha E_S$. The inset shows a snapshot the electric field distribution $|\vec{E}(x,z)|$ (lighter color corresponds to a higher field value in a.u.; here, $\Delta=0.2$), seed electrons are injected near $\vec{r}=0$.}
	\label{fig:GR_single_beam}
\end{figure}

To treat such fields, we adopt the model for a focused $E$-type CP laser beam developed by Narozhny and Fofanov \cite{narozhny2000scattering}. The focusing degree is controlled by the parameter $0<\Delta\lesssim 0.3$ (illustrated in the inset of Fig.~\ref{fig:GR_single_beam}). At $\Delta=0.1$, the beam is focused to the diffraction limit.

The short-time dynamics for a single electron injected in the focal center of such a beam was studied in detail in Ref.~\cite{Arseny2021}. The expressions for $\epsilon$ and $\oeff$ read:
\begin{equation}
	\label{eq:eps_oeff_single_pulse}
	\epsilon\approx 2\sqrt{2}\Delta E_0,\quad \omega_\mathrm{eff}=17\sqrt{2}\Delta^3\omega, 
\end{equation}
where $E_0$ is the electric field amplitude. By substituting these quantities into Eqs.~\eqref{eq:gamma_short_time}-\eqref{eq:chi_short_time}, one can obtain $\gamma_e(t)$ and $\chi_e(t)$ at $\omega t\ll 1$.

Let us discuss the particle dynamics qualitatively. Since the field is close to crossed, $(\vec{E}\vec{B})= O(\Delta^2)$, the initial and secondary $e^\pm$ are pushed by the laser in the direction of its propagation \cite{landau2013classical}. As a result, in a single beam of relatively low intensity, particles can escape from the focus before the quasi-steady state can form. This limits the applicability of our model in the low- and moderate-field regimes. Note that it contrasts with the CP standing wave $E$-antinode case. There, $e^-$ and $e^+$ are accelerated by the electric field in the opposite directions, so that newly created particles can return to the focus and sustain the cascade. As a result, a steady state can form even when it requires multiple periods. 

If the field is high and the cascade develops rapidly, we still can assume that the steady state can be established at a sub-cycle time. With this hypothesis, we apply the high-field model given in Eq.~\eqref{eq:growth_rate_high_cem}. While this can be checked numerically, it is reasonable to assume that the steady-state forms rapidly if $\Gamma \gg \omega/(2\pi)$.  The migration effects can be omitted in the high-field regime.

Recall that Eq.~\eqref{eq:Wcr_model_high_cem} is used to evaluate the average pair creation rate at high fields. Let us estimate $\gamma_\gamma$ in this formula. Following Ref.~\cite{Arseny2021}, for the chosen field model of a beam propagating along the $z$-axis, the non-zero field components at $t=0$, $\vec{r}=0$ read $E_y=-E_0$, $B_x=E_0(1-\Delta^2)$. As photons in the cascade co-propagate with the field, let us direct the photon momentum along the $z$-axis. Then, by using Eq.~\eqref{eq:chi}, it is straightforward to show that $\chi_\gamma=4\Delta \gamma_\gamma E_0/E_S$. Therefore, in view of Eq.~\eqref{eq:eps_oeff_single_pulse}, one should take $\beta=\sqrt{2}\Delta$  in Eq.~\eqref{eq:Wcr_model_high_cem}. With this, we can calculate $\Gamma$ by using Eq.~\eqref{eq:growth_rate_high_cem}. 

The resulting growth rate dependence on $E_0$ is presented in Fig.~\ref{fig:GR_single_beam}. We compare it to the results of Monte Carlo simulations. The high-field formula \eqref{eq:growth_rate_high_cem} matches the simulation results both at moderate and very strong focusing ($\Delta=0.1$ and $0.3$, respectively). Note that the characteristic field, at which the high-field model becomes applicable [given by Eq.~\eqref{eq:low_high_chi_field_correspondence}], reads $E_0\gg \alpha E_S/(2\sqrt{2}\Delta)$. It depends on the focusing parameter and differs from the simpler condition $E_0\gg\alpha E_S$ for the CP standing wave.

Triggering a prolific avalanche-type cascade in a single laser beam requires intensities that are by 1-2 orders of magnitudes higher than in the case of the standing wave configuration (compare the upper horizontal axes in Figs. \ref{fig:GR_Gauss} and \ref{fig:GR_single_beam}). Still, we conclude that this setup, on the one hand, is suitable for the cascade onset, and, on the other, at ultra-high intensities, it is crucial to take into account the possibility of cascade generation even in the single-beam geometry.

\subsection{Optimized multi-beam configuration }
\label{sec:v_c}
\begin{figure}%%%
	\includegraphics[width=\linewidth]{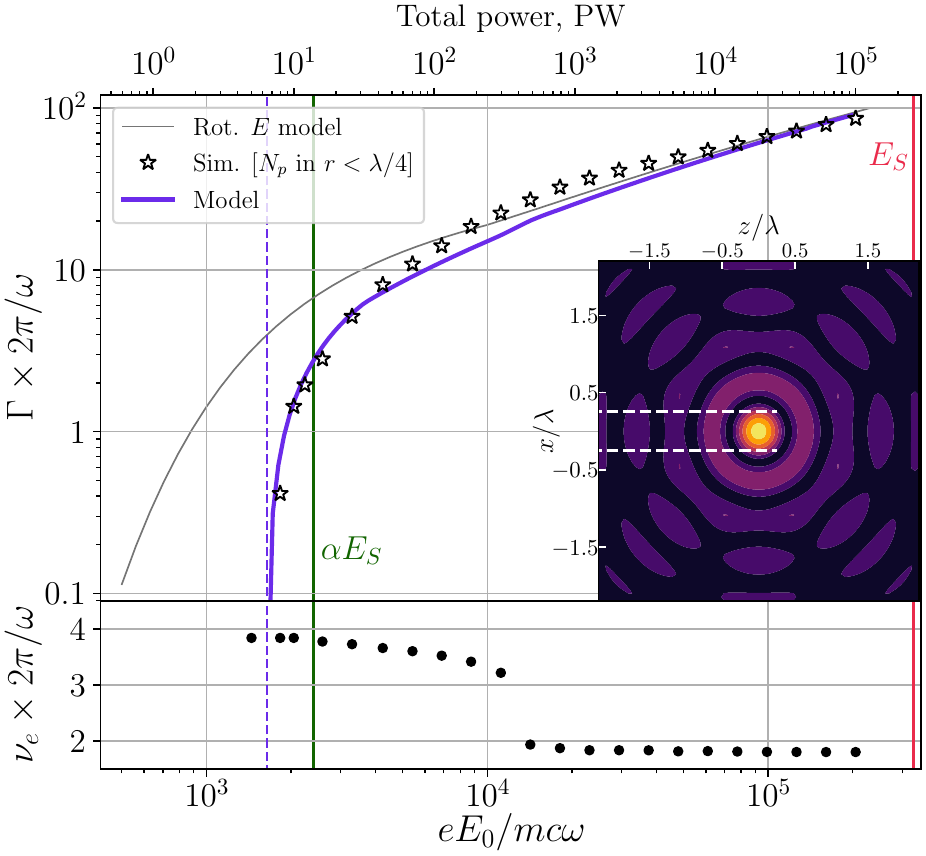}
	\caption{Particle number growth rate dependence on the peak field strength $E_0$ for a cascade developing in the mutual focus of 8 strongly focused laser beams with optimized polarization (as in Ref.~\cite{gelfer2015optimized}). The 3D Monte Carlo simulation results are shown with star-shaped markers. The solid lines show our model given by Eqs.~\eqref{eq:growth_rate_low_cem}-\eqref{eq:growth_rate_high_cem}.  The vertical dashed line correspond to the threshold field $E_0^*:$ $\Gamma(E_0^*)=0$ [see Eq.~\eqref{eq:eps_c_def}].  The upper horizontal axis corresponds to the total power delivered by 8 laser beams. The inset shows a snapshot of the electric field distribution $|\vec{E}(x,z)|$ (lighter color corresponds to a higher field value in a.u.), seed electrons are injected near $\vec{r}=0$. The white dashed lines at $x=\pm\lambda/4$ show the characteristic size of the focal spot. Bottom panel: the electron migration rate extracted from Monte Carlo simulations vs $E_0$.}
	\label{fig:GR_8_beams}
\end{figure}
As a final example, we consider the configuration proposed in Ref.~\cite{gelfer2015optimized},  a multi-beam setup that aims at optimizing the requirements on the total power of the laser setup for triggering avalanche-type cascades. In the best-case scenario, it was suggested to use 4 pairs of strongly focused counter-propagating beams. Each pair is tuned to have unique elliptic polarization, and the total electric field $E_0$ is maximal at the focal center due to constructive interference. For the optimal choice of polarization, the gain of $\chi_e(t)$ for injected electrons is maximized in the acceleration process.

Each pair of beams is tuned so that $B=0$ at the focal center, therefore, the corresponding invariant field is $\epsilon=E_0$. As for $\oeff$, since the analytic expression for the field model is cumbersome, we prefer to calculate it numerically with the aid of our dedicated package \cite{PairAvalanchesQED}. For the chosen parameters, we obtain $\oeff\approx 0.37\omega$.

The field structure is illustrated in Fig.~\ref{fig:GR_8_beams}.\footnote{We use the same field model and choice for polarization vectors as in Ref.~\cite{gelfer2015optimized}, see Eqs. (15) and (16) therein. The only difference is that here we choose $\lambda=0.8\mu$m. The focusing parameter is $\xi=8$, which corresponds to $\Delta\approx0.35$ in the notation used in Sec.~\ref{sec:v_b}.}
The focal spot has the characteristic radius $r_0\approx\lambda/4$ in all 3 dimensions and the migration rate for photons can be estimated as $\nu_\gamma\simeq 4c/\lambda$. This estimate is also reasonable for $\nu_e$ at moderate field strength. However, our simulations show that $\nu_e$ noticeably drops at high $E_0$. Therefore, we use the numerical data for $\nu_e(E_0)$ obtained in a similar way as for the CP standing wave configuration (see Appendix~\ref{sec:appendix_sim_setup}) with the only modification that a particle is considered as escaping if it reaches $r_0$.

With this, we can apply our model to calculate the cascade growth rate and the threshold field strength and compare with 3D Monte Carlo simulations. The results are presented in Fig.~\ref{fig:GR_8_beams}. To cover the full range of field strength, we combine the low- and high-field expressions from Eqs.~\eqref{eq:growth_rate_low_cem} and \eqref{eq:growth_rate_high_cem}, respectively. To extract the growth rate from simulations, we track $N_p(t)$ inside a sphere of radius $r_0$. We find excellent agreement between the simulation results and our model.
	
As in the case of two Gaussian laser beams, the presence of both electron and photon migration determines the cascade threshold. By evaluating it from Eq.~\eqref{eq:growth_rate_low_cem}, we obtain $eE_0^*/(mc\omega)\approx 1642$, corresponding to the total power of $P_{\text{8-beam}}^*\approx6.4$ PW.

At high fields, $\Gamma$ is comparable to the best-scenario case of a rotating $E$-field. With a multi-beam setup, reaching this regime requires significantly less laser radiation power than the 2-beam configuration. For example, at $P\approx 50$ PW, the growth rate $\Gamma\sim 10$, which ensures high particle gain in a sub-cycle time interval.

\section{Formation of dense electron-positron plasma in avalanche-type cascades}
\label{sec:vi}
\begin{figure}%%%
	\includegraphics[width=\linewidth]{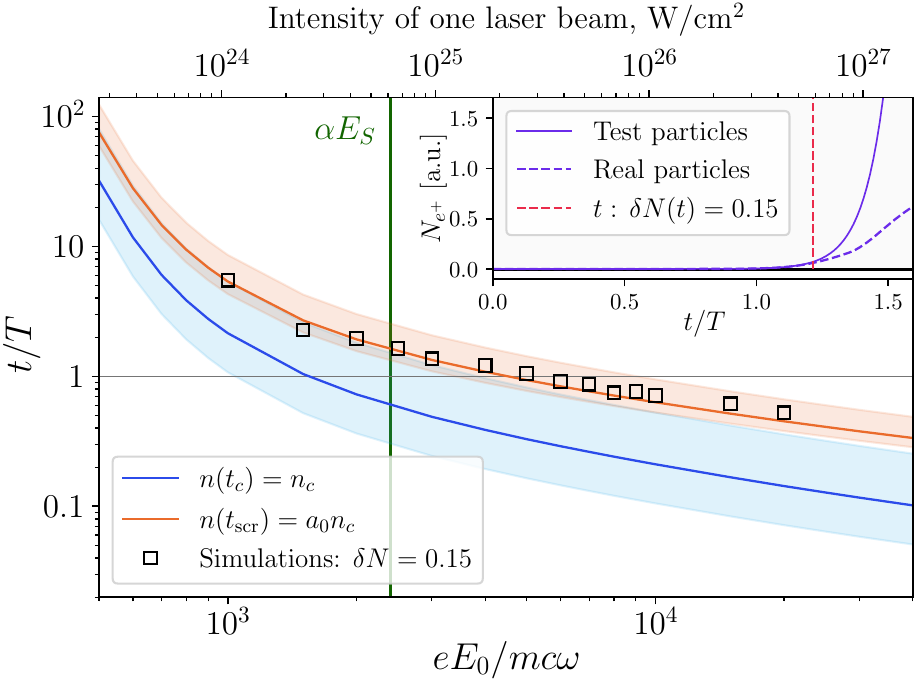}
	\caption{The field strength dependence of the time required to reach the electron critical plasma density $n_c$ (blue) and the relativistically opaque regime $a_0 n_c$ (orange) in a cascade in a uniform rotating electric field. The time depends on the initial electron density $n_0$, which is depicted as color bands: the lower bound corresponds to $n_0=10^{-5}n_c$, and the upper to $n_0=10^{-1}n_c$. Solid lines correspond to $n_0=10^{-2}n_c$. The inset shows two sample simulations in SMILEI: with real and test particles. Squares in the main plot correspond to a time point when  $\delta N=(N_{\mathrm{test}}-N_{\mathrm{real}})/N_{\mathrm{test}}$ reaches the value of 0.15. For reference purposes, we put a secondary horizontal axis at the top in the intensity units of a single laser beam of the amplitude $E_0/2$. The wavelength is set to $\lambda=0.8\mu$m.}
	\label{fig:screening}
\end{figure}
Our model can be applied to estimate the threshold for electron-positron plasma effects in a cascade in terms of field parameters. 
Let us assume that a cascade is triggered in a rotating $E$-field by electrons with initial density $n_0$. At time $t$ the particle density will reach $n(t)=n_0\exp(\Gamma t)$. For a field of an optical frequency, $\Gamma$ is well estimated by the asymptotic expressions \eqref{eq:growth_rate_low_eps_field_dep} and \eqref{eq:growth_rate_formula_high_field_explicit} matched at their crossing point. By using them, we can calculate the time required to reach density $n$:
\begin{widetext} 
\begin{equation}
	\label{eq:wait_time_plasma}
	\renewcommand{\arraystretch}{3.5}
	t \simeq \frac{2\pi}{\omega} \ln\left(\frac{n}{n_0}\right)\times\left\lbrace
	\begin{array}{ll}
		6.74\cdot 10^{-2}\,\sqrt{\dfrac{1000}{a_0}} \exp\left[2.89 \left(\dfrac{1000}{a_0}\right)\left(\dfrac{\lambda}{1\,\mu\text{m}}\right) \right], & a_0 < 3000 \left(\dfrac{\lambda}{1\,\mu\text{m}}\right),\\
		\dfrac{3.31\cdot 10^{-2}\left(\frac{1000}{a_0}\right)} {\sqrt{1+278.5\sqrt[4]{\left(\frac{1000}{a_0}\right)^3\left(\frac{\lambda}{1\,\mu\text{m}}\right) } } -1}, & a_0 > 3000 \left(\dfrac{\lambda}{1\,\mu\text{m}}\right)
	\end{array}
	\right.
\end{equation}
\end{widetext}
where we used the notation $a_0=eE_0/(mc\omega)$.
 
Eq.~\eqref{eq:wait_time_plasma} allows to identify two characteristic times. First, the time $t_c$ at which the produced particles reach the critical electron plasma density, $n(t_c)=n_c=\varepsilon_0 m \omega^2/e^2$, and second, $t_{\mathrm{scr}}$ corresponding to the time moment when the plasma starts to screen the external field, $n(t_{\mathrm{scr}})=a_0n_c$. The field dependence for these quantities is plotted in Fig.~\ref{fig:screening}. The time needed to reach an $e^-e^+$ plasma state as seen by the field is exponentially large at low field, but for $a_0\gtrsim 10^3$ it can be reached in several or even one field cycle depending on the initial density $n_0$. Backreaction from the produced plasma should become noticeable after one field cycle at $a_0\gtrsim 5\cdot 10^3$ (for optical lasers) if $n_0\gtrsim 10^{-2}n_c$. Note that the corresponding field strength is $E_0\gtrsim \alpha E_S$.

We calculated the characteristic screening time $t_{\mathrm{scr}}$ in the rotating $E$-field configuration through PIC simulations with SMILEI. The numerical setup is the same as for the previous simulations in Sections~\ref{sec:iv} and is detailed in Appendix~\ref{sec:appendix_sim_setup}. Here, we extended the time of each simulation so that the produced plasma could generate a noticeable field. 

To identify the significance of the plasma-generated field, we used the following convenient technical feature of SMILEI. We generated two sets of simulations under the same input parameters: with so-called test and real particles. The formers do not generate self-fields, therefore, a corresponding simulation is equivalent to a semiclassic Monte Carlo computation.\footnote{Technical details can be found in the SMILEI code manual \url{https://smileipic.github.io/Smilei/}.} The particle growth rate in such simulations remains constant in time once the steady state is reached. In contrast, when the density of real particles becomes high, the external field is screened, and the cascade growth rate can be reduced. The magnitude of screening can be guessed from the relative difference in the particle numbers obtained in the two simulations, $\delta N=(N_{\mathrm{test}}-N_{\mathrm{real}})/N_{\mathrm{test}}$. As $\delta N$ increases with time, $t_{\mathrm{scr}}$ can be associated with the time moment when $\delta N$ reaches a prescribed threshold value.

The results of PIC simulations are presented in Fig.~\ref{fig:screening}. Each of the (square) points corresponds to a time moment $t_{\mathrm{scr}}$ such that $\delta N(t_{\mathrm{scr}})=0.15$. The inset illustrates our method for extracting $t_{\mathrm{scr}}$ for one of the plotted points. Note that variation of the threshold value for $\delta N$ does not lead to a significant change in the results. The simulation results show good correspondence to the prediction of our model.  Let us remark that we did not account for particle migration. For $t_{\mathrm{scr}}$ at lower fields, that require several field revolutions to generate dense plasma, migration will lead to even longer times. However, at $E_0\gtrsim \alpha E_S$, $t_{\mathrm{scr}}$ becomes comparable to the field period, and at the same time the effect of migration becomes negligible (as we discussed in previous sections).

\begin{table}
	\caption{Estimated field strength and total laser power required to reach plasma effects in one field period in various field configurations if the initial seed electron density is $n_0=10^{-2}n_c$. $a_{0,n_c}$ and $a_{0,\mathrm{scr}}$ are defined by the conditions $t_{\mathrm{c}}(a_{0,n_c})=2\pi/\omega$, and $t_{\mathrm{scr}}(a_{0,\mathrm{scr}})=2\pi/\omega$. Here,  $\lambda=0.8$ $\mu$m.}
	\label{tab:plasma_threshold}
	\setlength{\tabcolsep}{5pt}
	\begin{tabular}{l|cc|cc}
		\hline\hline
		Configuration & $a_{0,n_c}$ & $P_{n_c}$, PW & $a_{0,\mathrm{scr}}$ & $P_{\mathrm{scr}}$, PW\\
		\hline
            Rotating $E$ & 1810 & -- & 5130 & -- \\
            CP standing wave & 1940 & -- & 5510 & -- \\
            2 beams, $w_0=1.5\lambda$ & 2320 & 260 & 6550 & 2070 \\
            2 beams, $w_0=\lambda$ & 2530 & 140 & 7140 & 1100 \\
            8 beams & 3010 & 22 & 8870 & 190 \\
		\hline\hline
	\end{tabular}
\end{table}

As we validated our approach with simulations for the rotating $E$-field, let us briefly discuss other field configurations. Table~\ref{tab:plasma_threshold} summarizes our estimates for the field strength required to reach the critical density $n_c$ and relativistic density $a_0n_c$ at the time scale of one field period. For this, we apply Eqs.~\eqref{eq:growth_rate_low_cem}-\eqref{eq:growth_rate_high_cem} with accounting for migration as discussed before, and use the equation $n_0\exp[2\pi\Gamma(a_0)/\omega]=n_f$ to calculate $a_0$ for the given initial and final densities $n_{0,f}$. As above, we pick $n_0=10^{-2}$. One can see that migration noticeably shifts the required field when the results for the standing wave and two Gaussian beam configurations are compared to the rotating $E$-field case. We also calculate the corresponding total power of the laser setup. To reach $n_c$ in the two Gaussian beam configuration, the minimal requirement exceeds 100 PW even in the best-case scenario ($w_0=\lambda$) for our choice of $n_0$. Creating relativistically opaque plasma seems unlikely for this type of field unless the initial target is dense. However, the optimized 8-beam setup \cite{gelfer2015optimized} lowers the requirements to feasible parameters in view of future multi-petawatt laser facilities.

\section{Summary and discussion}
\label{sec:conclusions}
In the present work, we considered electron-seeded avalanche-type (or self-sustained) QED cascades developing in the fields generated by ultra-intense laser beams. We proposed a general analytical model predicting the electron-positron pair yield for a wide class of electromagnetic fields. 
Our model is applied to several configurations that are discussed in the literature in the context of future experiments at the upcoming multi-petawatt laser facilities and shows predictive capabilities.

In past works, a large effort was put into (predominantly numerical) studies of cascades developing in CP standing waves \cite{Kirk2008, Kirk2009, fedotov2010limitations, elkina2011qed, nerush2011analytical, bashmakov2014effect, Grismayer2016, grismayer2017seeded, Jirka2016}. The conditions at the electric antinodes of such a field are optimal for the rapid acceleration of charged particles, which is decisive for the cascade onset. However, particles can migrate to the magnetic antinodes reducing the cascade efficiency. In focused laser fields, this effect is even stronger, as charges and photons can escape from the interaction area both transversely and longitudinally with respect to the laser optical axis. In our work, we highlight the importance of migration below a certain field threshold and show that it has an impact on the field strength necessary to trigger a cascade. This understanding is critical to make the optimal choice of field configuration in upcoming experiments at ultra-high-intensity laser facilities, in particular because the first tests will take place near the cascade onset threshold. We also show that on the contrary, at a very high field, migration becomes negligible and for a certain number of configurations the growth rate is given by the solution in the optimal case of a rotating $E$-field. 

At first, we treat the problem of cascade evolution in a general field. We build an analytical model applicable to the interaction region where the cascade can develop rapidly. At the onset stage, the cascade evolution is highly nonstationary. Nonetheless, the system can rapidly relax to the so-called steady state, in which the effective rates of photon emission $\Wrad$, pair creation $\Wcr$, and particle migration $\nu_{e,\gamma}$ are constant and depend only on the field parameters. Within LCFA, the steady-state master equations for the particle number can be cast in a simple form of Eqs.~\eqref{eq:const_rate_eq1}-\eqref{eq:const_rate_eq2}, which have an exponential solution $N_{e^\pm,\gamma}\sim e^{\Gamma t}$ with the growth rate $\Gamma$ given by Eq.~\eqref{eq:growth_rate_formula}.

The central result of this work is the analytical model for $\Gamma$ expressed in terms of field parameters and migration rates as given in Eqs.~\eqref{eq:growth_rate_low_cem}, \eqref{eq:growth_rate_high_cem}. The two equations correspond respectively to the (relatively) low-field regime when only the photons from the high-energy tail of the emission spectrum can create $e^-e^+$ pairs, and to the high-field regime, in which the pair creation process is dominated by the softer part of the photon spectrum ($\chi_\gamma\sim 1\ll \chi_e$). The related asymptotic representation given in Eqs.~\eqref{eq:growth_rate_low_eps_field_dep} and \eqref{eq:growth_rate_formula_high_field_explicit} provide an explicit dependence for $\Gamma$ on the field strength and local gradient, which appear to be robust in a wide range. These results are derived for a \textit{general} field configuration with the major requirement that a cascade can onset and then reach the steady state. We outline the procedure to calculate $\Gamma$ for a given field configuration at the end of Sec.~\ref{sec:iii_b}. For applications, we also supplement our work with a dedicated numerical code \cite{PairAvalanchesQED}.

The immediate outcome of the particle migration is the appearance of a well-defined critical-like threshold for the cascade formation. This happens if the outflow of photons and $e^\pm$ exceeds the pair production rate in the strong-field region. The threshold can be defined in terms of the field parameters, e.g. field strength $E^*$ as given in Eq.~\eqref{eq:eps_c_def}, and Eq.~\eqref{eq:eps_c_estimate} can be used to estimate $E^*$ in various field configurations. At vanishing migration, the threshold becomes smooth crossover-type.

The growth rate behavior near the threshold strongly depends on the particle migration rate and therefore on the field geometry, e.g. the laser beam waist $w_0$. The striking feature of the high-field regime is that migration becomes irrelevant when $\Wcr>\nu_{e,\gamma}$. From this perspective, field configurations that have different large-scale structures but equivalent $\epsilon$ and $\oeff$ in the peak field region (e.g. the focal center) can be grouped into the same universality class, as the cascade growth rate will be identical for them in the high-field regime.

The impact of migration in the low-field regime and the universality of the high-field regime can be seen for cascades developing in the $E$-antinode of two counter-propagating CP laser beams, which we studied in detail by applying our model and simulations. In particular, we revisited the relevant field models: a standing CP wave infinite in the direction transverse to the laser beam propagation, a uniform rotating electric field, and two focused Gaussian beams. 

At relatively low peak fields, $E_0\lesssim \alpha E_S$, the migration of particles significantly suppresses the growth rate already in the standing wave case due to the longitudinal migration of electrons (see Fig.~\ref{fig:model_simus_migration_rate}). For two CP Gaussian beams, the suppression is much stronger due to the additional migration of electrons and photons in the transverse direction (see Fig.~\ref{fig:GR_Gauss}). Both our model and simulations predict a well-defined cascade threshold, which depends on the Gaussian beam waist $w_0$. At increasing $w_0$ the threshold field strength decreases, however, the corresponding beam power grows, as shown in Table~\ref{tab:tab_gauss_threshold}. As expected, tight focusing is beneficial for reaching the cascade threshold when the output peak power is the limiting factor. At the diffraction limit, $w_0\approx\lambda$, the threshold is close to the anticipated parameters of the future multi-petawatt laser facilities.

At fields as high as $E_0\gtrsim \alpha E_S$, migration becomes negligible for both the standing wave and focused Gaussian beam configurations. The corresponding growth rates match the result for a uniform rotating $E$-field (see Fig.~\ref{fig:model_simus_migration_rate}), which confirms its universality for CP-standing-wave-like configurations at high fields. We suggest a new explicit formula \eqref{eq:growth_rate_formula_high_field_explicit} for the high-field growth rate that replaces the scaling $\Gamma\propto E_0^{1/4}$ proposed in Ref.~\cite{fedotov2010limitations}. Moreover, we demonstrate that the latter sets in only beyond the critical field of QED $E_S$ and is not practical for estimates in sub-critical fields of the optical frequency. The proposed here Eq.~\eqref{eq:growth_rate_formula_high_field_explicit} is reliable at $E_0> \alpha E_S$.

While the two-beam setup is one of the most discussed in the literature, we made use of our model generality to investigate two more intricate configurations: a single tightly focused beam beyond the paraxial approximation \cite{narozhny2000scattering}, and a multi-beam setup optimized for the cascade onset \cite{gelfer2015optimized}. The former drastically differs from standing-wave-like configurations, as the magnetic field component does not vanish at the focus. Still, an avalanche can develop at strong focusing, though the required intensity is beyond $10^{26}$ W/cm$^2$.

A multi-beam setup allows reaching high fields at much lower total intensity and power. For the 8-beam configuration with optimized polarization, the cascade threshold predicted by our model and simulations is as low as $P_{\text{8-beam}}^*\approx6.4$ PW, which is close to the result of Ref.~\cite{gelfer2015optimized} where a different definition for the threshold was suggested. The total power for 2 beams exceeds this value at least by an order of magnitude, $P_{\text{2-beam}}^*\approx34$ PW for $w_0=\lambda$ (see Table~\ref{tab:tab_gauss_threshold}).

Another important application of our model for the growth rate is the possibility to estimate the field strength at which plasma effects become important and 
self-consistent description is needed. For the rotating $E$-field, the produced electron-positron plasma can reach the critical density $n_c$ in one field cycle if two laser beams of intensity $I\approx 2\times10^{24}$ W/cm$^2$ each (at $\lambda=0.8$ $\mu$m) are used, meaning that some collective effects could be observable. However, this estimate implies a very large focal spot and hence extreme laser power. Our calculation shows (see Table~\ref{tab:plasma_threshold}) that at strong focusing ($w_0=\lambda$) the necessary intensity is higher, $I\approx 7\times10^{24}$ W/cm$^2$. The corresponding total power of two beams is $P\approx 140$ PW. At intensities $I\gtrsim 10^{25}$ W/cm$^2$, the external field can be fully screened for a short laser pulse, which however demands the power of an order magnitude higher. With the optimized 8-beam setup \cite{gelfer2015optimized}, the required power is lower by a substantial margin. It could be possible to create dense $e^-e^+$ plasma by illuminating an underdense target at 20-50 PW and create a relativistically opaque state at 200 PW.

In general our model allows to predict the pair density that is well-controllable by tuning the laser intensity.  Once available in the laboratory, avalanche-type cascades will open a path to studying relativistic $e^-e^+$ plasma, relevant to astrophysical phenomena.

In this work, we focused on field configurations that allow sustaining a steady state and hence a cascade for an extended time. This includes circularly polarized laser beams.
At the state of the art, ultra-high intensities are easier to access experimentally for linear polarization. Also, it was proposed that using the linear polarization can be more beneficial at lower field strength near the cascade onset threshold \cite{Jirka2016}. However, in such fields, the cascade steady state cannot extend for longer than one field period, and hence circular polarization should provide higher particle yield at higher fields. The full comparative study for both polarizations, as well as the generalization to single and multiple tightly focused laser beams and structured laser pulses (e.g. Laguerre-Gauss modes), is under progress and left for future work.

In the present model, we disregarded the effects of finite laser pulse duration. One may expect that the threshold field $E^*$ can rise for very short pulses. We also did not include the Sauter-Schwinger pair creation from vacuum, which might be important in the high-field limit \cite{fedotov2010limitations, fedotov2022high, bulanov2010multiple}. These and other related refinements should be done in future works.

\section*{Acknowledgements}
We express our gratitude to A.M. Fedotov, E.G. Gelfer, A. Gonoskov, T. Grismayer, M. Vranic, and S. Meuren for fruitful and elucidating discussions. This work used the open-source PIC code SMILEI, the authors are grateful to all SMILEI contributors and to the SMILEI-dev team for its support. Simulations were performed on the Irene-Joliot-Curie machine hosted at TGCC, France, using High-Performance Computing resources from GENCI-TGCC (Grant No. A0030507678). This work received financial support from the French state agency \textit{Agence Nationale de la Recherche}, in the context of the \textit{Investissements d'Avenir} program (reference ANR-18-EURE-0014). A.A.M. was supported by Sorbonne Universit\'e in the framework of the Initiative Physique des Infinis (IDEX SUPER).

\appendix
\section{Small-time dynamics of $e^\pm$ in a cascade}
\label{sec:appendix_weff}

Let us discuss the motion of electrons in between emissions of photons in a cascade developing in a general field, which varies in time and space with the characteristic frequency $\omega$ and the corresponding wavelength $\lambda$. We follow Ref.~\cite{Arseny2021} and provide only the details necessary for the current work. 

Consider an electron that is initially located at the origin at time $t=0$ in an external field. The electric and magnetic fields at the initial position are characterized by the EM field tensor $F^{\mu\nu}=F^{\mu\nu}(0)$ combining the electric $\vec{E}$ and magnetic $\vec{B}$ components. As mentioned in Section~\ref{sec:ii}, we assume that the electron motion is semiclassic, meaning that it is governed by the Lorentz equations:\footnote{The discussion of positron motion will be the same except the change of sign in the RHS of Eq.~\eqref{eq:Lorentz}.}
\begin{equation}
	\label{eq:Lorentz}
	\dfrac{\rmd  p^{\mu}}{\rmd  \tau} = -\dfrac{e}{mc}F^{\mu}_{\phantom{1} \nu}(x(\tau))p^{\nu},\quad p_\mu(0)=p_\mu^{(0)},
\end{equation}
where $p^\mu$ and $x^\mu$ are the 4-momentum and time-space coordinate of the electron and $\tau$ is the proper time.\footnote{In our notations, after every quantum event, an electron is assigned new initial conditions (this includes scattered electrons in the Compton process).} To solve this equation we impose the following assumptions:
\begin{enumerate}
\item[(i)] the time between photon emissions is small, hence, we consider time intervals $t\ll \omega^{-1}$. The electron propagates almost as in a constant field $F_{\mu\nu}(0)$, the corrections are provided by the field derivatives, $F_{\mu\nu,\sigma}(0)\equiv \partial F_{\mu\nu}(0)/\partial x^\sigma$;
\item[(ii)] due to acceleration by the field, $e^-$ rapidly becomes ultra-relativistic at time scale $ mc/eE \ll t$;
\item[(iii)] energy gained by $e^-$ during acceleration is much higher than the initial energy, $eEt/mc\gg \gamma_e(0)$;
\item[(iv)] The field is of electric type, namely, $E>cB$ (this is a wide class of fields, and the field of the electric antinode of a CP standing wave belongs to it).
\end{enumerate}

When solving the equations of motion, it is convenient to use the proper reference frame of the field, where the electric and magnetic fields are parallel (or at least one of them vanishes), and their magnitude is given by the invariants:
\begin{eqnarray}
	\label{eq:eps_definition}
	\epsilon &=& \sqrt{\sqrt{ \mathcal{F}^2 + \mathcal{G}^2 } + \mathcal{F}},\\
	\eta &=& \sqrt{\sqrt{ \mathcal{F}^2 + \mathcal{G}^2 } - \mathcal{F}} 
\end{eqnarray}
where $\mathcal{F}=(\vec{E}^2 - c^2\vec{B}^2)/2$ and $\mathcal{G}=c\vec{B}\cdot\vec{E}$ are the field invariants. Mathematically, $\alpha_k=\{\epsilon,-\epsilon, i\eta,-i\eta\}$ are the eigenvalues of the field tensor $F^\mu_{\hphantom\mu\nu}(0)$ with $u_k^\mu$ being the corresponding eigenvectors:
\begin{equation}
	F(0) u_{k} = \dfrac{\alpha_{k}}{c} u_{k},\quad k=1,\ldots,4.
\end{equation}
The solution to Eq.~\eqref{eq:Lorentz} at the 0th order in $\omega t\ll 1$ can be expanded in four terms $\propto u_k e^{e \alpha_k \tau/mc}$. However, in the field of electric type $\epsilon>0$ and $\epsilon>\eta$. At $\tau> mc/e\epsilon$, the dominating contribution corresponds to $\propto u_1 e^{e \epsilon \tau/mc}$, and the solution simplifies (see Ref.~\cite{Arseny2021} for more details). The eigenvector $u_1$ can be expressed in terms of the laboratory frame field components:
\begin{equation}
u_1^\mu=\left(1,\, \frac{c^2(\vec{E}\cdot\vec{B})\vec{B}+\epsilon c[\vec{E}\times\vec{B}]+\epsilon^2\vec{E}}{ \epsilon(c^2 B^2+\epsilon^2)}\right).
\end{equation}
It is worth noting that, although $\epsilon$ represents the electric field strength in the proper frame of the field, in the laboratory frame this quantity depends both on the electric and magnetic components.

Under the above listed conditions, we solve Eq.~\eqref{eq:Lorentz} to the order $O((\omega t)^3)$ to find the key parameters:
\begin{eqnarray}
	%\gamma_{e}(t)  &\simeq& \dfrac{\epsilon  t}{E_S\tau_C},\label{eq:gamma_short_time} \\
	\gamma_{e}(t)  &\simeq& \dfrac{e\epsilon  t}{mc},\label{eq:gamma_short_time_app} \\
	\chi_{e}^2(t) &\simeq&  \chi_e^2(0)+ \left(\dfrac{\hbar e^2\epsilon^2 \omega_{\rm eff}}{m^3 c^4} t^2\right)^2 \label{eq:chi_short_time_app},
\end{eqnarray}
where $\chi_e(0)= e \hbar\sqrt{-(p^{(0)}_\mu F^{\mu\nu}(0))^2}/m^3c^4$, and $\oeff$ is defined by
\begin{gather}
	\omega^2_{\rm eff} = F_{\mu \nu, \sigma }(0) u^{\mu}_{ 1} u^{\sigma}_{1} (J^{-1})^{\nu}_{\phantom{1}\lambda}F^{\lambda}_{\phantom{1}\kappa, \rho }(0) u^{\rho}_{1} u^{\kappa}_{1}, \label{eq:omega_eff} \\
	J = \left( 2 \dfrac{\epsilon}{c}\boldsymbol{1}- F(0)\right)^2,
\end{gather}
The initial value of the $\chi$-parameter in Eq.~\eqref{eq:chi_short_time_app} can be omitted too as the second term rapidly becomes dominating at the time scales of interest for us. As a result, we get Eqs.~\eqref{eq:gamma_short_time}, \eqref{eq:chi_short_time}. We remark here that the combination $\oeff t^2$ is Lorentz invariant, and hence Eq.~\eqref{eq:chi_short_time_app}.

\section{Electron dynamics and cascade growth rate in asymptotically weak and strong fields}
\label{sec:appendix_asymot}
For reference purposes, we collect all the asymptotic expressions in the low- and high-field regimes with full numerical coefficients for the quantities introduced in the main text. 

The full probability rates of photon emission and $e^-e^+$ pair creation are given by:
\begin{gather}
	\label{eq:rate_CS_asympt_full}
	\WCS(\gamma_{e}, \chi_{e}) \simeq \left\lbrace
	\begin{array}{ll}
		\dfrac{5}{2\sqrt{3}} \dfrac{\alpha \chi_e}{\tau_C\gamma_e}, & \chi_e\ll 1, \\
		\dfrac{14\Gamma(2/3)}{3^{7/3}}\dfrac{\alpha \chi_e^{2/3}}{\tau_C\gamma_e} , & \chi_e\gg 1,
	\end{array}\right.\\
	\label{eq:rate_BW_asympt_full}
	\WBW(\gamma_{\gamma}, \chi_{\gamma}) \simeq \left\lbrace
	\begin{array}{ll}
		\sqrt{\dfrac{3}{2}} \dfrac{3}{16} \dfrac{\alpha \chi_\gamma}{\tau_C\gamma_\gamma} e^{-8/3\chi_\gamma}, & \chi_\gamma\ll 1, \\
		\dfrac{3^{5/3} 5\Gamma^4(2/3)}{28\pi^2}\dfrac{\alpha \chi_\gamma^{2/3}}{\tau_C\gamma_\gamma} , & \chi_\gamma\gg 1.
	\end{array}\right.
\end{gather}
We used Eq.~\eqref{eq:rate_CS_asympt_full} to calculate the asymptotic values for $\tem$, $\gem$, and $\cem$ as described in Section~\ref{sec:ii_c}. Thus, the time of emission reads:
\begin{equation}
	\label{eq:tem_asympt_full}
	\tem \simeq \left\lbrace
	\begin{array}{ll}
		\dfrac{2\cdot \sqrt[4]{3}}{\sqrt{5}\alpha}\sqrt{\dfrac{\tau_C \alpha E_S}{\oeff\epsilon}}, & \cem\ll 1,\\
		\dfrac{3}{\alpha}\left(\dfrac{2}{7\Gamma(2/3)}\right)^{3/4}\sqrt{\dfrac{\tau_C}{\oeff}} \left(\dfrac{\alpha E_S}{\epsilon}\right)^{1/4}, & \cem\gg 1.\\
	\end{array}
	\right.
\end{equation}
Here, $\epsilon$ is the invariant field strength as defined in Eq.~\eqref{eq:eps_definition}, and $\oeff$ is given by Eq.~\eqref{eq:omega_eff}. The values for $\gem$ and $\cem$ are obtained straightforwardly by substituting $\tem$ to Eqs.~\eqref{eq:gamma_short_time}-\eqref{eq:chi_short_time}:
\begin{gather}
	\label{eq:gamma_tem_asympt_exact}
	\gem \simeq \left\lbrace
	\begin{array}{ll}
		\dfrac{2\cdot \sqrt[4]{3}}{\sqrt{5}}\sqrt{\dfrac{\epsilon}{\oeff \tau_C \alpha E_S}}, & \cem\ll 1,\\
		 \dfrac{3}{\sqrt{\oeff\tau_C}} \left[\dfrac{2}{7\Gamma(2/3)}\dfrac{\epsilon}{\alpha E_S}\right]^{3/4}, & \cem\gg 1,\\
	\end{array}
	\right.\\
	\label{eq:chi_tem_asympt_exact}
	\cem \simeq \left\lbrace
	\begin{array}{ll}
		\dfrac{4\sqrt{3}}{5} \dfrac{\epsilon }{\alpha E_S}, &  \epsilon\ll \alpha E_S ,\\
		9\left[ \dfrac{2}{7\Gamma(2/3)}  \dfrac{\epsilon}{\alpha E_S}\right]^{3/2}, &  \epsilon\gg \alpha E_S.\\
	\end{array}
	\right.
\end{gather} 
Note that in the last equation, we express the asymptotic conditions in terms of the invariant field $\epsilon$, so that they are consistent with the initial hypothesis $\cem\ll 1$ or $\cem\gg 1$, respectively. The threshold $\epsilon=\alpha E_S\approx E_S/137$ separates the regimes of the avalanche-type cascade development. 

The effective emission probability rate $\Wrad=\WCS(\gem,\cem)$ at $\tem$ at asymptotically low and high fields can be expressed as (see the corresponding discussion in Section~\ref{sec:iii_b}):

\begin{widetext} 
\begin{equation}
\label{eq:rate_CS_tem}
 \WCS(\gem,\cem)\simeq \dfrac{\alpha}{\tau_C}\times \left\lbrace
	\begin{array}{ll}
		\dfrac{\sqrt{5}}{\sqrt[4]{3}}  \sqrt{ \dfrac{\epsilon\oeff\tau_C}{\alpha E_S}}, &  \epsilon\ll \alpha E_S,\\
		\dfrac{\sqrt{2}}{9}\left[14\,\Gamma(2/3) \right]^{3/4} \sqrt{\oeff \tau_C}\left(\dfrac{\epsilon}{\alpha E_S}\right)^{1/4}, &  \epsilon\gg \alpha E_S.\\
	\end{array}
	\right.
\end{equation}

Finally, we collect the asymptotic expressions for the particle growth rate derived and discussed in Section~\ref{sec:iii_c} and the used numerical coefficients:
\begin{gather}
	\renewcommand{\arraystretch}{2.8}
	\label{eq:gamma_all_asympt}
	\Gamma(\epsilon) \simeq \left\lbrace 
	\begin{array}{ll}
		-\dfrac{\nu_e+\nu_\gamma-|\Delta\nu|}{2}
		+ \dfrac{\alpha}{\tau_C} \dfrac{ 3^{5/3}\, 5\Gamma^4\left(2/3\right)}{ 7\pi^2}\dfrac{\epsilon\oeff}{E_S |\Delta\nu|}\exp\left(-\dfrac{10 \alpha E_S}{3\sqrt{3}\epsilon}\right), & \epsilon\ll \alpha E_S,\,\,\Delta\nu>0,\\
		-\nu + \dfrac{\alpha}{\tau_C} \dfrac{ 3^{5/6}\sqrt{5}}{\sqrt{7}\pi}\Gamma^2\left(2/3\right) \sqrt{\dfrac{\epsilon\oeff\tau_C}{\alpha E_S} }\exp\left(-\dfrac{5 \alpha E_S}{3\sqrt{3}\epsilon}\right), & \epsilon\ll \alpha E_S,\,\,\nu_e=\nu_\gamma=\nu,\\
		\dfrac{\alpha}{\tau_C}c_1 \dfrac{\epsilon}{\alpha E_S} \left[ \sqrt{1+ c_2\sqrt{\oeff \tau_C}\left(\dfrac{\alpha E_S}{\epsilon} \right)^{3/4}} -1\right],& \alpha E_S<\epsilon<c_3\alpha(\oeff\tau_C)^{2/3},\\
		\dfrac{\alpha}{\tau_C}\dfrac{2\sqrt{2}}{9}\left[14\,\Gamma(2/3) \right]^{3/4} \sqrt{\oeff \tau_C}\left(\dfrac{\epsilon}{\alpha E_S}\right)^{1/4}, & c_3\alpha(\oeff\tau_C)^{2/3}\ll \epsilon.
	\end{array}
	\right.\\
	\label{eq:c1}
	c_1 = \dfrac{\tau_C \WBW(1,1)}{2}\approx 0.516\times 10^{-4},\\
	\label{eq:c2}
	c_2 = \dfrac{8\sqrt{2} \left(14 \Gamma(2/3) \right)^{3/4} }{9 \tau_C \WBW(1,1)}\approx 1.107\times 10^5,\\
	\label{eq:c3}
	c_3 =  \left(\dfrac{2}{3}\right)^{2/3}\dfrac{224 \Gamma(2/3)}{9(\tau_C \WBW(1,1))^{4/3}}\approx 5.31\times 10^6,\\
	\label{eq:WBW11}
	\tau_C \WBW(1,1)\approx 1.0318068870\times10^{-4}.
\end{gather}

\section{Particle number equations in quasi-steady state}
\label{sec:appendix_kinetic}
Cascade equations in general form read
\begin{gather}
	\label{eq:kin_e}
	\begin{split}
		\left[\frac{\partial }{\partial t} + \frac{c^2\vec{p}_e}{\varepsilon_e}\cdot\frac{\partial }{\partial \vec{r}}\pm\frac{\partial}{\partial \vec{p}_e}\cdot\vec{F}_L\right] f_\pm (\vec{r},\vec{p}_e,t)
		= & \int \rmd\vec{p}_\gamma\, W_{\mathrm{rad}}(\vec{p}_e+\vec{p}_\gamma \rightarrow \vec{p}_\gamma) f_\pm(\vec{r},\vec{p}_e+\vec{p}_\gamma,t)\\
		& - W_{\mathrm{rad}} (\vec{p}_e) f_\pm(\vec{r},\vec{p}_e,t)  + \int \rmd \vec{p}_\gamma \, W_{\mathrm{cr}}(\vec{p}_\gamma \rightarrow \vec{p}_e) f_\gamma(\vec{r},\vec{p}_\gamma,t),
	\end{split}\\
	\label{eq:kin_g}
	\begin{split}
		\left[\frac{\partial }{\partial t} + \frac{c^2\vec{p}_\gamma}{\varepsilon_\gamma}\cdot\frac{\partial }{\partial \vec{r}}\right] f_\gamma (\vec{r},\vec{p}_\gamma,t)
		= & \int \rmd\vec{p}_e\, W_{\mathrm{cr}}(\vec{p}_e \rightarrow \vec{p}_\gamma)\left[ f_-(\vec{r},\vec{p}_e,t)+f_+(\vec{r},\vec{p}_e,t)\right]\\
		& - W_{\mathrm{cr}} (\vec{p}_\gamma) f_\gamma(\vec{r},\vec{p}_\gamma,t),
	\end{split}
\end{gather}
\end{widetext} 
where $f_{a} (\vec{r},\vec{p},t)$, $a=\{\pm,\gamma\}$, are the electron, positron, and photon distribution functions. The LHS of the equations describes the continuous dynamics of particles with momenta $\vec{p}_{e,\gamma}$ and energies $\varepsilon_{e,\gamma}$ for $e^\pm$ and $\gamma$, respectively; $\pm\vec{F}_L=\pm e\left(\vec{E}+c^2\vec{p}_e\cross\vec{B}/\varepsilon_e\right)$ is the Lorentz force acting on a charged particle. In the RHS, we collect the source terms corresponding to the stochastic processes of photon emission and pair creation, where $W_{\mathrm{rad}}(\vec{p}_e \rightarrow \vec{p}_\gamma)$ is the differential probability rate for an electron or positron with momentum $\vec{p}_e$ to emit a photon with momentum $\vec{p}_\gamma$, and $W_{\mathrm{cr}}(\vec{p}_\gamma \rightarrow \vec{p}_e)$ is the differential probability rate for a photon with momentum $\vec{p}_\gamma$ to create a pair such that $e^-$ or $e^+$ gets momentum $\vec{p}_e$.  Equations~\eqref{eq:kin_e} and~\eqref{eq:kin_g} provide a starting point to model QED cascades in various scenarios, including electromagnetic air showers \cite{gaisser2016cosmic}, shower-type cascades in interaction of high-energy beams with laser pulses \cite{bulanov2013electromagnetic}, and avalanche-type cascades \cite{elkina2011qed, nerush2011analytical}. Let us point out here that the probability rates depend on the field strength, which in turn can depend on $\vec{r}$ and $t$, namely, $W_{\mathrm{cr,rad}}(\vec{p}_a \rightarrow \vec{p}_b)\equiv W_{\mathrm{cr,rad}}(\vec{p}_a \rightarrow \vec{p}_b;\vec{E}(\vec{r},t), \vec{B}(\vec{r},t))$, $W_{\mathrm{cr,rad}}(\vec{p}_a)\equiv W_{\mathrm{cr,rad}}(\vec{p}_a;\vec{E}(\vec{r},t), \vec{B}(\vec{r},t))$. Within the LCFA, the fields enter through the quantum parameter $\chi_{e,\gamma}$ of the incoming particle [see Eq.~\eqref{eq:chi}]. For brevity, we omit the explicit field dependence in our notation.

Let us consider an avalanche-type $e^-$-seeded cascade. We assume that $e^-$ is injected at $\vec{r}=0$ in the strong-field area optimal for the electron acceleration (e.g. the $E$-antinode center in a standing CP standing wave). We expect that in the triggered cascade, the number of produced pairs grows exponentially with time, $N_p\sim e^{\Gamma t}$. Also, we assume that the mean free path time of $e^\pm$ and $\gamma$ in the cascade is small (see Section~\ref{sec:ii}). Under these conditions, the majority of particles will be created in the small area, which we denote as domain $D$, where the process of continuous acceleration is optimized. Hence, to identify the dependence of $\Gamma$ on the field parameters, it is enough to study the particle distributions within $D$. Assuming that the field is almost homogeneous in $D$, let us integrate out the coordinate dependence $\vec{r}$. For Eq.~\eqref{eq:kin_e}, we obtain: 
\begin{widetext} 
\begin{equation}
	\label{eq:kin_e2}
	\begin{split}
		\frac{\partial }{\partial t} f_\pm (\vec{p}_e,t) + \frac{c^2\vec{p}_e}{\varepsilon_e}\cdot\int_{\partial D} \rmd\vec{S}\, f_\pm (\vec{r},\vec{p}_e,t)  \pm & \frac{\partial}{\partial \vec{p}_e}\cdot\int_D \rmd\vec{r}\, \vec{F}_L f_\pm (\vec{r},\vec{p}_e,t) \\
		= &\int \rmd\vec{p}_\gamma\, W_{\mathrm{rad}}(\vec{p}_e+\vec{p}_\gamma \rightarrow \vec{p}_\gamma) f_\pm(\vec{p}_e+\vec{p}_\gamma,t) \\
		& - W_{\mathrm{rad}} (\vec{p}_e) f_\pm(\vec{p}_e,t)  + \int \rmd\vec{p}_\gamma\, W_{\mathrm{cr}}(\vec{p}_\gamma \rightarrow \vec{p}_e) f_\gamma(\vec{p}_\gamma,t),
	\end{split}
\end{equation}
where we introduced
\[f_{\pm,\gamma} (\vec{p}_e,t) = \int_D \rmd\vec{r}\, f_{\pm,\gamma} (\vec{r},\vec{p}_e,t). \]
Here, to simplify the RHS, we approximate the field dependence of the probability rates by a uniform field in $D$, $W_{\mathrm{cr,rad}}(\vec{p}_a \rightarrow \vec{p}_b)\equiv W_{\mathrm{cr,rad}}(\vec{p}_a \rightarrow \vec{p}_b;\vec{E}(0,t), \vec{B}(0,t))$. For the second term in the LHS, we used the divergence theorem to cast it into the surface integral. It accounts for the particle flux through the surface $\partial D$. The flux can be caused by the migration of charged particles in the direction of the electric field gradient. In the field of a standing CP wave, particles migrate in the longitudinal direction to the $B$-antinodes. Then $D$ is formed by two transverse planes sandwiching the $E$-antinode (see Fig.~\ref{fig:fields_invariant_1D_standing}). In focused laser beams (e.g. Gaussian beams), migration in the transverse direction can be also naturally accounted for in Eq.~\eqref{eq:kin_e2} (see Fig.~\ref{fig:avalanche}), as in this case, $D$ is a closed surface containing the focus, e.g. a cylinder.

In the events of emission or pair creation, the incoming and outgoing particles are ultrarelativistic, thus $\varepsilon_e\approx c|\vec{p}_e|$. Following \cite{elkina2011qed,nerush2011analytical}, we assume that in each quantum process the momenta of the initial and secondary particles are collinear: 
\begin{gather*}
	\begin{split}
	W_{\mathrm{rad}}(\vec{p}_e \rightarrow \vec{p}_\gamma)=
	\int_0^1 \rmd\lambda\, \delta(\vec{p}_\gamma-\lambda \vec{p}_e) \varepsilon_e \left.\frac{\rmd W_{\mathrm{rad}}(\vec{p}_e,\varepsilon_\gamma)}{\rmd \varepsilon_\gamma}\right|_{\varepsilon_\gamma=\lambda\varepsilon_e},
	\end{split}\\
	\begin{split}
	W_{\mathrm{cr}}(\vec{p}_\gamma \rightarrow \vec{p}_e)=
	\int_0^1 \rmd\lambda\, \delta(\vec{p}-\lambda \vec{p}_\gamma) \varepsilon_\gamma \left.\frac{\rmd W_{\mathrm{cr}}(\vec{p}_\gamma,\varepsilon'_e)}{\rmd \varepsilon'_e}\right|_{\varepsilon'_e=\lambda\varepsilon_\gamma}.
	\end{split}
\end{gather*}
Note that within the LCFA, the probability rates explicitly depend only on the particle energy and $\chi$ parameter:
\begin{equation}
	\label{eq:diff_rates_full}
	\begin{split}
	\frac{\rmd W_{\mathrm{rad}}(\vec{p}_e,\varepsilon_\gamma)}{\rmd \varepsilon_\gamma} = \frac{\rmd W_{\text{rad}}(\varepsilon_e,\chi_e,\varepsilon_\gamma)}{\rmd \varepsilon_\gamma},\,\,
	\frac{\rmd W_{\mathrm{cr}}(\vec{p}_\gamma,\varepsilon_e)}{\rmd \varepsilon_e} = \frac{\rmd W_{\text{cr}}(\varepsilon_\gamma,\chi_\gamma,\varepsilon_e)}{\rmd \varepsilon_e},
	\end{split}
\end{equation}
[c.f. Eqs.~\eqref{eq:rate_CS} and~\eqref{eq:rate_BW}].\footnote{We use particle energy $\varepsilon_{e,\gamma}=\gamma_{e,\gamma}mc^2$ as the argument for the probability rates to simplify the notations within Appendix~\ref{sec:appendix_kinetic}. The transition to the notation used in Eqs.~\eqref{eq:rate_CS}, \eqref{eq:rate_BW} is straightforward.} By applying this to the RHS of Eq.~\eqref{eq:kin_e2}, we can integrate out the angular dependence in $\vec{p}_\gamma$. After repeating the same steps for Eq.~\eqref{eq:kin_g}, we get:
\begin{gather}
	\label{eq:kin_e3}
	\begin{split}
		\frac{\partial }{\partial t} &  f_\pm (\vec{p}_e,t) + \frac{c^2\vec{p}_e}{\varepsilon_e}\cdot\int_{\partial D} \rmd\vec{S}\, f_\pm (\vec{r},\vec{p}_e,t)  \pm  \frac{\partial}{\partial \vec{p}_e}\cdot\int_D \rmd\vec{r}\, \vec{F}_L\, f_\pm (\vec{r},\vec{p}_e,t) =
		- W_{\mathrm{rad}} (\vec{p}_e) f_\pm(\vec{p}_e,t)\\
		& + \int_{\varepsilon_e}^\infty \rmd\varepsilon' \frac{\varepsilon^{\prime 2}}{\varepsilon_e^2}
		\left.\left[ 
		\left.\frac{\rmd W_{\mathrm{rad}}(\vec{p}',\varepsilon_\gamma)}{\rmd \varepsilon_\gamma}\right|_{\varepsilon_\gamma=\varepsilon'-\varepsilon_e} f_\pm \left(\vec{p}',t\right) +
		\frac{\rmd W_{\mathrm{cr}}(\vec{p}',\varepsilon_e)}{\rmd \varepsilon_e} f_\gamma \left(\vec{p}',t\right) 
		\right] \right|_{\vec{p}'=\varepsilon'\frac{c\vec{p}_e}{\varepsilon_e}}.
	\end{split}\\
	\label{eq:kin_g3}
	\begin{split}
		\frac{\partial }{\partial t}   f_\gamma (\vec{p}_\gamma,t) + \frac{c^2\vec{p}_\gamma}{\varepsilon_\gamma}\cdot\int_{\partial D} \rmd\vec{S}\, f_\gamma (\vec{r},\vec{p}_\gamma,t)  = &
		- W_{\mathrm{cr}} (\vec{p}_\gamma) f_\gamma(\vec{p}_\gamma,t)\\
		& + \int_{\varepsilon_\gamma}^\infty \rmd\varepsilon' \frac{\varepsilon^{\prime 2}}{\varepsilon_\gamma^2}
		\frac{\rmd W_{\mathrm{rad}}(\vec{p}',\varepsilon_\gamma)}{\rmd \varepsilon_\gamma} 
		\left. 
		\left[  f_- \left(\vec{p}',t\right) + f_+ \left(\vec{p}',t\right) \right]
		\vphantom{\frac12}\right|_{\vec{p}'=\varepsilon'\frac{c\vec{p}_\gamma}{\varepsilon_\gamma}}.
	\end{split}
\end{gather}
For a uniform field, for example, if $\vec{F}_L=e\vec{E}(t)$ (this includes a uniform rotating electric field), the flux terms vanish and we recover the equations used in Refs.~\cite{elkina2011qed, nerush2011analytical}.

As we aim to obtain the equations for particle numbers, we integrate out the particle momenta from  Eqs.~\eqref{eq:kin_e3}-\eqref{eq:kin_g3}:
\begin{gather}
	\label{eq:kin_e4}
	\frac{\partial }{\partial t} N_\pm (t)+\nu_\pm N_\pm (t)= \int\rmd \vec{p}_e\int_{\varepsilon_e}^\infty \rmd\varepsilon' \frac{\varepsilon^{\prime 2}}{\varepsilon_e^2} 
	\left.
	\frac{\rmd W_{\mathrm{cr}}(\vec{p}',\varepsilon_e)}{\rmd \varepsilon_e} f_\gamma \left(\vec{p}',t\right)
	\right|_{\vec{p}'=\varepsilon'\frac{c\vec{p}_e}{\varepsilon_e}}
	, 
	\\ \label{eq:kin_g4}
	\frac{\partial }{\partial t} N_\gamma(t)+ \nu_\gamma N_\gamma (t)= \int\rmd\vec{p}_\gamma\left\lbrace- W_{\mathrm{cr}} (\vec{p}_\gamma) f_\gamma(\vec{p}_\gamma,t)
	+ \int_{\varepsilon_\gamma}^\infty \rmd\varepsilon' \frac{\varepsilon^{\prime 2}}{\varepsilon_\gamma^2}
	\frac{\rmd W_{\mathrm{rad}}(\vec{p}',\varepsilon_\gamma)}{\rmd \varepsilon_\gamma} 
	\left. 
	\left[  f_- \left(\vec{p}',t\right) + f_+ \left(\vec{p}',t\right) \right]
	\vphantom{\frac12}\right|_{\vec{p}'=\varepsilon'\frac{c\vec{p}_\gamma}{\varepsilon_\gamma}} 
	\right\rbrace,
\end{gather}
where 
\begin{eqnarray}
	\label{eq:N_def}
	N_{a}(t) = \int\rmd\vec{p}\,f_{a}(\vec{p},t),\\
	\vec{j}_a(\vec{r},t)=\int\rmd\vec{p}_a\, \frac{c^2\vec{p}_a}{\varepsilon_a} f_a (\vec{r},\vec{p}_a,t),\\
	\label{eq:nu_def}
	\nu_a(t) = \frac{1}{N_\pm(t)}  \int_{\partial D} \rmd\vec{S}\, \vec{j}_a(\vec{r},t)
\end{eqnarray}
are the particle numbers, flows, and effective migration rates, respectively. Motivated by the symmetry argumentation, let us assume that the migration rates for electrons and positrons are equal: $\nu_-(t)=\nu_+(t)\equiv\nu_e(t)$. Note that in Eq.~\eqref{eq:kin_e4} we used that 
\[
\int\rmd \vec{p}_e\left\lbrace
\int_{\varepsilon}^\infty \rmd\varepsilon' \frac{\varepsilon^{\prime 2}}{\varepsilon_e^2}
\left. 
\left[
\left.\frac{\rmd W_{\mathrm{rad}}(\vec{p}',\varepsilon_\gamma)}{\rmd \varepsilon_\gamma}\right|_{\varepsilon_\gamma=\varepsilon'-\varepsilon_e} f_\pm \left(\vec{p}',t\right) 
\right] \right|_{\vec{p}'=\varepsilon'\frac{c\vec{p}_e}{\varepsilon_e}} - W_{\mathrm{rad}} (\vec{p}_e) f_\pm(\vec{p}_e,t)\right\rbrace=0,
\]
which can be shown by interchanging the integrals in the first term and noticing that $\rmd W_{\mathrm{rad}}(\vec{p}',\varepsilon_\gamma-\varepsilon_e)/\rmd\varepsilon_\gamma=-\rmd W_{\mathrm{rad}}(\vec{p}',\varepsilon_\gamma-\varepsilon_e)/\rmd\varepsilon_e$ under the integral. Physically, it corresponds to the balance of radiation.

To simplify the equations further, let us consider the following term in the RHS of Eq.~\eqref{eq:kin_g4}:
\begin{equation}
\int\rmd\vec{p}_\gamma \int_{\varepsilon_\gamma}^\infty \rmd\varepsilon' \frac{\varepsilon^{\prime 2}}{\varepsilon_\gamma^2}
	\frac{\rmd W_{\mathrm{rad}}(\vec{p}',\varepsilon_\gamma)}{\rmd \varepsilon_\gamma} 
	\left. f_\pm \left(\vec{p}',t\right) \vphantom{\frac12}\right|_{\vec{p}'=\varepsilon'\frac{c\vec{p}_\gamma}{\varepsilon_\gamma}} 
\end{equation}
It is convenient to pass to spherical coordinates in momentum space, $\vec{p}_a\rightarrow \{\varepsilon_a/c,\vec{n}_a\}$, where $\vec{n}_a=\vec{p}_a/|\vec{p}_a|$ is the unit vector. The emission collinearity condition $\vec{p}'=\varepsilon'\frac{c\vec{p}_\gamma}{\varepsilon_\gamma}$ simply means that the integration over $\vec{n}_\gamma$ in the outer integral can be replaced by $\vec{n}'$:
\begin{equation}
\label{eq:eq:kin_g4_RHS}
\frac{1}{c^3}\int_0^\infty\rmd \varepsilon_\gamma d\vec{n}'\,\varepsilon_\gamma^2\int_{\varepsilon_e}^\infty \rmd\varepsilon' \frac{\varepsilon^{\prime 2}}{\varepsilon_\gamma ^2} 
	\frac{\rmd W_{\mathrm{rad}}(\varepsilon',\chi',\varepsilon_\gamma)}{\rmd \varepsilon_\gamma} f_\pm \left(\varepsilon',\vec{n}',t\right),
\end{equation}
Note that we wrote the argument of $d\Wrad/d\varepsilon_\gamma$ so that it corresponds to Eq.~\eqref{eq:diff_rates_full}, and $\chi'$ depends on $\vec{n}'$ (e.g. in the reference frame co-moving with the electric field component \cite{nerush2011analytical}). Now we can interchange the $d\varepsilon_\gamma$ and $d\varepsilon'$ integrals:
\begin{equation}
\frac{1}{c^3}\int_0^\infty\rmd \varepsilon' d\vec{n}'\, \varepsilon'^2 f_\pm \left(\varepsilon',\vec{n}', t\right)\int_{0}^{\varepsilon'} \rmd\varepsilon_\gamma 
	\frac{\rmd W_{\mathrm{rad}}(\varepsilon',\chi',\varepsilon_\gamma)}{\rmd \varepsilon_\gamma} 
\end{equation}
The rightmost integral gives $\Wrad(\varepsilon',\chi')$, and the whole expression can be written in a compact form:
\begin{equation}
\label{eq:eq:kin_g4_RHS_3}
\int\rmd \vec{p}'\, \Wrad(\vec{p}') f_\pm(\vec{p}',t).
\end{equation}
After repeating the same steps for the RHS of Eq.~\eqref{eq:kin_e4}, we can rewrite the particle number equations as follows:
\begin{align}
\label{eq:kin_e5}
\frac{\partial }{\partial t} N_\pm (t) &=-\nu_e(t) N_\pm (t) + \Wcr(t)N_\gamma(t),\\
\label{eq:kin_g5}
\frac{\partial }{\partial t} N_\gamma(t) &=-\nu_\gamma(t) N_\gamma(t)+ \Wrad(t)\left[N_-(t)+N_+(t)\right]-\Wcr(t)N_\gamma(t),
\end{align}
where we introduced time-dependent effective rates:
\begin{gather}
\label{eq:Wrad_t}
\Wrad(t) = \frac{1}{N_-(t)+N_+(t)} \int\rmd\vec{p}'\Wrad(\vec{p}') \left[f_-(\vec{p}',t)+f_+(\vec{p}',t)\right],\\
\label{eq:Wcr_t}
\Wcr(t) = \frac{1}{N_\gamma(t)}\int\rmd \vec{p}'\, \Wcr(\vec{p}') f_\gamma(\vec{p}',t).
\end{gather}
Recall that, here, the probability rates also depend on time, $W_{\mathrm{rad,cr}} (\vec{p}')\equiv W_{\mathrm{rad,cr}} (\vec{p}'; \vec{E}(0,t), \vec{B}(0,t))$.

To advance, we assume now that the cascade rapidly reaches the (quasi-)steady state, and that the time dependence in $f_a$ factorizes: $f_{a}(\vec{p}_a,t)=f_{a}(\vec{p}_a)g_a(t)$ [anticipating that $g_a(t)\sim e^{\Gamma t}$]. For a general field, this assumption can be justified at a timescale $\delta t$: $\tem \ll \delta t \ll \omega^{-1}$, namely, if the cascade becomes abundant at a sub-cycle scale. However, for a CP standing wave, this hypothesis can be extended to larger timescales, as the probability rates are independent of time in the frame co-rotating with the field. Transition to such the frame can be done for Eqs.~\eqref{eq:kin_e}, \eqref{eq:kin_g} (see Ref.~\cite{nerush2011analytical}), however, does not change the form of Eqs.~\eqref{eq:kin_e5}, \eqref{eq:kin_g5}. In addition, we rely on LCFA, so that $\Wrad(\vec{p}')=\Wrad(\varepsilon',\chi')$, $\Wcr(\vec{p}')=\Wcr(\varepsilon',\chi')$, hence, $\chi'$ can be used as an independent integration variable in Eqs.~\eqref{eq:Wrad_t}, \eqref{eq:Wcr_t}. Passing to the new variables $\vec{p}'\rightarrow\{\varepsilon',\chi',\theta'\}$ casts the distribution function to
\begin{equation}
\label{eq:distrib_f_e_spherical}
f_\pm \left(\vec{p}'\right) d\vec{p}'= \tilde{f}_\pm(\varepsilon',\chi',\theta')d\varepsilon'd\chi'\sin\theta' d\theta',
\end{equation}
where $-\pi/2<\theta'\leq \pi/2$ is the azimuthal angle in the system such that the electric field is in the plane $\theta'=0$. As a result of the distribution function factorisation $g_a(t)$ cancels out in Eqs.~\eqref{eq:Wrad_t}, \eqref{eq:Wcr_t} [with $N_a(t)$ written explicitly, see Eq.~\eqref{eq:N_def}], and the effective rates become stationary:
\begin{gather}
\label{eq:Wrad_stationary}
\Wrad = \frac{\displaystyle \int d\varepsilon'd\chi'\sin\theta' d\theta' \,\Wrad(\varepsilon',\chi') \left[\tilde{f}_-(\varepsilon',\chi',\theta')+\tilde{f}_+(\varepsilon',\chi',\theta')\right]}{\displaystyle \int d\varepsilon'd\chi'\sin\theta' d\theta' \left[\tilde{f}_-(\varepsilon',\chi',\theta')+\tilde{f}_+(\varepsilon',\chi',\theta')\right]},\\
\label{eq:Wcr_stationary}
\Wcr = \frac{\displaystyle \int d\varepsilon'd\chi'\sin\theta' d\theta'\, \Wcr(\varepsilon',\chi') \tilde{f}_\gamma(\varepsilon',\chi',\theta')}{\displaystyle \int d\varepsilon'd\chi'\sin\theta' d\theta'\, \tilde{f}_\gamma(\varepsilon',\chi',\theta')}.
\end{gather}
\end{widetext} 
We can apply the same reasoning and pass to the stationary migration rate, $\nu_e(t)\rightarrow\nu_e$, see Eq.~\eqref{eq:nu_def}. By using these expressions in Eqs.~\eqref{eq:kin_e5}, \eqref{eq:kin_g5}, we finally arrive at the particle number equations in the steady state:
\begin{eqnarray}
	\label{eq:steady_state_equation_e}
	\frac{\rmd N_\pm(t)}{\rmd t} = W_{\mathrm{cr}} N_\gamma(t) - \nu_e N_\pm(t),\\
	\label{eq:steady_state_equation_g}
	\begin{split}
	\frac{\rmd N_\gamma(t)}{\rmd t} = &W_{\mathrm{rad}}[N_-(t) + N_+(t)] \\
	&	-W_{\mathrm{cr}} N_\gamma(t)  - \nu_\gamma N_\gamma(t).
	\end{split}
\end{eqnarray}
By introducing $N_p=(N_-+N_+)/2$ we cast these relations into Eqs.~\eqref{eq:const_rate_eq1}-\eqref{eq:const_rate_eq2}. At $\nu_e=\nu_\gamma=0$, we get the equations in the form used in Refs.~\cite{grismayer2017seeded, bashmakov2014effect, luo2018qed}. 

The effective rate of photon emission given in Eqs.~\eqref{eq:Wrad_stationary} can be simplified further if we take into account the electron dynamics. Let us consider the integral:
\[
\int d\varepsilon'd\chi'\sin\theta' d\theta' \,\Wrad(\varepsilon',\chi') \tilde{f}_\pm(\varepsilon',\chi',\theta')
\]
The short-time expressions \eqref{eq:gamma_short_time}-\eqref{eq:chi_short_time} relate $\varepsilon_e$ and $\chi_e$ as they are parametrised. We can express $\chi_e$ in terms of $\varepsilon_e$: $\chi_e(\varepsilon_e)=(\varepsilon_e^2/mc^2)^2 \oeff\tau_C$, meaning that the distribution function can be written as
\begin{equation}
\label{eq:distrib_f_e_chi}
\tilde{f}_\pm(\varepsilon',\chi',\theta')=\tilde{f}_\pm(\varepsilon',\theta')\delta(\chi'-\chi_e(\varepsilon')).
\end{equation}
Notably, in the uniform rotating electric field configuration, we can make an additional simplification: $\tilde{f}_\pm(\varepsilon',\theta')=\tilde{f}_\pm(\varepsilon')\delta(\theta')/\sin\theta'$. Then the integral reads
\[
\int d\varepsilon'\sin\theta' d\theta' \,\Wrad(\varepsilon',\chi_e(\varepsilon')) \tilde{f}_\pm(\varepsilon',\theta').
\]
In Section~\ref{sec:iii_b}, we propose to approximate the rate in this integral by a constant $\Wrad(\varepsilon',\chi_e(\varepsilon'))\approx\WCS(\gem,\cem)$, where $\cem=\chi_e(\gem mc^2)$. Then the effective rate in Eq.~\eqref{eq:Wrad_stationary} simplifies to $\Wrad\approx\WCS(\gem,\cem)$.

Developing a similar line of reasoning applied to the RHS of Eq.~\eqref{eq:Wcr_stationary}, requires coupling $\varepsilon_\gamma$ and $\chi_\gamma$.
Since photons propagate along straight lines, and their energy is not altered, let us estimate $\chi_\gamma(\varepsilon_\gamma)\sim \sin(\theta_\gamma) \epsilon \varepsilon_\gamma/(E_S mc^2)$, where $\theta_\gamma$ is some characteristic angle between the photon momentum and the electric field direction. Then we can rewrite the photon distribution function as in Eq.~\eqref{eq:distrib_f_e_chi}, substitute it into the effective rate \eqref{eq:Wcr_stationary}, and obtain $\Wcr\approx \WBW(\varepsilon_\gamma/(mc^2), \sin(\theta_\gamma) \epsilon \varepsilon_\gamma/(E_S mc^2))$. Here, $\sin(\theta_\gamma)$ can be used as a fitting parameter when comparing the resulting growth rate expressions to numerical data.

Let us make several remarks. First, our derivation shows that the steady-state hypothesis, namely, the factorization of the distribution function time dependence, is equivalent to taking the limit $W_\mathrm{rad,cr}(t\gg\tem)\rightarrow W_\mathrm{rad,cr}$. The solution to the steady-state master equations is exponential in time, as we discuss in Section~\ref{sec:iii}.

Second, the particular time required for this limit to set in, i.e. the steady state formation time, depends on the particular field configuration and its parameters. As of now, the formation of the steady distribution has to be confirmed with numerical simulations for each field configuration, which was done in the past works \cite{elkina2011qed,grismayer2017seeded,gelfer2015optimized,Jirka2016,Arseny2021} and the present study. A rigorous proof of the steady state existence would be very interesting, as well as the calculation of the relaxation time, and a systematic classification of the field configurations allowing such a state.

Third, as we mentioned, the approach implies that the steady state can be formed rapidly (when compared to time scale $\omega^{-1}$). It can be extended to larger timescales beyond $\omega^{-1}$ by assuming that the factorized part of the distribution function $f_a(\vec{p})$ changes with time adiabatically. As a result, the effective rates and the resulting cascade growth rate will exhibit `slow' time dependence at the scale $\omega^{-1}$. However, if the field configuration (e.g. a rotating $E$-field or a CP standing wave $E$-antinode) has some symmetry allowing for uniform acceleration at any time scale, effective rates can remain constant at this time scale. The same argumentation can be applied to take into account the spatial variation of the field at the scale $c/\omega$.

And last, the cascade growth rate $\Gamma$ is extracted from the solution to Eqs.~\eqref{eq:steady_state_equation_e}-\eqref{eq:steady_state_equation_g}. However, the effective rates $\Wradcr$ and the migration coefficient $\nu_e$ are defined via the distribution functions, which are unknown \textit{a priori}. While solving the full distribution function analytically is intricate, the rates can be estimated under additional approximations, as we propose in Sec.~\ref{sec:iii_b}, or extracted from numerical simulations, as we do in Sec.~\ref{sec:iv}.

\section{The model of Grismayer et al \cite{grismayer2017seeded}}
\label{sec:appendix_grismayer}
The cascade growth rate model proposed by Grismayer et al \cite{grismayer2017seeded} [see Eqs.~(9)--(10) therein] for the rotating $E$-field was advancing the models of Fedotov et al \cite{fedotov2010limitations} and Bashmakov et al \cite{bashmakov2014effect}. The improvement was achieved by reconsidering the assumption that almost all of the energy of a radiating electron is transferred to a photon. Instead, the electron radiation spectrum $d\WCS/d\chi_\gamma$ with $0<\chi_\gamma<\chi_e$ was fully accounted for in Ref.~\cite{grismayer2017seeded}. At the same time, Grismayer et al follow the preceding works by assuming that the electron mean free path time is given by $t_e\sim W_{\mathrm{rad}}^{-1}$:
\[W_\mathrm{rad}\equiv \WCS(\bar{\gamma}_e,\bar{\chi}_e)=\int_0^{\bar{\chi}_e} d\chi_\gamma \frac{d\WCS(\bar{\gamma}_e,\bar{\chi}_e,\chi_\gamma)}{d\chi_\gamma},\]
where $\bar{\gamma}_e$ and $\bar{\chi}_e$ are the characteristic values of $\gamma_e$ and $\chi_e$ at the moment of emission, which depend only on the field parameters. The corresponding model for the number of produced pairs was introduced in Refs.~\cite{Grismayer2016, grismayer2017seeded} by using a phenomenological approach. We show, how this model arises from the kinetic approach and discuss the approximations that are required to obtain it. 

Let us take Eq.~\eqref{eq:kin_e4} coupled to Eq.~\eqref{eq:kin_g3} as the starting point. We set the migration term to zero in the context of the uniform rotating electric field model. As in Appendix~\ref{sec:appendix_kinetic}, we assume that the cascade reaches the steady state.
For Eq.~\eqref{eq:kin_e4}, we repeat the steps proposed in Appendix~\ref{sec:appendix_kinetic}, and arrive at:
\begin{equation}
	\label{eq:kin_e6}
	\frac{\partial }{\partial t} N_\pm (t) = \int\rmd \vec{p}'  W_{\mathrm{cr}}(\varepsilon',\chi') f_\gamma \left(\vec{p}', t\right).
\end{equation}
We aim at obtaining a closed equation on $N_\pm (t)$. For this, we will use Eq.~\eqref{eq:kin_g3} to express $f_\gamma\left(\varepsilon_e,\chi_e,t\right)$: we approximate the integral over $\varepsilon'$ in the RHS in order to pass to the particle numbers $N_\pm (t)$, and then integrate the whole equation in time. 

As we aim at keeping the differential rate in the expression, $dW_{\mathrm{rad}}/d\chi_\gamma=d\WCS/d\chi_\gamma$, the second integral in the RHS  Eq.~\eqref{eq:kin_g3} can be simplified only by imposing an additional assumption for the distribution functions $f_\pm(\vec{p}_e,t)$. In the steady state, the normalized electron spectrum does not change with time and has a prominent peak, as was noted by Grismayer et al~\cite{grismayer2017seeded} and confirmed in the current work with numerical simulations. We rewrite the steady-state distribution function in the form:
\begin{equation}
	f_\pm(\vec{p}_e,t) = N_\pm(t) \frac{1}{\mathcal{N}} e^{g(\varepsilon_e)},
\end{equation} 
where $\mathcal{N}$ normalizes the electron spectrum, $\mathcal{N}=\int d\varepsilon_e e^{g(\varepsilon_e)}$. Let us assume for the moment that $g(\varepsilon_e)$ has a single maximum located at $\bar{\varepsilon}_e$ (see also the discussion below). Therefore, we can use the Laplace method to approximate the integral:  
\begin{widetext} 
\begin{equation}
	\label{eq:laplace}
	\int_{\varepsilon_\gamma}^\infty \rmd\varepsilon' \frac{\varepsilon^{\prime 2}}{\varepsilon_\gamma^2}
	\frac{\rmd W_{\mathrm{rad}}(\vec{p}',\varepsilon_\gamma)}{\rmd \varepsilon_\gamma}  
	f_\pm \left(\vec{p}',t\right)\approx \frac{N_\pm(t)}{4\pi \varepsilon_\gamma^2} \frac{d\WCS(\bar{\gamma}_e,\bar{\chi}_e,\varepsilon_\gamma)}{d\varepsilon_\gamma},
\end{equation}
where we used that $\mathcal{N}\approx 4\pi c^{-3}\bar{\varepsilon}_e^2 e^{g(\bar{\varepsilon}_e)} \sqrt{2\pi/|g''(\bar{\varepsilon}_e)|}$. As a result, we get the equation
\begin{equation}
	\label{eq:kin_g6}
	\frac{\partial }{\partial t}   f_\gamma (\vec{p}_\gamma,t)  = 
	- W_{\mathrm{cr}} (\vec{p}_\gamma) f_\gamma(\vec{p}_\gamma,t) + \frac{[N_+(t)+N_-(t)]}{4\pi \varepsilon_\gamma^2} \frac{d\WCS(\bar{\gamma}_e,\bar{\chi}_e,\varepsilon_\gamma)}{d\varepsilon_\gamma},
\end{equation}
which can be solved explicitly:
\begin{equation}
	f_\gamma (\vec{p}_\gamma,t) = \frac{c^3}{4\pi \varepsilon_\gamma^2} \frac{d\WCS(\bar{\gamma}_e,\bar{\chi}_e,\varepsilon_\gamma)}{d\varepsilon_\gamma} \int_0^t dt'\, [N_+(t')+N_-(t')]e^{-W_{\mathrm{cr}}(\varepsilon_\gamma,\chi_\gamma)(t-t')}.
\end{equation}
Here, we used the initial condition $f_\gamma (\vec{p}_\gamma,0)=0$. Finally, after substituting $f_\gamma (\vec{p}_\gamma,t)$, putting $W_{\mathrm{cr}}=\WBW$ into Eq.~\eqref{eq:kin_e6} and rearranging the expression we get:
\begin{equation}
	\label{eq:grismayer_t}
	\frac{\partial }{\partial t} N_p (t) = 2\int_0^t dt'\, N_p(t')\int_0^{\bar{\chi}_e} d\chi'\,  \frac{d\WCS(\bar{\gamma}_e,\bar{\chi}_e,\chi')}{d\chi'}\WBW(\varepsilon',\chi') e^{-\WBW(\varepsilon',\chi')(t-t')}
\end{equation}
\end{widetext} 
where we passed to the number of pairs $N_p=N_-+N_+$, changed the integration variable $\varepsilon'\rightarrow\chi'$, and put the upper limit of the $\chi'$ integral as $d\WCS(\bar{\gamma}_e,\bar{\chi}_e,\chi')/d\chi'=0$  for $\chi'>\bar{\chi}_e$. Following Ref.~\cite{grismayer2017seeded}, we formally solve Eq.~\eqref{eq:grismayer_t} by using the Laplace transform 

\[N_p (s) = \int_0^{\infty} dt\, N_p (t) e^{-st}.
\]
When Eq.~\eqref{eq:grismayer_t} is rewritten in the image space, one can show that the function $N_p (s)$ has poles at
\begin{equation}
	\label{eq:grismayer}
	s - 2\int_0^{\bar{\chi}_e} d\chi'\, 
	\frac{ \frac{d\WCS(\bar{\gamma}_e,\bar{\chi}_e,\chi')}{d\chi'}\WBW(\varepsilon',\chi')} {s+\WBW(\varepsilon',\chi')}=0.
\end{equation}
A positive solution $s_+>0$ of this equation provides the cascade growth rate in the steady state, $N_p (t)\simeq N_p (0)e^{\Gamma t}$,  $\Gamma=s_+$. Eq.~\eqref{eq:grismayer} can be solved numerically. For more details, see Refs.~\cite{grismayer2017seeded, Grismayer2016}.

Equations~\eqref{eq:grismayer_t} and\eqref{eq:grismayer} reproduce the model of Grismayer et al [see Eq.~(9) in Ref.~\cite{grismayer2017seeded}]. It should be noted that in our derivation, we had to assume that the position of the electron spectrum coincides with $\bar{\varepsilon}_e$, namely, at the characteristic energy of the electron at the time of emission. This particular supposition (i) allowed decoupling $f_\pm$ and $dW_{\mathrm{rad}}/d\varepsilon_\gamma$ under the integral in Eq.~\eqref{eq:laplace}, and (ii) naturally cuts off the $\chi'$-integral in Eq.~\eqref{eq:grismayer} at $\bar{\chi}_e$. However, our simulations show that for $\varepsilon_e\gtrsim\bar{\varepsilon}_e$ the spectrum decays exponentially, $f_\pm(\varepsilon_e,t)\propto \exp[-(\varepsilon_e/\bar{\varepsilon}_e)^n]$. Such behavior is expected as according to the initial hypothesis electrons emit photons as they reach such energies. Therefore, accounting for this feature may result in a refined version of Eq.~\eqref{eq:grismayer}. 

Let us point out here, that we demonstrated in Appendix~\ref{sec:appendix_kinetic} that Eqs.~\eqref{eq:const_rate_eq1}-\eqref{eq:const_rate_eq2} for $N_{\pm,\gamma}(t)$ are \textit{exact} in terms of the particle spectra, as we integrate them out without additional approximations (except the LCFA). The approximations that were applied in past works \cite{fedotov2010limitations, elkina2011qed, bashmakov2014effect} and the new model that we propose in this work in Section~\ref{sec:iii}, also inexplicitly account for the particle spectrum precisely. The possible inaccuracy of the mentioned models is contained in the approximation for $\bar{\varepsilon}_{e,\gamma}$, $\bar{\chi}_{e,\gamma}$, that are used to estimate the emission and pair creation rates.

\section{Comparison of the growth rate model for cascades in a uniform rotating electric field with the results of past works}
\label{sec:appendix_comparison}
\begin{table*}
	\caption{Summary of the particle growth rate models in avalanche-type QED cascades developing in a uniform rotating electric field. The table collects the results of Refs.~\cite{fedotov2010limitations, elkina2011qed, bashmakov2014effect, grismayer2017seeded, Grismayer2016}, and this paper.}
	\label{tab:rotating_E_models}
	\setlength{\tabcolsep}{4.pt}
	\begin{tabularx}{\textwidth}{lcccccc}
		\hline\hline
		Reference & $\chi_e$ & $\gamma_e$ & $\chi_\gamma$  & $W_{\mathrm{rad}}$ & $W_{\mathrm{cr}}$ & $\Gamma$ \\
		\hline
		\cite{fedotov2010limitations}${}^{\text{Fedotov}}$, \cite{elkina2011qed}${}^{\text{Elkina}}$
		& 
		$ \mu^{3/2}$,\footnotemark[1] & $\dfrac{\mu^{3/4}}{\sqrt{ \omegadimless}}$ & $\sim\chi_e\gg1$ & $\sim \dfrac{\alpha \chi_e^{2/3}}{\tau_C\gamma_e}$ & $W_{\mathrm{rad}}$  & $\sim \dfrac{\alpha \mu^{1/4}}{\tau_C\sqrt{\omegadimless}}$ \\
		\cite{bashmakov2014effect}${}^{\text{Bashmakov}}$
		& $  0.57\mu^{3/2}$ & $\approx\dfrac{\mu^{3/4}}{\sqrt{\omegadimless}}$ & $\sim\chi_e\gg1$  & $\WCS(\chi_e\gg1)$\footnotemark[2] & $\WBW(\chi_\gamma\gg1)$ & Eq.~\eqref{eq:growth_rate_formula} at $\nu_e=0$ \\
		\cite{grismayer2017seeded}${}^{\text{Grismayer}}$ low $E_0$ 
		& $\dfrac{(\alpha\mu)^2}{\omegadimless}$ & $\dfrac{4 \alpha\mu}{\pi \omegadimless}$
		& $0<\chi_\gamma\leq\chi_e$\footnotemark[3] 
		& -- & $\WBW(\chi_\gamma\ll 1)$  & $C_1\dfrac{\alpha^2\mu}{\tau_C} e^{-\frac{8\omegadimless }{3\alpha^2\mu^2}}$,\footnotemark[4]  \\ 
		\cite{grismayer2017seeded}${}^{\text{Grismayer}}$ high $E_0$
		&  $ 1.24\mu^{3/2}$ & $\dfrac{\mu^{3/4}}{\sqrt{\omegadimless}}$ & $0<\chi_\gamma\leq\chi_e$\footnotemark[3]  & $\WCS(\chi_e\gg1)$
		& $\WBW(\chi_\gamma)$
		& Root of Eq.~\eqref{eq:grismayer}  \\
		This work $E_0<\alpha E_S$ 
		& $C_2 \mu$ & $\sqrt{2C_2} \sqrt{\dfrac{\mu}{\omegadimless}}$ & $\sim\cem$&  $\WCS(\cem\ll 1)$ & $\WBW(\cem\ll 1)$  & $C_3 \dfrac{\alpha \sqrt{\mu \omegadimless }}{\tau_C}e^{-\frac{5}{3\sqrt{3}\mu}}$ \\
		This work $E_0>\alpha E_S$ 
		& $C_4\mu^{3/2}$ & $\sqrt{2C_4} \dfrac{\mu^{3/4}}{\sqrt{\omegadimless}} $ & $\sim 1$ 
		&  $\WCS(\cem\gg 1)$ & $\WBW(1)$ & $\dfrac{c_1 \alpha \mu}{\tau_C} \left[ \sqrt{1+ c_2\sqrt{\omegadimless}\mu^{3/4}} -1\right]$
		\\ \hline\hline
	\end{tabularx}
	\footnotetext[1]{For brevity, we define the dimensionless quantities $\mu =E_0/(\alpha E_S)$, $\omegadimless=\omega\tau_C=\hbar\omega/(mc^2)$.}
	\footnotetext[2]{To keep the table concise, we omit $\gamma_{e,\gamma}$ in the argument of $\WCSBW$ [see Eqs.~\eqref{eq:rate_CS_tot}, \eqref{eq:rate_BW_tot}]; for the asymptotic expressions see Eqs.~\eqref{eq:rate_CS_asympt}-\eqref{eq:rate_BW_asympt}.}
	\footnotetext[3]{The $\chi_\gamma$ distribution is given by $d\WCS/d\chi_\gamma$, see Eq.~\eqref{eq:rate_CS}.}
	\footnotetext[4]{The numerical constants: $C_1=\pi^{3/2}/(20\cdot6^{1/4})\approx0.18$, $C_2=4\sqrt{3}/5\approx1.39$, $C_3=3^{5/6}\sqrt{5}\,\Gamma^2\left(2/3\right)/(\sqrt{14}\pi)\approx0.87$, $C_4=9\left(2/[7\Gamma(2/3)]\right)^{3/2} \approx0.87$, for $c_{1,2}$ see Eqs.~\eqref{eq:c1}, \eqref{eq:c2}.}
\end{table*}

Let us discuss how our model given in Eqs.~\eqref{eq:growth_rate_low_cem}-\eqref{eq:growth_rate_high_cem} compares to the results of previous works when applied to cascades in a rotating $E$-field configuration. The available growth rate models are plotted in Fig.~\ref{fig:growth_rate_rotating} and summarized in Table~\ref{tab:rotating_E_models}. The scaling $\Gamma \propto E_0^{1/4}$ at $E_0\gg\alpha E_S$ was first obtained by Fedotov et al \cite{fedotov2010limitations,elkina2011qed}. The coefficient for this scaling can be refined by using the result of Bashmakov et al \cite{bashmakov2014effect}, where it was initially proposed to use the analogue of Eq.~\eqref{eq:growth_rate_formula}, which, however, does not account for the migration effect and implies using $\Wradcr$ estimated in the spirit of Refs.~\cite{fedotov2010limitations, elkina2011qed}. Both results are plotted in Fig.~\ref{fig:growth_rate_rotating}. Although they give the right order of magnitude estimate at high $E_0$, the $E_0^{1/4}$ scaling sets in only beyond $E_S$ (for a laser field of the optical frequency), as we mentioned at the end of Sec.~\ref{sec:iii_c}. In contrast to that, our formula~\eqref{eq:growth_rate_formula_high_field_explicit} provides reliable high-field scaling at $E_0>\alpha E_S$.

In our notations, the scaling of Fedotov et al \cite{fedotov2010limitations,elkina2011qed} is based on the assumption that $\chi_\gamma\sim\cem$, therefore, it does not account for the emission spectrum shape. For instance, the probability rate of two successive events of photon emission and pair creation by this photon (in the steady state) would be proportional to $P_{e^-\rightarrow e^{-}e^{+}}\propto \Wrad \Wcr\sim \Wrad^2$. Grismayer et al \cite{Grismayer2016,grismayer2017seeded} proposed a phenomenological model improving on that front by convoluting the photon spectrum $d\WCS/d\chi_\gamma$ with pair creation rate $\WBW(\varepsilon_\gamma,\chi_\gamma)$. We re-derive this model in Appendix~\ref{sec:appendix_grismayer} and clarify the used approximations. The growth rate results from solving the transcendental equation  \eqref{eq:grismayer}. It can be solved numerically once the expressions for $\bar{\gamma}_e$, $\bar{\chi}_e$ are known. In Ref.~\cite{grismayer2017seeded}, this calculation was carried out at asymptotically low and high fields. 

Recall that in Section~\ref{sec:iii_a} we argue that our model based on Eqs.~\eqref{eq:const_rate_eq1}-\eqref{eq:growth_rate_formula} incorporates the emission spectra without additional approximations. Let us compare our results against the model of Grismayer et al. 

For a low field, Grismayer et al propose setting $\bar{\gamma}_e\propto E_0$, $\bar{\chi}_e\propto E_0^2$  (see Table~\ref{tab:rotating_E_models}) extracted from classical dynamics of an electron averaged over the field cycle. The thin dashed line in Fig.~\ref{fig:growth_rate_rotating} illustrates the resulting growth rate. Notably, the asymptotic decays as $\propto \exp[-8/(3\bar{\chi}_e)]$, which is steeper than in our model Eq.~\eqref{eq:growth_rate_low_eps_nu_zero} [see also Table~\ref{tab:rotating_E_models}]. Though, the numerical data falls in between the two asymptotes.

In the high field regime,  Grismayer et al use the $E_0$-scaling for $\bar{\gamma}_e$, $\bar{\chi}_e$ derived by Fedotov et al from the short time dynamics [we write the corresponding expressions valid for a general field in Eqs.~\eqref{eq:gamma_tem_asympt_exact}, \eqref{eq:chi_tem_asympt_exact}] with one additional modification, such that $\bar{\chi}_e$ is defined as $\bar{\chi}_e:= A[E_0/(\alpha E_S)]^{3/2}$, where $A:=1.24$ is a free parameter used to fit the numerical data. The fit allows matching the data by the asymptote at very high $E_0$, as shown in Fig.~\ref{fig:growth_rate_rotating}. The results of of Grismayer et al are close to our predictions. The advantage of our model is that it does not require fitting parameters.

The study in Ref.~\cite{grismayer2017seeded} did not cover the intermediate regime due to the lack of the corresponding expressions for $\bar{\gamma}_e$, $\bar{\chi}_e$. As we propose them in Section~\ref{sec:ii_c}, namely, $\gem$ and $\cem$, we tested the model of Grismayer et al in the full range. The result is shown in Fig.~\ref{fig:growth_rate_rotating} (the sparse-dashed line).\footnote{We assigned $\bar{\gamma}_e:=\gem$, $\bar{\chi}_e:=\cem$ without the fitting parameter used in Ref.~\cite{grismayer2017seeded}.} Unfortunately, the result significantly underestimates the particle growth rate, which is, presumably, because of the overall pair production rate undercount. 
The radiation and pair creation events are coupled in Eq.~\eqref{eq:grismayer_t}: $\chi'$ of the emitted photon also enters $\WBW$, namely, the photon supposedly decays into a pair shortly after the emission event. In our model, these processes are decoupled, as rates $\Wrad, \Wcr$ are calculated independently. The possible variation of $\chi_{e,\gamma}$ in between the quantum processes is taken into account implicitly. As a result, we were able to reproduce the numerical data with Eqs.~\eqref{eq:growth_rate_low_cem} and \eqref{eq:growth_rate_high_cem} in the full range of $E_0$ without using free parameters (apart from the model matching point choice).

\section{Simulation setup}
\label{sec:appendix_sim_setup}
\begin{figure}%%%
	\includegraphics[width=\linewidth]{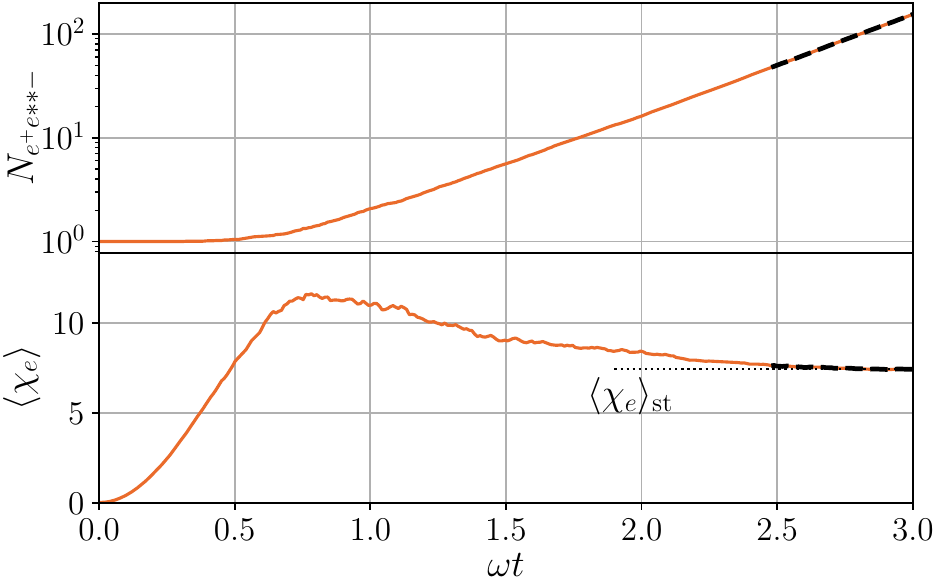}
	\caption{Example of one simulation in a rotating electric field. Top panel: the time dependence of the $e^-e^+$ pair number. Bottom panel: the average $\chi_e$ value of the electrons in the cascade. In the steady state, the growth rate $\Gamma$ and $\langle \chi_e \rangle$ become stationary (see the corresponding horizontal line $\langle \chi_e \rangle_{\mathrm{st}}$ in the bottom panel). The growth rate is extracted by using the exponential fitting of the dependence $N_{e^-e^+}(t)\propto e^{\Gamma t}$ at the data points depicted with the thick dashed line. The plot results from a simulation at $a_0=5275$.}
	\label{fig:growth_rate_extraction}
\end{figure}

For the numerical study, we use the PIC code SMILEI \cite{derouillat2018smilei} and perform simulations in the 1D3V geometry (1 spatial dimension in the direction transverse to the field and 3 momentum projections). The code consistently solves the equations of motion for particles and Maxwell's equations for EM fields, and includes the quantum processes of photon emission by electrons and positrons and pair creation by photons within the LCFA as given by Eqs.~\eqref{eq:rate_CS}-\eqref{eq:rate_BW_tot}.

For the standing wave configuration, we pick the simulation box of length $4 \lambda$, and preliminary fill it with the field, so that the box is centered at the electric antinode. We initialize $\approx100$ seed electrons at rest in the vicinity of the center. The spatial and temporal resolution is set to $\lambda/128$ and $\min(T/ 500,T / a_0)$, respectively. We performed the convergence checks to validate the choice of temporal resolution. For the rotating field configuration, we use a $6\lambda$ simulation box. The field is prescribed by Eq.~\eqref{eq:rotating_E}, and seed electrons are distributed in the middle $2\lambda$-waist. For both configurations, we pick the initial electron density of $0.01n_c$, where $n_c=\varepsilon_0 m \omega^2/e^2$ is the critical plasma density. Since our aim is testing the cascade growth model presented in Section~\ref{sec:iii}, we stay in the regime of relatively low particle densities when plasma effects are not significant (we discuss the field screening by the produced plasma in Section~\ref{sec:conclusions}).

We performed several full 3D PIC simulations in a realistic configuration when the standing wave is formed by two focused Gaussian laser beams of circular polarization. The beam waist $w_0$ was varied to study the effect of transverse migration. The seeding conditions are the same as for the standing wave configuration, however, in this case, we start the simulation by injecting the beams from two opposite sides of an empty box, and seed electrons later, when the standing wave is formed. The EM field is calculated consistently in the PIC loop. We keep the total field amplitude at $E_0$ to match simulations in the other field configurations.

\textbf{Extraction of the growth rate.} In each simulation run, we observe an avalanche exhibiting exponential growth of the particle number, see an example in Fig.~\ref{fig:growth_rate_extraction}.
We continue each run for a long enough time to ensure that the cascade reaches a steady state. The data collected in the steady state can be used to extract the growth rate $\Gamma$. The corresponding time points can be found by studying the time-dependence of the average electron Lorentz factor $\langle\gamma_e\rangle$ and $\chi$-parameter $\langle\chi_e\rangle$. When the cascade reaches the steady state, $\langle\chi_e\rangle$ relaxes to the constant value $\langle\chi_e\rangle_{\mathrm{st}}$, as illustrated in Fig.~\ref{fig:growth_rate_extraction}. Notably, the characteristic relaxation time depends on the field strength and decreases for higher $E_0$. We select the steady-state time points by the following procedure: (i) equate value of $\langle\chi_e\rangle$ reached at the end of the simulation to $\langle\chi_e\rangle_{\mathrm{st}}$, and (ii) pick time points from the data set tail that satisfy the condition $|\langle\chi_e\rangle(t)-\langle\chi_e\rangle_{\mathrm{st}}|/\langle\chi_e\rangle_{\mathrm{st}}<0.02$. The growth rate $\Gamma$ can be extracted by fitting the data for $N_{e^-e^+}(t)$ with the exponential function $A\exp(\Gamma t)$ for the selected points, where $A$ and $\Gamma$ are the fitting parameters.

\textbf{Evaluation of the migration rate $\nu_e$.}
Let us consider seed electrons in the field of a standing wave. If they are injected in spatial regions where $\chi_e(t)$ grows fast, they can trigger a cascade. Recall that $\chi_e(t)$ is proportional to the value of $\epsilon^2 \omega_{\rm eff}$ [see Eq.~\eqref{eq:chi_short_time}], which we plot in Fig.~\ref{fig:fields_invariant_1D_standing}. It peaks in the electric node centers and rapidly falls off near the magnetic antinodes, where it vanishes. It is therefore beneficial to seed particles near the electric field peak, however, they will migrate to the magnetic antinodes, as the latter are spiralling attractors for charges \cite{lehman2012,Esirkepov2014,Gong2016_attractors,Kirk2016,King2016classical}. Let us calculate the migration rate $\nu_e$ of $e^-e^+$ from the $E$-antinode center as a function of $E_0$ to evaluate its effect on the cascade growth rate.

The motion of an electron in a standing wave is known to be chaotic already in the classical regime. In our case, it is also affected by radiation. As general analytic expressions for the trajectories for this case are unknown, we perform statistical simulation of particle migration using SMILEI PIC code \cite{derouillat2018smilei}. The simulation setup is the same as described above for the standing wave configuration, except that we seed $\sim 10^5$ macro-particles in the vicinity of the $E$-antinode and switch off pair creation for the photon species in the code, as we are interested particularly in the trajectories of the seed particles. Nevertheless, we fully account for the recoil due to photon emissions by the electrons. We define $\nu_e$ as the inverse time at which half of the particles propagate at distance $\lambda/8$ from the origin, meaning that they reach the spatial region where $\epsilon^2\oeff=0$ (one of the shaded regions in Fig. \ref{fig:fields_invariant_1D_standing}). The resulting dependence $\nu_e(a_0)$ is presented in the bottom panel of Fig. \ref{fig:model_simus_migration_rate}. At low intensities, the migration rate is close to $\nu_e\approx T^{-1}=\omega/2\pi$, which is consistent with the analytical study in Ref.~\cite{Gong2016_attractors}. However as the amplitude grows, radiation effects become important and $\nu_e$ slowly decreases.

\twocolumngrid
%\bibliography{biblio2} 

\begin{thebibliography}{113}%
	\makeatletter
	\providecommand \@ifxundefined [1]{%
		\@ifx{#1\undefined}
	}%
	\providecommand \@ifnum [1]{%
		\ifnum #1\expandafter \@firstoftwo
		\else \expandafter \@secondoftwo
		\fi
	}%
	\providecommand \@ifx [1]{%
		\ifx #1\expandafter \@firstoftwo
		\else \expandafter \@secondoftwo
		\fi
	}%
	\providecommand \natexlab [1]{#1}%
	\providecommand \enquote  [1]{``#1''}%
	\providecommand \bibnamefont  [1]{#1}%
	\providecommand \bibfnamefont [1]{#1}%
	\providecommand \citenamefont [1]{#1}%
	\providecommand \href@noop [0]{\@secondoftwo}%
	\providecommand \href [0]{\begingroup \@sanitize@url \@href}%
	\providecommand \@href[1]{\@@startlink{#1}\@@href}%
	\providecommand \@@href[1]{\endgroup#1\@@endlink}%
	\providecommand \@sanitize@url [0]{\catcode `\\12\catcode `\$12\catcode
		`\&12\catcode `\#12\catcode `\^12\catcode `\_12\catcode `\%12\relax}%
	\providecommand \@@startlink[1]{}%
	\providecommand \@@endlink[0]{}%
	\providecommand \url  [0]{\begingroup\@sanitize@url \@url }%
	\providecommand \@url [1]{\endgroup\@href {#1}{\urlprefix }}%
	\providecommand \urlprefix  [0]{URL }%
	\providecommand \Eprint [0]{\href }%
	\providecommand \doibase [0]{https://doi.org/}%
	\providecommand \selectlanguage [0]{\@gobble}%
	\providecommand \bibinfo  [0]{\@secondoftwo}%
	\providecommand \bibfield  [0]{\@secondoftwo}%
	\providecommand \translation [1]{[#1]}%
	\providecommand \BibitemOpen [0]{}%
	\providecommand \bibitemStop [0]{}%
	\providecommand \bibitemNoStop [0]{.\EOS\space}%
	\providecommand \EOS [0]{\spacefactor3000\relax}%
	\providecommand \BibitemShut  [1]{\csname bibitem#1\endcsname}%
	\let\auto@bib@innerbib\@empty
	%</preamble>
	\bibitem [{\citenamefont {Blandford}\ and\ \citenamefont
		{Znajek}(1977)}]{blandford1977electromagnetic}%
	\BibitemOpen
	\bibfield  {author} {\bibinfo {author} {\bibfnamefont {R.~D.}\ \bibnamefont
			{Blandford}}\ and\ \bibinfo {author} {\bibfnamefont {R.~L.}\ \bibnamefont
			{Znajek}},\ }\bibfield  {title} {\bibinfo {title} {Electromagnetic extraction
			of energy from {K}err black holes},\ }\href
	{https://doi.org/https://doi.org/10.1093/mnras/179.3.433} {\bibfield
		{journal} {\bibinfo  {journal} {MNRAS}\ }\textbf {\bibinfo {volume} {179}},\
		\bibinfo {pages} {433} (\bibinfo {year} {1977})}\BibitemShut {NoStop}%
	\bibitem [{\citenamefont {Ford}\ \emph {et~al.}(2018)\citenamefont {Ford},
		\citenamefont {Keenan},\ and\ \citenamefont {Medvedev}}]{ford2018electron}%
	\BibitemOpen
	\bibfield  {author} {\bibinfo {author} {\bibfnamefont {A.~L.}\ \bibnamefont
			{Ford}}, \bibinfo {author} {\bibfnamefont {B.~D.}\ \bibnamefont {Keenan}},\
		and\ \bibinfo {author} {\bibfnamefont {M.~V.}\ \bibnamefont {Medvedev}},\
	}\bibfield  {title} {\bibinfo {title} {Electron-positron cascade in
			magnetospheres of spinning black holes},\ }\href
	{https://doi.org/https://doi.org/10.1103/PhysRevD.98.063016} {\bibfield
		{journal} {\bibinfo  {journal} {Phys. Rev. D}\ }\textbf {\bibinfo {volume}
			{98}},\ \bibinfo {pages} {063016} (\bibinfo {year} {2018})}\BibitemShut
	{NoStop}%
	\bibitem [{\citenamefont {Michel}\ and\ \citenamefont
		{Li}(1999)}]{michel1999electrodynamics}%
	\BibitemOpen
	\bibfield  {author} {\bibinfo {author} {\bibfnamefont {F.~C.}\ \bibnamefont
			{Michel}}\ and\ \bibinfo {author} {\bibfnamefont {H.}~\bibnamefont {Li}},\
	}\bibfield  {title} {\bibinfo {title} {Electrodynamics of neutron stars},\
	}\href {https://doi.org/https://doi.org/10.1016/S0370-1573(99)00002-2}
	{\bibfield  {journal} {\bibinfo  {journal} {Phys. Rep.}\ }\textbf {\bibinfo
			{volume} {318}},\ \bibinfo {pages} {227} (\bibinfo {year}
		{1999})}\BibitemShut {NoStop}%
	\bibitem [{\citenamefont {Goldreich}\ and\ \citenamefont
		{Julian}(1969)}]{goldreich1969pulsar}%
	\BibitemOpen
	\bibfield  {author} {\bibinfo {author} {\bibfnamefont {P.}~\bibnamefont
			{Goldreich}}\ and\ \bibinfo {author} {\bibfnamefont {W.~H.}\ \bibnamefont
			{Julian}},\ }\bibfield  {title} {\bibinfo {title} {Pulsar electrodynamics},\
	}\href {https://doi.org/10.1086/150119} {\bibfield  {journal} {\bibinfo
			{journal} {Astrophys. J., vol. 157, p. 869}\ }\textbf {\bibinfo {volume}
			{157}},\ \bibinfo {pages} {869} (\bibinfo {year} {1969})}\BibitemShut
	{NoStop}%
	\bibitem [{\citenamefont {Arons}(1979)}]{arons1979some}%
	\BibitemOpen
	\bibfield  {author} {\bibinfo {author} {\bibfnamefont {J.}~\bibnamefont
			{Arons}},\ }\bibfield  {title} {\bibinfo {title} {Some problems of pulsar
			physics or {I}'m madly in love with electricity},\ }\href
	{https://doi.org/10.1007/BF00172212} {\bibfield  {journal} {\bibinfo
			{journal} {Space Sci. Rev.}\ }\textbf {\bibinfo {volume} {24}},\ \bibinfo
		{pages} {437} (\bibinfo {year} {1979})}\BibitemShut {NoStop}%
	\bibitem [{\citenamefont {Gueroult}\ \emph {et~al.}(2019)\citenamefont
		{Gueroult}, \citenamefont {Shi}, \citenamefont {Rax},\ and\ \citenamefont
		{Fisch}}]{gueroult2019determining}%
	\BibitemOpen
	\bibfield  {author} {\bibinfo {author} {\bibfnamefont {R.}~\bibnamefont
			{Gueroult}}, \bibinfo {author} {\bibfnamefont {Y.}~\bibnamefont {Shi}},
		\bibinfo {author} {\bibfnamefont {J.-M.}\ \bibnamefont {Rax}},\ and\ \bibinfo
		{author} {\bibfnamefont {N.~J.}\ \bibnamefont {Fisch}},\ }\bibfield  {title}
	{\bibinfo {title} {Determining the rotation direction in pulsars},\ }\href
	{https://doi.org/https://doi.org/10.1038/s41467-019-11243-4} {\bibfield
		{journal} {\bibinfo  {journal} {Nat. Commun.}\ }\textbf {\bibinfo {volume}
			{10}},\ \bibinfo {pages} {3232} (\bibinfo {year} {2019})}\BibitemShut
	{NoStop}%
	\bibitem [{\citenamefont {Piran}(2005)}]{piran2005physics}%
	\BibitemOpen
	\bibfield  {author} {\bibinfo {author} {\bibfnamefont {T.}~\bibnamefont
			{Piran}},\ }\bibfield  {title} {\bibinfo {title} {The physics of gamma-ray
			bursts},\ }\href {https://doi.org/https://doi.org/10.1103/RevModPhys.76.1143}
	{\bibfield  {journal} {\bibinfo  {journal} {Rev. Mod. Phys.}\ }\textbf
		{\bibinfo {volume} {76}},\ \bibinfo {pages} {1143} (\bibinfo {year}
		{2005})}\BibitemShut {NoStop}%
	\bibitem [{\citenamefont {Chang}\ \emph {et~al.}(2008)\citenamefont {Chang},
		\citenamefont {Spitkovsky},\ and\ \citenamefont {Arons}}]{chang2008long}%
	\BibitemOpen
	\bibfield  {author} {\bibinfo {author} {\bibfnamefont {P.}~\bibnamefont
			{Chang}}, \bibinfo {author} {\bibfnamefont {A.}~\bibnamefont {Spitkovsky}},\
		and\ \bibinfo {author} {\bibfnamefont {J.}~\bibnamefont {Arons}},\ }\bibfield
	{title} {\bibinfo {title} {Long-term evolution of magnetic turbulence in
			relativistic collisionless shocks: electron-positron plasmas},\ }\href
	{https://doi.org/10.1086/524764} {\bibfield  {journal} {\bibinfo  {journal}
			{Astrophys. J.}\ }\textbf {\bibinfo {volume} {674}},\ \bibinfo {pages} {378}
		(\bibinfo {year} {2008})}\BibitemShut {NoStop}%
	\bibitem [{\citenamefont {Spitkovsky}(2008)}]{spitkovsky2008particle}%
	\BibitemOpen
	\bibfield  {author} {\bibinfo {author} {\bibfnamefont {A.}~\bibnamefont
			{Spitkovsky}},\ }\bibfield  {title} {\bibinfo {title} {Particle acceleration
			in relativistic collisionless shocks: {F}ermi process at last?},\ }\href
	{https://doi.org/10.1086/590248} {\bibfield  {journal} {\bibinfo  {journal}
			{Astrophys. J.}\ }\textbf {\bibinfo {volume} {682}},\ \bibinfo {pages} {L5}
		(\bibinfo {year} {2008})}\BibitemShut {NoStop}%
	\bibitem [{\citenamefont {Kumar}\ and\ \citenamefont
		{Zhang}(2015)}]{kumar2015physics}%
	\BibitemOpen
	\bibfield  {author} {\bibinfo {author} {\bibfnamefont {P.}~\bibnamefont
			{Kumar}}\ and\ \bibinfo {author} {\bibfnamefont {B.}~\bibnamefont {Zhang}},\
	}\bibfield  {title} {\bibinfo {title} {The physics of gamma-ray bursts \&
			relativistic jets},\ }\href
	{https://doi.org/https://doi.org/10.1016/j.physrep.2014.09.008} {\bibfield
		{journal} {\bibinfo  {journal} {Phys. Rep.}\ }\textbf {\bibinfo {volume}
			{561}},\ \bibinfo {pages} {1} (\bibinfo {year} {2015})}\BibitemShut {NoStop}%
	\bibitem [{\citenamefont {Timokhin}\ and\ \citenamefont
		{Harding}(2019)}]{timokhin2019maximum}%
	\BibitemOpen
	\bibfield  {author} {\bibinfo {author} {\bibfnamefont {A.~N.}\ \bibnamefont
			{Timokhin}}\ and\ \bibinfo {author} {\bibfnamefont {A.~K.}\ \bibnamefont
			{Harding}},\ }\bibfield  {title} {\bibinfo {title} {On the maximum pair
			multiplicity of pulsar cascades},\ }\href
	{https://doi.org/10.3847/1538-4357/aaf050} {\bibfield  {journal} {\bibinfo
			{journal} {Astrophys. J.}\ }\textbf {\bibinfo {volume} {871}},\ \bibinfo
		{pages} {12} (\bibinfo {year} {2019})}\BibitemShut {NoStop}%
	\bibitem [{\citenamefont {Cruz}\ \emph {et~al.}(2021)\citenamefont {Cruz},
		\citenamefont {Grismayer},\ and\ \citenamefont {Silva}}]{cruz2021kinetic}%
	\BibitemOpen
	\bibfield  {author} {\bibinfo {author} {\bibfnamefont {F.}~\bibnamefont
			{Cruz}}, \bibinfo {author} {\bibfnamefont {T.}~\bibnamefont {Grismayer}},\
		and\ \bibinfo {author} {\bibfnamefont {L.~O.}\ \bibnamefont {Silva}},\
	}\bibfield  {title} {\bibinfo {title} {Kinetic model of large-amplitude
			oscillations in neutron star pair cascades},\ }\href
	{https://doi.org/10.3847/1538-4357/abd2c0} {\bibfield  {journal} {\bibinfo
			{journal} {Astrophys. J.}\ }\textbf {\bibinfo {volume} {908}},\ \bibinfo
		{pages} {149} (\bibinfo {year} {2021})}\BibitemShut {NoStop}%
	\bibitem [{\citenamefont {Cruz}\ \emph {et~al.}(2022)\citenamefont {Cruz},
		\citenamefont {Grismayer}, \citenamefont {Iteanu}, \citenamefont {Tortone},\
		and\ \citenamefont {Silva}}]{cruz2022model}%
	\BibitemOpen
	\bibfield  {author} {\bibinfo {author} {\bibfnamefont {F.}~\bibnamefont
			{Cruz}}, \bibinfo {author} {\bibfnamefont {T.}~\bibnamefont {Grismayer}},
		\bibinfo {author} {\bibfnamefont {S.}~\bibnamefont {Iteanu}}, \bibinfo
		{author} {\bibfnamefont {P.}~\bibnamefont {Tortone}},\ and\ \bibinfo {author}
		{\bibfnamefont {L.~O.}\ \bibnamefont {Silva}},\ }\bibfield  {title} {\bibinfo
		{title} {Model of pulsar pair cascades in non-uniform electric fields:
			{G}rowth rate, density profile, and screening time},\ }\bibfield  {journal}
	{\bibinfo  {journal} {Phys. Plasmas}\ }\textbf {\bibinfo {volume} {29}},\
	\href {https://doi.org/https://doi.org/10.1063/5.0085847}
	{https://doi.org/10.1063/5.0085847} (\bibinfo {year} {2022})\BibitemShut
	{NoStop}%
	\bibitem [{\citenamefont {Cerutti}\ \emph {et~al.}(2013)\citenamefont
		{Cerutti}, \citenamefont {Werner}, \citenamefont {Uzdensky},\ and\
		\citenamefont {Begelman}}]{cerutti2013simulations}%
	\BibitemOpen
	\bibfield  {author} {\bibinfo {author} {\bibfnamefont {B.}~\bibnamefont
			{Cerutti}}, \bibinfo {author} {\bibfnamefont {G.~R.}\ \bibnamefont {Werner}},
		\bibinfo {author} {\bibfnamefont {D.~A.}\ \bibnamefont {Uzdensky}},\ and\
		\bibinfo {author} {\bibfnamefont {M.~C.}\ \bibnamefont {Begelman}},\
	}\bibfield  {title} {\bibinfo {title} {Simulations of particle acceleration
			beyond the classical synchrotron burnoff limit in magnetic reconnection: an
			explanation of the {C}rab flares},\ }\href
	{https://doi.org/10.1088/0004-637X/770/2/147} {\bibfield  {journal} {\bibinfo
			{journal} {Astrophys. J.}\ }\textbf {\bibinfo {volume} {770}},\ \bibinfo
		{pages} {147} (\bibinfo {year} {2013})}\BibitemShut {NoStop}%
	\bibitem [{\citenamefont {Philippov}\ and\ \citenamefont
		{Spitkovsky}(2014)}]{philippov2014ab}%
	\BibitemOpen
	\bibfield  {author} {\bibinfo {author} {\bibfnamefont {A.~A.}\ \bibnamefont
			{Philippov}}\ and\ \bibinfo {author} {\bibfnamefont {A.}~\bibnamefont
			{Spitkovsky}},\ }\bibfield  {title} {\bibinfo {title} {Ab initio pulsar
			magnetosphere: three-dimensional particle-in-cell simulations of axisymmetric
			pulsars},\ }\href {https://doi.org/10.1088/2041-8205/785/2/L33} {\bibfield
		{journal} {\bibinfo  {journal} {Astrophys. J. Lett.}\ }\textbf {\bibinfo
			{volume} {785}},\ \bibinfo {pages} {L33} (\bibinfo {year}
		{2014})}\BibitemShut {NoStop}%
	\bibitem [{\citenamefont {Hu}\ and\ \citenamefont
		{Beloborodov}(2022)}]{hu2022axisymmetric}%
	\BibitemOpen
	\bibfield  {author} {\bibinfo {author} {\bibfnamefont {R.}~\bibnamefont
			{Hu}}\ and\ \bibinfo {author} {\bibfnamefont {A.~M.}\ \bibnamefont
			{Beloborodov}},\ }\bibfield  {title} {\bibinfo {title} {Axisymmetric pulsar
			magnetosphere revisited},\ }\href {https://doi.org/10.3847/1538-4357/ac961d}
	{\bibfield  {journal} {\bibinfo  {journal} {Astrophys. J.}\ }\textbf
		{\bibinfo {volume} {939}},\ \bibinfo {pages} {42} (\bibinfo {year}
		{2022})}\BibitemShut {NoStop}%
	\bibitem [{\citenamefont {Chen}\ and\ \citenamefont
		{Fiuza}(2023)}]{chen2023perspectives}%
	\BibitemOpen
	\bibfield  {author} {\bibinfo {author} {\bibfnamefont {H.}~\bibnamefont
			{Chen}}\ and\ \bibinfo {author} {\bibfnamefont {F.}~\bibnamefont {Fiuza}},\
	}\bibfield  {title} {\bibinfo {title} {Perspectives on relativistic
			electron--positron pair plasma experiments of astrophysical relevance using
			high-power lasers},\ }\bibfield  {journal} {\bibinfo  {journal} {Phys.
			Plasmas}\ }\textbf {\bibinfo {volume} {30}},\ \href
	{https://doi.org/https://doi.org/10.1063/5.0134819}
	{https://doi.org/10.1063/5.0134819} (\bibinfo {year} {2023})\BibitemShut
	{NoStop}%
	\bibitem [{\citenamefont {Qu}\ \emph {et~al.}(2023)\citenamefont {Qu},
		\citenamefont {Meuren},\ and\ \citenamefont {Fisch}}]{qu2023creating}%
	\BibitemOpen
	\bibfield  {author} {\bibinfo {author} {\bibfnamefont {K.}~\bibnamefont
			{Qu}}, \bibinfo {author} {\bibfnamefont {S.}~\bibnamefont {Meuren}},\ and\
		\bibinfo {author} {\bibfnamefont {N.~J.}\ \bibnamefont {Fisch}},\ }\bibfield
	{title} {\bibinfo {title} {Creating pair plasmas with observable collective
			effects},\ }\href {https://doi.org/10.1088/1361-6587/acb080} {\bibfield
		{journal} {\bibinfo  {journal} {Plasma Phys. Control. Fusion}\ }\textbf
		{\bibinfo {volume} {65}},\ \bibinfo {pages} {034007} (\bibinfo {year}
		{2023})}\BibitemShut {NoStop}%
	\bibitem [{\citenamefont {Sarri}\ \emph
		{et~al.}(2015{\natexlab{a}})\citenamefont {Sarri}, \citenamefont {Poder},
		\citenamefont {Cole}, \citenamefont {Schumaker}, \citenamefont {Di~Piazza},
		\citenamefont {Reville}, \citenamefont {Dzelzainis}, \citenamefont {Doria},
		\citenamefont {Gizzi}, \citenamefont {Grittani} \emph
		{et~al.}}]{sarri2015generation}%
	\BibitemOpen
	\bibfield  {author} {\bibinfo {author} {\bibfnamefont {G.}~\bibnamefont
			{Sarri}}, \bibinfo {author} {\bibfnamefont {K.}~\bibnamefont {Poder}},
		\bibinfo {author} {\bibfnamefont {J.}~\bibnamefont {Cole}}, \bibinfo {author}
		{\bibfnamefont {W.}~\bibnamefont {Schumaker}}, \bibinfo {author}
		{\bibfnamefont {A.}~\bibnamefont {Di~Piazza}}, \bibinfo {author}
		{\bibfnamefont {B.}~\bibnamefont {Reville}}, \bibinfo {author} {\bibfnamefont
			{T.}~\bibnamefont {Dzelzainis}}, \bibinfo {author} {\bibfnamefont
			{D.}~\bibnamefont {Doria}}, \bibinfo {author} {\bibfnamefont
			{L.}~\bibnamefont {Gizzi}}, \bibinfo {author} {\bibfnamefont
			{G.}~\bibnamefont {Grittani}}, \emph {et~al.},\ }\bibfield  {title} {\bibinfo
		{title} {Generation of neutral and high-density electron--positron pair
			plasmas in the laboratory},\ }\href
	{https://doi.org/https://doi.org/10.1038/ncomms7747} {\bibfield  {journal}
		{\bibinfo  {journal} {Nat. Commun.}\ }\textbf {\bibinfo {volume} {6}},\
		\bibinfo {pages} {6747} (\bibinfo {year} {2015}{\natexlab{a}})}\BibitemShut
	{NoStop}%
	\bibitem [{\citenamefont {Arrowsmith}\ \emph {et~al.}(2023)\citenamefont
		{Arrowsmith}, \citenamefont {Bott}, \citenamefont {Cruz}, \citenamefont
		{Froula} \emph {et~al.}}]{arrowsmith2023laboratory}%
	\BibitemOpen
	\bibfield  {author} {\bibinfo {author} {\bibfnamefont {C.~D.}\ \bibnamefont
			{Arrowsmith}}, \bibinfo {author} {\bibfnamefont {A.~F.~A.}\ \bibnamefont
			{Bott}}, \bibinfo {author} {\bibfnamefont {F.~D.}\ \bibnamefont {Cruz}},
		\bibinfo {author} {\bibfnamefont {D.~H.}\ \bibnamefont {Froula}}, \emph
		{et~al.},\ }\bibfield  {title} {\bibinfo {title} {Laboratory realization of
			relativistic pair-plasma beams},\ }\href {https://arxiv.org/abs/2312.05244}
	{\bibfield  {journal} {\bibinfo  {journal} {arXiv preprint arXiv:2312.05244}\
		} (\bibinfo {year} {2023})}\BibitemShut {NoStop}%
	\bibitem [{\citenamefont {Ridgers}\ \emph {et~al.}(2013)\citenamefont
		{Ridgers}, \citenamefont {Brady}, \citenamefont {Duclous}, \citenamefont
		{Kirk}, \citenamefont {Bennett}, \citenamefont {Arber},\ and\ \citenamefont
		{Bell}}]{ridgers2013dense}%
	\BibitemOpen
	\bibfield  {author} {\bibinfo {author} {\bibfnamefont {C.~P.}\ \bibnamefont
			{Ridgers}}, \bibinfo {author} {\bibfnamefont {C.~S.}\ \bibnamefont {Brady}},
		\bibinfo {author} {\bibfnamefont {R.}~\bibnamefont {Duclous}}, \bibinfo
		{author} {\bibfnamefont {J.~G.}\ \bibnamefont {Kirk}}, \bibinfo {author}
		{\bibfnamefont {K.}~\bibnamefont {Bennett}}, \bibinfo {author} {\bibfnamefont
			{T.~D.}\ \bibnamefont {Arber}},\ and\ \bibinfo {author} {\bibfnamefont
			{A.~R.}\ \bibnamefont {Bell}},\ }\bibfield  {title} {\bibinfo {title} {Dense
			electron-positron plasmas and bursts of gamma-rays from laser-generated
			quantum electrodynamic plasmas},\ }\href {https://doi.org/10.1063/1.4801513}
	{\bibfield  {journal} {\bibinfo  {journal} {Phys. Plasmas}\ }\textbf
		{\bibinfo {volume} {20}},\ \bibinfo {pages} {056701} (\bibinfo {year}
		{2013})}\BibitemShut {NoStop}%
	\bibitem [{\citenamefont {Narozhny}\ and\ \citenamefont
		{Fedotov}(2014)}]{narozhny2014creation}%
	\BibitemOpen
	\bibfield  {author} {\bibinfo {author} {\bibfnamefont {N.~B.}\ \bibnamefont
			{Narozhny}}\ and\ \bibinfo {author} {\bibfnamefont {A.~M.}\ \bibnamefont
			{Fedotov}},\ }\bibfield  {title} {\bibinfo {title} {Creation of
			electron-positron plasma with superstrong laser field},\ }\href
	{https://doi.org/10.1140/epjst/e2014-02159-1} {\bibfield  {journal} {\bibinfo
			{journal} {Eur. Phys. J. Spec. Top.}\ }\textbf {\bibinfo {volume} {223}},\
		\bibinfo {pages} {1083} (\bibinfo {year} {2014})}\BibitemShut {NoStop}%
	\bibitem [{\citenamefont {Sarri}\ \emph
		{et~al.}(2015{\natexlab{b}})\citenamefont {Sarri}, \citenamefont {Dieckmann},
		\citenamefont {Kourakis}, \citenamefont {Di~Piazza}, \citenamefont {Reville},
		\citenamefont {Keitel},\ and\ \citenamefont {Zepf}}]{sarri2015overview}%
	\BibitemOpen
	\bibfield  {author} {\bibinfo {author} {\bibfnamefont {G.}~\bibnamefont
			{Sarri}}, \bibinfo {author} {\bibfnamefont {M.~E.}\ \bibnamefont
			{Dieckmann}}, \bibinfo {author} {\bibfnamefont {I.}~\bibnamefont {Kourakis}},
		\bibinfo {author} {\bibfnamefont {A.}~\bibnamefont {Di~Piazza}}, \bibinfo
		{author} {\bibfnamefont {B.}~\bibnamefont {Reville}}, \bibinfo {author}
		{\bibfnamefont {C.}~\bibnamefont {Keitel}},\ and\ \bibinfo {author}
		{\bibfnamefont {M.}~\bibnamefont {Zepf}},\ }\bibfield  {title} {\bibinfo
		{title} {Overview of laser-driven generation of electron--positron beams},\
	}\href {https://doi.org/10.1017/S002237781500046X} {\bibfield  {journal}
		{\bibinfo  {journal} {J. Plasma Phys.}\ }\textbf {\bibinfo {volume} {81}},\
		\bibinfo {pages} {455810401} (\bibinfo {year}
		{2015}{\natexlab{b}})}\BibitemShut {NoStop}%
	\bibitem [{\citenamefont {Zhang}\ \emph {et~al.}(2020)\citenamefont {Zhang},
		\citenamefont {Bulanov}, \citenamefont {Seipt}, \citenamefont {Arefiev},\
		and\ \citenamefont {Thomas}}]{zhang2020relativistic}%
	\BibitemOpen
	\bibfield  {author} {\bibinfo {author} {\bibfnamefont {P.}~\bibnamefont
			{Zhang}}, \bibinfo {author} {\bibfnamefont {S.}~\bibnamefont {Bulanov}},
		\bibinfo {author} {\bibfnamefont {D.}~\bibnamefont {Seipt}}, \bibinfo
		{author} {\bibfnamefont {A.~V.}\ \bibnamefont {Arefiev}},\ and\ \bibinfo
		{author} {\bibfnamefont {A.~G.~R.}\ \bibnamefont {Thomas}},\ }\bibfield
	{title} {\bibinfo {title} {Relativistic plasma physics in supercritical
			fields},\ }\bibfield  {journal} {\bibinfo  {journal} {Phys. Plasmas}\
	}\textbf {\bibinfo {volume} {27}},\ \href
	{https://doi.org/https://doi.org/10.1063/1.5144449}
	{https://doi.org/10.1063/1.5144449} (\bibinfo {year} {2020})\BibitemShut
	{NoStop}%
	\bibitem [{\citenamefont {Qu}\ \emph {et~al.}(2021)\citenamefont {Qu},
		\citenamefont {Meuren},\ and\ \citenamefont {Fisch}}]{qu2021signature}%
	\BibitemOpen
	\bibfield  {author} {\bibinfo {author} {\bibfnamefont {K.}~\bibnamefont
			{Qu}}, \bibinfo {author} {\bibfnamefont {S.}~\bibnamefont {Meuren}},\ and\
		\bibinfo {author} {\bibfnamefont {N.~J.}\ \bibnamefont {Fisch}},\ }\bibfield
	{title} {\bibinfo {title} {Signature of collective plasma effects in
			beam-driven {QED} cascades},\ }\href
	{https://doi.org/10.1103/PhysRevLett.127.095001} {\bibfield  {journal}
		{\bibinfo  {journal} {Phys. Rev. Lett.}\ }\textbf {\bibinfo {volume} {127}},\
		\bibinfo {pages} {095001} (\bibinfo {year} {2021})}\BibitemShut {NoStop}%
	\bibitem [{\citenamefont {Fedotov}\ \emph {et~al.}(2023)\citenamefont
		{Fedotov}, \citenamefont {Ilderton}, \citenamefont {Karbstein}, \citenamefont
		{King}, \citenamefont {Seipt}, \citenamefont {Taya},\ and\ \citenamefont
		{Torgrimsson}}]{fedotov2022high}%
	\BibitemOpen
	\bibfield  {author} {\bibinfo {author} {\bibfnamefont {A.}~\bibnamefont
			{Fedotov}}, \bibinfo {author} {\bibfnamefont {A.}~\bibnamefont {Ilderton}},
		\bibinfo {author} {\bibfnamefont {F.}~\bibnamefont {Karbstein}}, \bibinfo
		{author} {\bibfnamefont {B.}~\bibnamefont {King}}, \bibinfo {author}
		{\bibfnamefont {D.}~\bibnamefont {Seipt}}, \bibinfo {author} {\bibfnamefont
			{H.}~\bibnamefont {Taya}},\ and\ \bibinfo {author} {\bibfnamefont
			{G.}~\bibnamefont {Torgrimsson}},\ }\bibfield  {title} {\bibinfo {title}
		{Advances in {QED} with intense background fields},\ }\href
	{https://doi.org/10.1016/j.physrep.2023.01.003} {\bibfield  {journal}
		{\bibinfo  {journal} {Phys. Rep.}\ }\textbf {\bibinfo {volume} {1010}},\
		\bibinfo {pages} {1} (\bibinfo {year} {2023})}\BibitemShut {NoStop}%
	\bibitem [{\citenamefont {Gonoskov}\ \emph {et~al.}(2022)\citenamefont
		{Gonoskov}, \citenamefont {Blackburn}, \citenamefont {Marklund},\ and\
		\citenamefont {Bulanov}}]{gonoskov2022charged}%
	\BibitemOpen
	\bibfield  {author} {\bibinfo {author} {\bibfnamefont {A.}~\bibnamefont
			{Gonoskov}}, \bibinfo {author} {\bibfnamefont {T.~G.}\ \bibnamefont
			{Blackburn}}, \bibinfo {author} {\bibfnamefont {M.}~\bibnamefont
			{Marklund}},\ and\ \bibinfo {author} {\bibfnamefont {S.~S.}\ \bibnamefont
			{Bulanov}},\ }\bibfield  {title} {\bibinfo {title} {Charged particle motion
			and radiation in strong electromagnetic fields},\ }\href
	{https://doi.org/10.1103/RevModPhys.94.045001} {\bibfield  {journal}
		{\bibinfo  {journal} {Rev. Mod. Phys.}\ }\textbf {\bibinfo {volume} {94}},\
		\bibinfo {pages} {045001} (\bibinfo {year} {2022})}\BibitemShut {NoStop}%
	\bibitem [{\citenamefont {Popruzhenko}\ and\ \citenamefont
		{Fedotov}(2023)}]{popruzhenko2023dynamics}%
	\BibitemOpen
	\bibfield  {author} {\bibinfo {author} {\bibfnamefont {S.~V.}\ \bibnamefont
			{Popruzhenko}}\ and\ \bibinfo {author} {\bibfnamefont {A.~M.}\ \bibnamefont
			{Fedotov}},\ }\bibfield  {title} {\bibinfo {title} {Dynamics and radiation of
			charged particles in ultra-intense laser fields},\ }\href
	{https://doi.org/10.3367/UFNr.2023.03.039335} {\bibfield  {journal} {\bibinfo
			{journal} {Phys. Usp.}\ }\textbf {\bibinfo {volume} {66}},\ \bibinfo {pages}
		{460} (\bibinfo {year} {2023})}\BibitemShut {NoStop}%
	\bibitem [{\citenamefont {Di~Piazza}\ \emph {et~al.}(2012)\citenamefont
		{Di~Piazza}, \citenamefont {M{\"u}ller}, \citenamefont {Hatsagortsyan},\ and\
		\citenamefont {Keitel}}]{di2012extremely}%
	\BibitemOpen
	\bibfield  {author} {\bibinfo {author} {\bibfnamefont {A.}~\bibnamefont
			{Di~Piazza}}, \bibinfo {author} {\bibfnamefont {C.}~\bibnamefont
			{M{\"u}ller}}, \bibinfo {author} {\bibfnamefont {K.}~\bibnamefont
			{Hatsagortsyan}},\ and\ \bibinfo {author} {\bibfnamefont {C.~H.}\
			\bibnamefont {Keitel}},\ }\bibfield  {title} {\bibinfo {title} {Extremely
			high-intensity laser interactions with fundamental quantum systems},\ }\href
	{https://doi.org/10.1103/RevModPhys.84.1177} {\bibfield  {journal} {\bibinfo
			{journal} {Rev. Mod. Phys.}\ }\textbf {\bibinfo {volume} {84}},\ \bibinfo
		{pages} {1177} (\bibinfo {year} {2012})}\BibitemShut {NoStop}%
	\bibitem [{\citenamefont {Ruffini}\ \emph {et~al.}(2010)\citenamefont
		{Ruffini}, \citenamefont {Vereshchagin},\ and\ \citenamefont
		{Xue}}]{ruffini2010electron}%
	\BibitemOpen
	\bibfield  {author} {\bibinfo {author} {\bibfnamefont {R.}~\bibnamefont
			{Ruffini}}, \bibinfo {author} {\bibfnamefont {G.}~\bibnamefont
			{Vereshchagin}},\ and\ \bibinfo {author} {\bibfnamefont {S.-S.}\ \bibnamefont
			{Xue}},\ }\bibfield  {title} {\bibinfo {title} {Electron--positron pairs in
			physics and astrophysics: {F}rom heavy nuclei to black holes},\ }\href
	{https://doi.org/https://doi.org/10.1016/j.physrep.2009.10.004} {\bibfield
		{journal} {\bibinfo  {journal} {Phys. Rep.}\ }\textbf {\bibinfo {volume}
			{487}},\ \bibinfo {pages} {1} (\bibinfo {year} {2010})}\BibitemShut {NoStop}%
	\bibitem [{\citenamefont {Gaisser}\ \emph {et~al.}(2016)\citenamefont
		{Gaisser}, \citenamefont {Engel},\ and\ \citenamefont
		{Resconi}}]{gaisser2016cosmic}%
	\BibitemOpen
	\bibfield  {author} {\bibinfo {author} {\bibfnamefont {T.~K.}\ \bibnamefont
			{Gaisser}}, \bibinfo {author} {\bibfnamefont {R.}~\bibnamefont {Engel}},\
		and\ \bibinfo {author} {\bibfnamefont {E.}~\bibnamefont {Resconi}},\
	}\href@noop {} {\emph {\bibinfo {title} {Cosmic rays and particle physics}}}\
	(\bibinfo  {publisher} {Cambridge University Press, Cambridge},\ \bibinfo
	{year} {2016})\BibitemShut {NoStop}%
	\bibitem [{\citenamefont {Sokolov}\ \emph {et~al.}(2010)\citenamefont
		{Sokolov}, \citenamefont {Naumova}, \citenamefont {Nees},\ and\ \citenamefont
		{Mourou}}]{sokolov2010pair}%
	\BibitemOpen
	\bibfield  {author} {\bibinfo {author} {\bibfnamefont {I.~V.}\ \bibnamefont
			{Sokolov}}, \bibinfo {author} {\bibfnamefont {N.~M.}\ \bibnamefont
			{Naumova}}, \bibinfo {author} {\bibfnamefont {J.~A.}\ \bibnamefont {Nees}},\
		and\ \bibinfo {author} {\bibfnamefont {G.~A.}\ \bibnamefont {Mourou}},\
	}\bibfield  {title} {\bibinfo {title} {Pair creation in {QED}-strong pulsed
			laser fields interacting with electron beams},\ }\href
	{https://doi.org/10.1103/PhysRevLett.105.195005} {\bibfield  {journal}
		{\bibinfo  {journal} {Phys. Rev. Lett.}\ }\textbf {\bibinfo {volume} {105}},\
		\bibinfo {pages} {195005} (\bibinfo {year} {2010})}\BibitemShut {NoStop}%
	\bibitem [{\citenamefont {Bulanov}\ \emph {et~al.}(2013)\citenamefont
		{Bulanov}, \citenamefont {Schroeder}, \citenamefont {Esarey},\ and\
		\citenamefont {Leemans}}]{bulanov2013electromagnetic}%
	\BibitemOpen
	\bibfield  {author} {\bibinfo {author} {\bibfnamefont {S.~S.}\ \bibnamefont
			{Bulanov}}, \bibinfo {author} {\bibfnamefont {C.~B.}\ \bibnamefont
			{Schroeder}}, \bibinfo {author} {\bibfnamefont {E.}~\bibnamefont {Esarey}},\
		and\ \bibinfo {author} {\bibfnamefont {W.~P.}\ \bibnamefont {Leemans}},\
	}\bibfield  {title} {\bibinfo {title} {Electromagnetic cascade in high-energy
			electron, positron, and photon interactions with intense laser pulses},\
	}\href {https://doi.org/10.1103/PhysRevA.87.062110} {\bibfield  {journal}
		{\bibinfo  {journal} {Phys. Rev. A}\ }\textbf {\bibinfo {volume} {87}},\
		\bibinfo {pages} {062110} (\bibinfo {year} {2013})}\BibitemShut {NoStop}%
	\bibitem [{\citenamefont {Blackburn}\ and\ \citenamefont
		{Murphy}(2017)}]{blackburn2017scaling}%
	\BibitemOpen
	\bibfield  {author} {\bibinfo {author} {\bibfnamefont {T.}~\bibnamefont
			{Blackburn}}\ and\ \bibinfo {author} {\bibfnamefont {C.}~\bibnamefont
			{Murphy}},\ }\bibfield  {title} {\bibinfo {title} {Scaling laws for positron
			production in laser--electron-beam collisions},\ }\href
	{https://doi.org/https://doi.org/10.1103/PhysRevA.96.022128} {\bibfield
		{journal} {\bibinfo  {journal} {Phys. Rev. A}\ }\textbf {\bibinfo {volume}
			{96}},\ \bibinfo {pages} {022128} (\bibinfo {year} {2017})}\BibitemShut
	{NoStop}%
	\bibitem [{\citenamefont {Mercuri-Baron}\ \emph {et~al.}(2021)\citenamefont
		{Mercuri-Baron}, \citenamefont {Grech}, \citenamefont {Niel}, \citenamefont
		{Grassi}, \citenamefont {Lobet}, \citenamefont {Di~Piazza},\ and\
		\citenamefont {Riconda}}]{mercuri2021impact}%
	\BibitemOpen
	\bibfield  {author} {\bibinfo {author} {\bibfnamefont {A.}~\bibnamefont
			{Mercuri-Baron}}, \bibinfo {author} {\bibfnamefont {M.}~\bibnamefont
			{Grech}}, \bibinfo {author} {\bibfnamefont {F.}~\bibnamefont {Niel}},
		\bibinfo {author} {\bibfnamefont {A.}~\bibnamefont {Grassi}}, \bibinfo
		{author} {\bibfnamefont {M.}~\bibnamefont {Lobet}}, \bibinfo {author}
		{\bibfnamefont {A.}~\bibnamefont {Di~Piazza}},\ and\ \bibinfo {author}
		{\bibfnamefont {C.}~\bibnamefont {Riconda}},\ }\bibfield  {title} {\bibinfo
		{title} {Impact of the laser spatio-temporal shape on {B}reit--{W}heeler pair
			production},\ }\href {https://doi.org/10.1088/1367-2630/ac1975} {\bibfield
		{journal} {\bibinfo  {journal} {New J. Phys.}\ }\textbf {\bibinfo {volume}
			{23}},\ \bibinfo {pages} {085006} (\bibinfo {year} {2021})}\BibitemShut
	{NoStop}%
	\bibitem [{\citenamefont {Zerby}\ and\ \citenamefont
		{Moran}(1963)}]{zerby1963studies}%
	\BibitemOpen
	\bibfield  {author} {\bibinfo {author} {\bibfnamefont {C.~D.}\ \bibnamefont
			{Zerby}}\ and\ \bibinfo {author} {\bibfnamefont {H.~S.}\ \bibnamefont
			{Moran}},\ }\bibfield  {title} {\bibinfo {title} {Studies of the longitudinal
			development of electron—photon cascade showers},\ }\href
	{https://doi.org/10.1063/1.1702764} {\bibfield  {journal} {\bibinfo
			{journal} {J. Appl. Phys.}\ }\textbf {\bibinfo {volume} {34}},\ \bibinfo
		{pages} {2445} (\bibinfo {year} {1963})}\BibitemShut {NoStop}%
	\bibitem [{\citenamefont {Ritus}(1985)}]{ritus1985}%
	\BibitemOpen
	\bibfield  {author} {\bibinfo {author} {\bibfnamefont {V.~I.}\ \bibnamefont
			{Ritus}},\ }\bibfield  {title} {\bibinfo {title} {Quantum effects of the
			interaction of elementary particles with an intense electromagnetic field},\
	}\href {https://doi.org/https://doi.org/10.1007/BF01120220} {\bibfield
		{journal} {\bibinfo  {journal} {J. Russ. Laser Res.}\ }\textbf {\bibinfo
			{volume} {6}},\ \bibinfo {pages} {497} (\bibinfo {year} {1985})}\BibitemShut
	{NoStop}%
	\bibitem [{\citenamefont {Bell}\ and\ \citenamefont {Kirk}(2008)}]{Kirk2008}%
	\BibitemOpen
	\bibfield  {author} {\bibinfo {author} {\bibfnamefont {A.~R.}\ \bibnamefont
			{Bell}}\ and\ \bibinfo {author} {\bibfnamefont {J.~G.}\ \bibnamefont
			{Kirk}},\ }\bibfield  {title} {\bibinfo {title} {Possibility of prolific pair
			production with high-power lasers},\ }\href
	{https://doi.org/10.1103/PhysRevLett.101.200403} {\bibfield  {journal}
		{\bibinfo  {journal} {Phys. Rev. Lett.}\ }\textbf {\bibinfo {volume} {101}},\
		\bibinfo {pages} {200403} (\bibinfo {year} {2008})}\BibitemShut {NoStop}%
	\bibitem [{\citenamefont {Kirk}\ \emph {et~al.}(2009)\citenamefont {Kirk},
		\citenamefont {Bell},\ and\ \citenamefont {Arka}}]{Kirk2009}%
	\BibitemOpen
	\bibfield  {author} {\bibinfo {author} {\bibfnamefont {J.~G.}\ \bibnamefont
			{Kirk}}, \bibinfo {author} {\bibfnamefont {A.~R.}\ \bibnamefont {Bell}},\
		and\ \bibinfo {author} {\bibfnamefont {I.}~\bibnamefont {Arka}},\ }\bibfield
	{title} {\bibinfo {title} {Pair production in counter-propagating laser
			beams},\ }\href {https://doi.org/10.1088/0741-3335/51/8/085008} {\bibfield
		{journal} {\bibinfo  {journal} {Plasma Phys. Control. Fusion}\ }\textbf
		{\bibinfo {volume} {51}},\ \bibinfo {pages} {085008} (\bibinfo {year}
		{2009})}\BibitemShut {NoStop}%
	\bibitem [{\citenamefont {Fedotov}\ \emph {et~al.}(2010)\citenamefont
		{Fedotov}, \citenamefont {Narozhny}, \citenamefont {Mourou},\ and\
		\citenamefont {Korn}}]{fedotov2010limitations}%
	\BibitemOpen
	\bibfield  {author} {\bibinfo {author} {\bibfnamefont {A.~M.}\ \bibnamefont
			{Fedotov}}, \bibinfo {author} {\bibfnamefont {N.~B.}\ \bibnamefont
			{Narozhny}}, \bibinfo {author} {\bibfnamefont {G.}~\bibnamefont {Mourou}},\
		and\ \bibinfo {author} {\bibfnamefont {G.}~\bibnamefont {Korn}},\ }\bibfield
	{title} {\bibinfo {title} {Limitations on the attainable intensity of high
			power lasers},\ }\href {https://doi.org/10.1103/PhysRevLett.105.080402}
	{\bibfield  {journal} {\bibinfo  {journal} {Phys. Rev. Lett.}\ }\textbf
		{\bibinfo {volume} {105}},\ \bibinfo {pages} {080402} (\bibinfo {year}
		{2010})}\BibitemShut {NoStop}%
	\bibitem [{\citenamefont {Gonoskov}\ \emph {et~al.}(2017)\citenamefont
		{Gonoskov}, \citenamefont {Bashinov}, \citenamefont {Bastrakov},
		\citenamefont {Efimenko}, \citenamefont {Ilderton}, \citenamefont {Kim},
		\citenamefont {Marklund}, \citenamefont {Meyerov}, \citenamefont {Muraviev},\
		and\ \citenamefont {Sergeev}}]{gonoskov2017ultrabright}%
	\BibitemOpen
	\bibfield  {author} {\bibinfo {author} {\bibfnamefont {A.}~\bibnamefont
			{Gonoskov}}, \bibinfo {author} {\bibfnamefont {A.}~\bibnamefont {Bashinov}},
		\bibinfo {author} {\bibfnamefont {S.}~\bibnamefont {Bastrakov}}, \bibinfo
		{author} {\bibfnamefont {E.}~\bibnamefont {Efimenko}}, \bibinfo {author}
		{\bibfnamefont {A.}~\bibnamefont {Ilderton}}, \bibinfo {author}
		{\bibfnamefont {A.}~\bibnamefont {Kim}}, \bibinfo {author} {\bibfnamefont
			{M.}~\bibnamefont {Marklund}}, \bibinfo {author} {\bibfnamefont
			{I.}~\bibnamefont {Meyerov}}, \bibinfo {author} {\bibfnamefont
			{A.}~\bibnamefont {Muraviev}},\ and\ \bibinfo {author} {\bibfnamefont
			{A.}~\bibnamefont {Sergeev}},\ }\bibfield  {title} {\bibinfo {title}
		{Ultrabright gev photon source via controlled electromagnetic cascades in
			laser-dipole waves},\ }\href
	{https://doi.org/https://doi.org/10.1103/PhysRevX.7.041003} {\bibfield
		{journal} {\bibinfo  {journal} {Phys. Rev. X}\ }\textbf {\bibinfo {volume}
			{7}},\ \bibinfo {pages} {041003} (\bibinfo {year} {2017})}\BibitemShut
	{NoStop}%
	\bibitem [{\citenamefont {Mironov}\ \emph {et~al.}(2021)\citenamefont
		{Mironov}, \citenamefont {Gelfer},\ and\ \citenamefont
		{Fedotov}}]{Arseny2021}%
	\BibitemOpen
	\bibfield  {author} {\bibinfo {author} {\bibfnamefont {A.~A.}\ \bibnamefont
			{Mironov}}, \bibinfo {author} {\bibfnamefont {E.~G.}\ \bibnamefont
			{Gelfer}},\ and\ \bibinfo {author} {\bibfnamefont {A.~M.}\ \bibnamefont
			{Fedotov}},\ }\bibfield  {title} {\bibinfo {title} {Onset of electron-seeded
			cascades in generic electromagnetic fields},\ }\href
	{https://doi.org/10.1103/PhysRevA.104.012221} {\bibfield  {journal} {\bibinfo
			{journal} {Phys. Rev. A}\ }\textbf {\bibinfo {volume} {104}},\ \bibinfo
		{pages} {012221} (\bibinfo {year} {2021})}\BibitemShut {NoStop}%
	\bibitem [{\citenamefont {Elkina}\ \emph {et~al.}(2011)\citenamefont {Elkina},
		\citenamefont {Fedotov}, \citenamefont {Kostyukov}, \citenamefont {Legkov},
		\citenamefont {Narozhny}, \citenamefont {Nerush},\ and\ \citenamefont
		{Ruhl}}]{elkina2011qed}%
	\BibitemOpen
	\bibfield  {author} {\bibinfo {author} {\bibfnamefont {N.~V.}\ \bibnamefont
			{Elkina}}, \bibinfo {author} {\bibfnamefont {A.~M.}\ \bibnamefont {Fedotov}},
		\bibinfo {author} {\bibfnamefont {I.~Y.}\ \bibnamefont {Kostyukov}}, \bibinfo
		{author} {\bibfnamefont {M.~V.}\ \bibnamefont {Legkov}}, \bibinfo {author}
		{\bibfnamefont {N.~B.}\ \bibnamefont {Narozhny}}, \bibinfo {author}
		{\bibfnamefont {E.~N.}\ \bibnamefont {Nerush}},\ and\ \bibinfo {author}
		{\bibfnamefont {H.}~\bibnamefont {Ruhl}},\ }\bibfield  {title} {\bibinfo
		{title} {{QED} cascades induced by circularly polarized laser fields},\
	}\href {https://doi.org/10.1103/PhysRevSTAB.14.054401} {\bibfield  {journal}
		{\bibinfo  {journal} {Phys. Rev. STAB}\ }\textbf {\bibinfo {volume} {14}},\
		\bibinfo {pages} {054401} (\bibinfo {year} {2011})}\BibitemShut {NoStop}%
	\bibitem [{\citenamefont {Mironov}\ \emph {et~al.}(2014)\citenamefont
		{Mironov}, \citenamefont {Narozhny},\ and\ \citenamefont
		{Fedotov}}]{mironov2014collapse}%
	\BibitemOpen
	\bibfield  {author} {\bibinfo {author} {\bibfnamefont {A.~A.}\ \bibnamefont
			{Mironov}}, \bibinfo {author} {\bibfnamefont {N.~B.}\ \bibnamefont
			{Narozhny}},\ and\ \bibinfo {author} {\bibfnamefont {A.~M.}\ \bibnamefont
			{Fedotov}},\ }\bibfield  {title} {\bibinfo {title} {Collapse and revival of
			electromagnetic cascades in focused intense laser pulses},\ }\href
	{https://doi.org/10.1016/j.physleta.2014.09.058} {\bibfield  {journal}
		{\bibinfo  {journal} {Phys. Lett. A}\ }\textbf {\bibinfo {volume} {378}},\
		\bibinfo {pages} {3254} (\bibinfo {year} {2014})}\BibitemShut {NoStop}%
	\bibitem [{\citenamefont {Grismayer}\ \emph {et~al.}(2017)\citenamefont
		{Grismayer}, \citenamefont {Vranic}, \citenamefont {Martins}, \citenamefont
		{Fonseca},\ and\ \citenamefont {Silva}}]{grismayer2017seeded}%
	\BibitemOpen
	\bibfield  {author} {\bibinfo {author} {\bibfnamefont {T.}~\bibnamefont
			{Grismayer}}, \bibinfo {author} {\bibfnamefont {M.}~\bibnamefont {Vranic}},
		\bibinfo {author} {\bibfnamefont {J.~L.}\ \bibnamefont {Martins}}, \bibinfo
		{author} {\bibfnamefont {R.~A.}\ \bibnamefont {Fonseca}},\ and\ \bibinfo
		{author} {\bibfnamefont {L.~O.}\ \bibnamefont {Silva}},\ }\bibfield  {title}
	{\bibinfo {title} {Seeded {QED} cascades in counterpropagating laser
			pulses},\ }\href {https://doi.org/10.1103/PhysRevE.95.023210} {\bibfield
		{journal} {\bibinfo  {journal} {Phys. Rev. E}\ }\textbf {\bibinfo {volume}
			{95}},\ \bibinfo {pages} {023210} (\bibinfo {year} {2017})}\BibitemShut
	{NoStop}%
	\bibitem [{\citenamefont {Bashmakov}\ \emph {et~al.}(2014)\citenamefont
		{Bashmakov}, \citenamefont {Nerush}, \citenamefont {Kostyukov}, \citenamefont
		{Fedotov},\ and\ \citenamefont {Narozhny}}]{bashmakov2014effect}%
	\BibitemOpen
	\bibfield  {author} {\bibinfo {author} {\bibfnamefont {V.~F.}\ \bibnamefont
			{Bashmakov}}, \bibinfo {author} {\bibfnamefont {E.~N.}\ \bibnamefont
			{Nerush}}, \bibinfo {author} {\bibfnamefont {I.~Y.}\ \bibnamefont
			{Kostyukov}}, \bibinfo {author} {\bibfnamefont {A.~M.}\ \bibnamefont
			{Fedotov}},\ and\ \bibinfo {author} {\bibfnamefont {N.~B.}\ \bibnamefont
			{Narozhny}},\ }\bibfield  {title} {\bibinfo {title} {Effect of laser
			polarization on quantum electrodynamical cascading},\ }\href
	{https://doi.org/10.1063/1.4861863} {\bibfield  {journal} {\bibinfo
			{journal} {Phys. Plasmas}\ }\textbf {\bibinfo {volume} {21}},\ \bibinfo
		{pages} {013105} (\bibinfo {year} {2014})}\BibitemShut {NoStop}%
	\bibitem [{\citenamefont {Jirka}\ \emph {et~al.}(2016)\citenamefont {Jirka},
		\citenamefont {Klimo}, \citenamefont {Bulanov}, \citenamefont {Esirkepov},
		\citenamefont {Gelfer}, \citenamefont {Bulanov}, \citenamefont {Weber},\ and\
		\citenamefont {Korn}}]{Jirka2016}%
	\BibitemOpen
	\bibfield  {author} {\bibinfo {author} {\bibfnamefont {M.}~\bibnamefont
			{Jirka}}, \bibinfo {author} {\bibfnamefont {O.}~\bibnamefont {Klimo}},
		\bibinfo {author} {\bibfnamefont {S.~V.}\ \bibnamefont {Bulanov}}, \bibinfo
		{author} {\bibfnamefont {T.~Z.}\ \bibnamefont {Esirkepov}}, \bibinfo {author}
		{\bibfnamefont {E.}~\bibnamefont {Gelfer}}, \bibinfo {author} {\bibfnamefont
			{S.~S.}\ \bibnamefont {Bulanov}}, \bibinfo {author} {\bibfnamefont
			{S.}~\bibnamefont {Weber}},\ and\ \bibinfo {author} {\bibfnamefont
			{G.}~\bibnamefont {Korn}},\ }\bibfield  {title} {\bibinfo {title} {Electron
			dynamics and $\ensuremath{\gamma}$ and ${e}^{\ensuremath{-}}{e}^{+}$
			production by colliding laser pulses},\ }\href
	{https://doi.org/10.1103/PhysRevE.93.023207} {\bibfield  {journal} {\bibinfo
			{journal} {Phys. Rev. E}\ }\textbf {\bibinfo {volume} {93}},\ \bibinfo
		{pages} {023207} (\bibinfo {year} {2016})}\BibitemShut {NoStop}%
	\bibitem [{\citenamefont {Gelfer}\ \emph {et~al.}(2015)\citenamefont {Gelfer},
		\citenamefont {Mironov}, \citenamefont {Fedotov}, \citenamefont {Bashmakov},
		\citenamefont {Nerush}, \citenamefont {Kostyukov},\ and\ \citenamefont
		{Narozhny}}]{gelfer2015optimized}%
	\BibitemOpen
	\bibfield  {author} {\bibinfo {author} {\bibfnamefont {E.~G.}\ \bibnamefont
			{Gelfer}}, \bibinfo {author} {\bibfnamefont {A.~A.}\ \bibnamefont {Mironov}},
		\bibinfo {author} {\bibfnamefont {A.~M.}\ \bibnamefont {Fedotov}}, \bibinfo
		{author} {\bibfnamefont {V.~F.}\ \bibnamefont {Bashmakov}}, \bibinfo {author}
		{\bibfnamefont {E.~N.}\ \bibnamefont {Nerush}}, \bibinfo {author}
		{\bibfnamefont {I.~Y.}\ \bibnamefont {Kostyukov}},\ and\ \bibinfo {author}
		{\bibfnamefont {N.~B.}\ \bibnamefont {Narozhny}},\ }\bibfield  {title}
	{\bibinfo {title} {Optimized multibeam configuration for observation of {QED}
			cascades},\ }\href {https://doi.org/10.1103/PhysRevA.92.022113} {\bibfield
		{journal} {\bibinfo  {journal} {Phys. Rev. A}\ }\textbf {\bibinfo {volume}
			{92}},\ \bibinfo {pages} {022113} (\bibinfo {year} {2015})}\BibitemShut
	{NoStop}%
	\bibitem [{\citenamefont {Marklund}\ \emph {et~al.}(2023)\citenamefont
		{Marklund}, \citenamefont {Blackburn}, \citenamefont {Gonoskov},
		\citenamefont {Magnusson}, \citenamefont {Bulanov},\ and\ \citenamefont
		{Ilderton}}]{marklund2023towards}%
	\BibitemOpen
	\bibfield  {author} {\bibinfo {author} {\bibfnamefont {M.}~\bibnamefont
			{Marklund}}, \bibinfo {author} {\bibfnamefont {T.~G.}\ \bibnamefont
			{Blackburn}}, \bibinfo {author} {\bibfnamefont {A.}~\bibnamefont {Gonoskov}},
		\bibinfo {author} {\bibfnamefont {J.}~\bibnamefont {Magnusson}}, \bibinfo
		{author} {\bibfnamefont {S.~S.}\ \bibnamefont {Bulanov}},\ and\ \bibinfo
		{author} {\bibfnamefont {A.}~\bibnamefont {Ilderton}},\ }\bibfield  {title}
	{\bibinfo {title} {Towards critical and supercritical electromagnetic
			fields},\ }\href {https://doi.org/https://doi.org/10.1017/hpl.2022.46}
	{\bibfield  {journal} {\bibinfo  {journal} {High Power Laser Sci. Eng.}\
		}\textbf {\bibinfo {volume} {11}},\ \bibinfo {pages} {e19} (\bibinfo {year}
		{2023})}\BibitemShut {NoStop}%
	\bibitem [{\citenamefont {Vincenti}\ \emph {et~al.}(2023)\citenamefont
		{Vincenti}, \citenamefont {Clark}, \citenamefont {Fedeli}, \citenamefont
		{Martin}, \citenamefont {Sainte-Marie},\ and\ \citenamefont
		{Zaim}}]{vincenti2023plasma}%
	\BibitemOpen
	\bibfield  {author} {\bibinfo {author} {\bibfnamefont {H.}~\bibnamefont
			{Vincenti}}, \bibinfo {author} {\bibfnamefont {T.}~\bibnamefont {Clark}},
		\bibinfo {author} {\bibfnamefont {L.}~\bibnamefont {Fedeli}}, \bibinfo
		{author} {\bibfnamefont {P.}~\bibnamefont {Martin}}, \bibinfo {author}
		{\bibfnamefont {A.}~\bibnamefont {Sainte-Marie}},\ and\ \bibinfo {author}
		{\bibfnamefont {N.}~\bibnamefont {Zaim}},\ }\bibfield  {title} {\bibinfo
		{title} {Plasma mirrors as a path to the {S}chwinger limit: theoretical and
			numerical developments},\ }\href
	{https://doi.org/https://doi.org/10.1140/epjs/s11734-023-00909-2} {\bibfield
		{journal} {\bibinfo  {journal} {Eur. Phys. J.: Spec. Top.}\ }\textbf
		{\bibinfo {volume} {232}},\ \bibinfo {pages} {2303} (\bibinfo {year}
		{2023})}\BibitemShut {NoStop}%
	\bibitem [{\citenamefont {Ridgers}\ \emph {et~al.}(2012)\citenamefont
		{Ridgers}, \citenamefont {Brady}, \citenamefont {Duclous}, \citenamefont
		{Kirk}, \citenamefont {Bennett}, \citenamefont {Arber}, \citenamefont
		{Robinson},\ and\ \citenamefont {Bell}}]{ridgers2012dense}%
	\BibitemOpen
	\bibfield  {author} {\bibinfo {author} {\bibfnamefont {C.~P.}\ \bibnamefont
			{Ridgers}}, \bibinfo {author} {\bibfnamefont {C.~S.}\ \bibnamefont {Brady}},
		\bibinfo {author} {\bibfnamefont {R.}~\bibnamefont {Duclous}}, \bibinfo
		{author} {\bibfnamefont {J.~G.}\ \bibnamefont {Kirk}}, \bibinfo {author}
		{\bibfnamefont {K.}~\bibnamefont {Bennett}}, \bibinfo {author} {\bibfnamefont
			{T.~D.}\ \bibnamefont {Arber}}, \bibinfo {author} {\bibfnamefont {A.~P.~L.}\
			\bibnamefont {Robinson}},\ and\ \bibinfo {author} {\bibfnamefont {A.~R.}\
			\bibnamefont {Bell}},\ }\bibfield  {title} {\bibinfo {title} {Dense
			electron-positron plasmas and ultraintense $\gamma$ rays from
			laser-irradiated solids},\ }\href
	{https://doi.org/https://doi.org/10.1103/PhysRevLett.108.165006} {\bibfield
		{journal} {\bibinfo  {journal} {Phys. Rev. Lett.}\ }\textbf {\bibinfo
			{volume} {108}},\ \bibinfo {pages} {165006} (\bibinfo {year}
		{2012})}\BibitemShut {NoStop}%
	\bibitem [{\citenamefont {Jirka}\ \emph {et~al.}(2017)\citenamefont {Jirka},
		\citenamefont {Klimo}, \citenamefont {Vranic}, \citenamefont {Weber},\ and\
		\citenamefont {Korn}}]{jirka2017qed}%
	\BibitemOpen
	\bibfield  {author} {\bibinfo {author} {\bibfnamefont {M.}~\bibnamefont
			{Jirka}}, \bibinfo {author} {\bibfnamefont {O.}~\bibnamefont {Klimo}},
		\bibinfo {author} {\bibfnamefont {M.}~\bibnamefont {Vranic}}, \bibinfo
		{author} {\bibfnamefont {S.}~\bibnamefont {Weber}},\ and\ \bibinfo {author}
		{\bibfnamefont {G.}~\bibnamefont {Korn}},\ }\bibfield  {title} {\bibinfo
		{title} {{QED} cascade with 10 {PW}-class lasers},\ }\href
	{https://doi.org/10.1038/s41598-017-15747-1} {\bibfield  {journal} {\bibinfo
			{journal} {Sci. Rep.}\ }\textbf {\bibinfo {volume} {7}},\ \bibinfo {pages}
		{15302} (\bibinfo {year} {2017})}\BibitemShut {NoStop}%
	\bibitem [{\citenamefont {Slade-Lowther}\ \emph {et~al.}(2019)\citenamefont
		{Slade-Lowther}, \citenamefont {Sorbo},\ and\ \citenamefont
		{Ridgers}}]{Slade2019}%
	\BibitemOpen
	\bibfield  {author} {\bibinfo {author} {\bibfnamefont {C.}~\bibnamefont
			{Slade-Lowther}}, \bibinfo {author} {\bibfnamefont {D.~D.}\ \bibnamefont
			{Sorbo}},\ and\ \bibinfo {author} {\bibfnamefont {C.~P.}\ \bibnamefont
			{Ridgers}},\ }\bibfield  {title} {\bibinfo {title} {Identifying the
			electron–positron cascade regimes in high-intensity laser-matter
			interactions},\ }\href {https://doi.org/10.1088/1367-2630/aafa39} {\bibfield
		{journal} {\bibinfo  {journal} {New J. Phys.}\ }\textbf {\bibinfo {volume}
			{21}},\ \bibinfo {pages} {013028} (\bibinfo {year} {2019})}\BibitemShut
	{NoStop}%
	\bibitem [{\citenamefont {Zhu}\ \emph {et~al.}(2016)\citenamefont {Zhu},
		\citenamefont {Yu}, \citenamefont {Sheng}, \citenamefont {Yin}, \citenamefont
		{Turcu},\ and\ \citenamefont {Pukhov}}]{zhu2016dense}%
	\BibitemOpen
	\bibfield  {author} {\bibinfo {author} {\bibfnamefont {X.-L.}\ \bibnamefont
			{Zhu}}, \bibinfo {author} {\bibfnamefont {T.-P.}\ \bibnamefont {Yu}},
		\bibinfo {author} {\bibfnamefont {Z.-M.}\ \bibnamefont {Sheng}}, \bibinfo
		{author} {\bibfnamefont {Y.}~\bibnamefont {Yin}}, \bibinfo {author}
		{\bibfnamefont {I.~C.~E.}\ \bibnamefont {Turcu}},\ and\ \bibinfo {author}
		{\bibfnamefont {A.}~\bibnamefont {Pukhov}},\ }\bibfield  {title} {\bibinfo
		{title} {Dense {GeV} electron--positron pairs generated by lasers in
			near-critical-density plasmas},\ }\bibfield  {journal} {\bibinfo  {journal}
		{Nat. Commun.}\ }\textbf {\bibinfo {volume} {7}},\ \href
	{https://doi.org/https://doi.org/10.1038/ncomms13686}
	{https://doi.org/10.1038/ncomms13686} (\bibinfo {year} {2016})\BibitemShut
	{NoStop}%
	\bibitem [{\citenamefont {Samsonov}\ \emph {et~al.}(2021)\citenamefont
		{Samsonov}, \citenamefont {Kostyukov},\ and\ \citenamefont
		{Nerush}}]{samsonov2021hydrodynamical}%
	\BibitemOpen
	\bibfield  {author} {\bibinfo {author} {\bibfnamefont {A.~S.}\ \bibnamefont
			{Samsonov}}, \bibinfo {author} {\bibfnamefont {I.~Y.}\ \bibnamefont
			{Kostyukov}},\ and\ \bibinfo {author} {\bibfnamefont {E.~N.}\ \bibnamefont
			{Nerush}},\ }\bibfield  {title} {\bibinfo {title} {Hydrodynamical model of
			{QED} cascade expansion in an extremely strong laser pulse},\ }\bibfield
	{journal} {\bibinfo  {journal} {Matter Radiat. Extremes}\ }\textbf {\bibinfo
		{volume} {6}},\ \href {https://doi.org/https://doi.org/10.1063/5.0035347}
	{https://doi.org/10.1063/5.0035347} (\bibinfo {year} {2021})\BibitemShut
	{NoStop}%
	\bibitem [{\citenamefont {Artemenko}\ and\ \citenamefont
		{Kostyukov}(2017)}]{artemenko2017ionization}%
	\BibitemOpen
	\bibfield  {author} {\bibinfo {author} {\bibfnamefont {I.~I.}\ \bibnamefont
			{Artemenko}}\ and\ \bibinfo {author} {\bibfnamefont {I.~Y.}\ \bibnamefont
			{Kostyukov}},\ }\bibfield  {title} {\bibinfo {title} {Ionization-induced
			laser-driven {QED} cascade in noble gases},\ }\href
	{https://doi.org/10.1103/PhysRevA.96.032106} {\bibfield  {journal} {\bibinfo
			{journal} {Phys. Rev. A}\ }\textbf {\bibinfo {volume} {96}},\ \bibinfo
		{pages} {032106} (\bibinfo {year} {2017})}\BibitemShut {NoStop}%
	\bibitem [{\citenamefont {Tamburini}\ \emph {et~al.}(2017)\citenamefont
		{Tamburini}, \citenamefont {Di~Piazza},\ and\ \citenamefont
		{Keitel}}]{Tamburini_2017}%
	\BibitemOpen
	\bibfield  {author} {\bibinfo {author} {\bibfnamefont {M.}~\bibnamefont
			{Tamburini}}, \bibinfo {author} {\bibfnamefont {A.}~\bibnamefont
			{Di~Piazza}},\ and\ \bibinfo {author} {\bibfnamefont {C.~H.}\ \bibnamefont
			{Keitel}},\ }\bibfield  {title} {\bibinfo {title} {Laser-pulse-shape control
			of seeded {QED} cascades},\ }\bibfield  {journal} {\bibinfo  {journal} {Sci.
			Rep.}\ }\textbf {\bibinfo {volume} {7}},\ \href
	{https://doi.org/http://doi.org/10.1038/s41598-017-05891-z}
	{http://doi.org/10.1038/s41598-017-05891-z} (\bibinfo {year}
	{2017})\BibitemShut {NoStop}%
	\bibitem [{\citenamefont {Jirka}\ and\ \citenamefont
		{Bulanov}(2024)}]{jirka2024effects}%
	\BibitemOpen
	\bibfield  {author} {\bibinfo {author} {\bibfnamefont {M.}~\bibnamefont
			{Jirka}}\ and\ \bibinfo {author} {\bibfnamefont {S.}~\bibnamefont
			{Bulanov}},\ }\bibfield  {title} {\bibinfo {title} {Effects of colliding
			laser pulses polarization on $ e^-e^+ $ cascade development in extreme
			focusing},\ }\href {https://arxiv.org/abs/2401.08410} {\bibfield  {journal}
		{\bibinfo  {journal} {arXiv preprint arXiv:2401.08410}\ } (\bibinfo {year}
		{2024})}\BibitemShut {NoStop}%
	\bibitem [{\citenamefont {Lobet}\ \emph {et~al.}(2015)\citenamefont {Lobet},
		\citenamefont {Ruyer}, \citenamefont {Debayle}, \citenamefont {d'Humi\`eres},
		\citenamefont {Grech}, \citenamefont {Lemoine},\ and\ \citenamefont
		{Gremillet}}]{lobet2015shock}%
	\BibitemOpen
	\bibfield  {author} {\bibinfo {author} {\bibfnamefont {M.}~\bibnamefont
			{Lobet}}, \bibinfo {author} {\bibfnamefont {C.}~\bibnamefont {Ruyer}},
		\bibinfo {author} {\bibfnamefont {A.}~\bibnamefont {Debayle}}, \bibinfo
		{author} {\bibfnamefont {E.}~\bibnamefont {d'Humi\`eres}}, \bibinfo {author}
		{\bibfnamefont {M.}~\bibnamefont {Grech}}, \bibinfo {author} {\bibfnamefont
			{M.}~\bibnamefont {Lemoine}},\ and\ \bibinfo {author} {\bibfnamefont
			{L.}~\bibnamefont {Gremillet}},\ }\bibfield  {title} {\bibinfo {title}
		{Ultrafast synchrotron-enhanced thermalization of laser-driven colliding pair
			plasmas},\ }\href {https://doi.org/10.1103/PhysRevLett.115.215003} {\bibfield
		{journal} {\bibinfo  {journal} {Phys. Rev. Lett.}\ }\textbf {\bibinfo
			{volume} {115}},\ \bibinfo {pages} {215003} (\bibinfo {year}
		{2015})}\BibitemShut {NoStop}%
	\bibitem [{\citenamefont {Yu}\ \emph {et~al.}(2018)\citenamefont {Yu},
		\citenamefont {Yuan}, \citenamefont {Liu}, \citenamefont {Chen},
		\citenamefont {Luo}, \citenamefont {Weng},\ and\ \citenamefont
		{Sheng}}]{yu2018qed}%
	\BibitemOpen
	\bibfield  {author} {\bibinfo {author} {\bibfnamefont {J.~Y.}\ \bibnamefont
			{Yu}}, \bibinfo {author} {\bibfnamefont {T.}~\bibnamefont {Yuan}}, \bibinfo
		{author} {\bibfnamefont {W.~Y.}\ \bibnamefont {Liu}}, \bibinfo {author}
		{\bibfnamefont {M.}~\bibnamefont {Chen}}, \bibinfo {author} {\bibfnamefont
			{W.}~\bibnamefont {Luo}}, \bibinfo {author} {\bibfnamefont {S.~M.}\
			\bibnamefont {Weng}},\ and\ \bibinfo {author} {\bibfnamefont {Z.~M.}\
			\bibnamefont {Sheng}},\ }\bibfield  {title} {\bibinfo {title} {{QED} effects
			induced harmonics generation in extreme intense laser foil interaction},\
	}\href {https://doi.org/10.1088/1361-6587/aaae35} {\bibfield  {journal}
		{\bibinfo  {journal} {Plasma Phys. Control. Fusion}\ }\textbf {\bibinfo
			{volume} {60}},\ \bibinfo {pages} {044011} (\bibinfo {year}
		{2018})}\BibitemShut {NoStop}%
	\bibitem [{\citenamefont {Luo}\ \emph {et~al.}(2018)\citenamefont {Luo},
		\citenamefont {Liu}, \citenamefont {Yuan}, \citenamefont {Chen},
		\citenamefont {Yu}, \citenamefont {Li}, \citenamefont {Del~Sorbo},
		\citenamefont {Ridgers},\ and\ \citenamefont {Sheng}}]{luo2018qed}%
	\BibitemOpen
	\bibfield  {author} {\bibinfo {author} {\bibfnamefont {W.}~\bibnamefont
			{Luo}}, \bibinfo {author} {\bibfnamefont {W.-Y.}\ \bibnamefont {Liu}},
		\bibinfo {author} {\bibfnamefont {T.}~\bibnamefont {Yuan}}, \bibinfo {author}
		{\bibfnamefont {M.}~\bibnamefont {Chen}}, \bibinfo {author} {\bibfnamefont
			{J.-Y.}\ \bibnamefont {Yu}}, \bibinfo {author} {\bibfnamefont {F.-Y.}\
			\bibnamefont {Li}}, \bibinfo {author} {\bibfnamefont {D.}~\bibnamefont
			{Del~Sorbo}}, \bibinfo {author} {\bibfnamefont {C.~P.}\ \bibnamefont
			{Ridgers}},\ and\ \bibinfo {author} {\bibfnamefont {Z.-M.}\ \bibnamefont
			{Sheng}},\ }\bibfield  {title} {\bibinfo {title} {{QED} cascade saturation in
			extreme high fields},\ }\href {https://doi.org/10.1038/s41598-018-26785-8}
	{\bibfield  {journal} {\bibinfo  {journal} {Sci. Rep.}\ }\textbf {\bibinfo
			{volume} {8}},\ \bibinfo {pages} {8400} (\bibinfo {year} {2018})}\BibitemShut
	{NoStop}%
	\bibitem [{\citenamefont {Nerush}\ \emph {et~al.}(2011)\citenamefont {Nerush},
		\citenamefont {Bashmakov},\ and\ \citenamefont
		{Kostyukov}}]{nerush2011analytical}%
	\BibitemOpen
	\bibfield  {author} {\bibinfo {author} {\bibfnamefont {E.~N.}\ \bibnamefont
			{Nerush}}, \bibinfo {author} {\bibfnamefont {V.~F.}\ \bibnamefont
			{Bashmakov}},\ and\ \bibinfo {author} {\bibfnamefont {I.~Y.}\ \bibnamefont
			{Kostyukov}},\ }\bibfield  {title} {\bibinfo {title} {Analytical model for
			electromagnetic cascades in rotating electric field},\ }\href
	{https://doi.org/10.1063/1.3624481} {\bibfield  {journal} {\bibinfo
			{journal} {Phys. Plasmas}\ }\textbf {\bibinfo {volume} {18}},\ \bibinfo
		{pages} {083107} (\bibinfo {year} {2011})}\BibitemShut {NoStop}%
	\bibitem [{\citenamefont {Grismayer}\ \emph {et~al.}(2016)\citenamefont
		{Grismayer}, \citenamefont {Vranic}, \citenamefont {Martins}, \citenamefont
		{Fonseca},\ and\ \citenamefont {Silva}}]{Grismayer2016}%
	\BibitemOpen
	\bibfield  {author} {\bibinfo {author} {\bibfnamefont {T.}~\bibnamefont
			{Grismayer}}, \bibinfo {author} {\bibfnamefont {M.}~\bibnamefont {Vranic}},
		\bibinfo {author} {\bibfnamefont {J.~L.}\ \bibnamefont {Martins}}, \bibinfo
		{author} {\bibfnamefont {R.~A.}\ \bibnamefont {Fonseca}},\ and\ \bibinfo
		{author} {\bibfnamefont {L.~O.}\ \bibnamefont {Silva}},\ }\bibfield  {title}
	{\bibinfo {title} {Laser absorption via quantum electrodynamics cascades in
			counter propagating laser pulses},\ }\href
	{https://doi.org/10.1063/1.4950841} {\bibfield  {journal} {\bibinfo
			{journal} {Phys. Plasmas}\ }\textbf {\bibinfo {volume} {23}},\ \bibinfo
		{pages} {056706} (\bibinfo {year} {2016})}\BibitemShut {NoStop}%
	\bibitem [{\citenamefont {Efimenko}\ \emph {et~al.}(2018)\citenamefont
		{Efimenko}, \citenamefont {Bashinov}, \citenamefont {Bastrakov},
		\citenamefont {Gonoskov}, \citenamefont {Muraviev}, \citenamefont {Meyerov},
		\citenamefont {Kim},\ and\ \citenamefont {Sergeev}}]{efimenko2018extreme}%
	\BibitemOpen
	\bibfield  {author} {\bibinfo {author} {\bibfnamefont {E.~S.}\ \bibnamefont
			{Efimenko}}, \bibinfo {author} {\bibfnamefont {A.~V.}\ \bibnamefont
			{Bashinov}}, \bibinfo {author} {\bibfnamefont {S.~I.}\ \bibnamefont
			{Bastrakov}}, \bibinfo {author} {\bibfnamefont {A.~A.}\ \bibnamefont
			{Gonoskov}}, \bibinfo {author} {\bibfnamefont {A.~A.}\ \bibnamefont
			{Muraviev}}, \bibinfo {author} {\bibfnamefont {I.~B.}\ \bibnamefont
			{Meyerov}}, \bibinfo {author} {\bibfnamefont {A.~V.}\ \bibnamefont {Kim}},\
		and\ \bibinfo {author} {\bibfnamefont {A.~M.}\ \bibnamefont {Sergeev}},\
	}\bibfield  {title} {\bibinfo {title} {Extreme plasma states in
			laser-governed vacuum breakdown},\ }\href
	{https://doi.org/10.1038/s41598-018-20745-y} {\bibfield  {journal} {\bibinfo
			{journal} {Sci. Rep.}\ }\textbf {\bibinfo {volume} {8}},\ \bibinfo {pages}
		{2329} (\bibinfo {year} {2018})}\BibitemShut {NoStop}%
	\bibitem [{\citenamefont {Papadopoulos}\ \emph {et~al.}(2019)\citenamefont
		{Papadopoulos}, \citenamefont {Zou}, \citenamefont {Le~Blanc}, \citenamefont
		{Ranc}, \citenamefont {Druon}, \citenamefont {Martin}, \citenamefont
		{Fr{\'e}neaux}, \citenamefont {Beluze}, \citenamefont {Lebas}, \citenamefont
		{Chabanis} \emph {et~al.}}]{papadopoulos2019first}%
	\BibitemOpen
	\bibfield  {author} {\bibinfo {author} {\bibfnamefont {D.}~\bibnamefont
			{Papadopoulos}}, \bibinfo {author} {\bibfnamefont {J.-P.}\ \bibnamefont
			{Zou}}, \bibinfo {author} {\bibfnamefont {C.}~\bibnamefont {Le~Blanc}},
		\bibinfo {author} {\bibfnamefont {L.}~\bibnamefont {Ranc}}, \bibinfo {author}
		{\bibfnamefont {F.}~\bibnamefont {Druon}}, \bibinfo {author} {\bibfnamefont
			{L.}~\bibnamefont {Martin}}, \bibinfo {author} {\bibfnamefont
			{A.}~\bibnamefont {Fr{\'e}neaux}}, \bibinfo {author} {\bibfnamefont
			{A.}~\bibnamefont {Beluze}}, \bibinfo {author} {\bibfnamefont
			{N.}~\bibnamefont {Lebas}}, \bibinfo {author} {\bibfnamefont
			{M.}~\bibnamefont {Chabanis}}, \emph {et~al.},\ }\bibfield  {title} {\bibinfo
		{title} {First commissioning results of the apollon laser on the 1 {PW} beam
			line},\ }in\ \href
	{https://doi.org/https://doi.org/10.1364/CLEO_SI.2019.STu3E.4} {\emph
		{\bibinfo {booktitle} {CLEO: Science and Innovations}}}\ (\bibinfo
	{organization} {Optica Publishing Group},\ \bibinfo {year} {2019})\ pp.\
	\bibinfo {pages} {STu3E--4}\BibitemShut {NoStop}%
	\bibitem [{\citenamefont {Weber}\ \emph {et~al.}(2017)\citenamefont {Weber},
		\citenamefont {Bechet}, \citenamefont {Borneis}, \citenamefont {Brabec},
		\citenamefont {Bu{\v{c}}ka}, \citenamefont {Chacon-Golcher}, \citenamefont
		{Ciappina}, \citenamefont {DeMarco}, \citenamefont {Fajstavr}, \citenamefont
		{Falk} \emph {et~al.}}]{weber_mre2017}%
	\BibitemOpen
	\bibfield  {author} {\bibinfo {author} {\bibfnamefont {S.}~\bibnamefont
			{Weber}}, \bibinfo {author} {\bibfnamefont {S.}~\bibnamefont {Bechet}},
		\bibinfo {author} {\bibfnamefont {S.}~\bibnamefont {Borneis}}, \bibinfo
		{author} {\bibfnamefont {L.}~\bibnamefont {Brabec}}, \bibinfo {author}
		{\bibfnamefont {M.}~\bibnamefont {Bu{\v{c}}ka}}, \bibinfo {author}
		{\bibfnamefont {E.}~\bibnamefont {Chacon-Golcher}}, \bibinfo {author}
		{\bibfnamefont {M.}~\bibnamefont {Ciappina}}, \bibinfo {author}
		{\bibfnamefont {M.}~\bibnamefont {DeMarco}}, \bibinfo {author} {\bibfnamefont
			{A.}~\bibnamefont {Fajstavr}}, \bibinfo {author} {\bibfnamefont
			{K.}~\bibnamefont {Falk}}, \emph {et~al.},\ }\bibfield  {title} {\bibinfo
		{title} {P3: An installation for high-energy density plasma physics and
			ultra-high intensity laser--matter interaction at {ELI-B}eamlines},\ }\href
	{https://doi.org/10.1016/j.mre.2017.03.003} {\bibfield  {journal} {\bibinfo
			{journal} {Matter Radiat. at Extremes}\ }\textbf {\bibinfo {volume} {2}},\
		\bibinfo {pages} {149} (\bibinfo {year} {2017})}\BibitemShut {NoStop}%
	\bibitem [{cor()}]{corels}%
	\BibitemOpen
	\href@noop {} {\bibinfo {title} {Center for {R}elativistic {L}aser
			{S}cience}},\ \bibinfo {note}
	{\url{https://corels.ibs.re.kr/html/corels_en/}}\BibitemShut {NoStop}%
	\bibitem [{\citenamefont {Tanaka}\ \emph {et~al.}(2020)\citenamefont {Tanaka},
		\citenamefont {Spohr}, \citenamefont {Balabanski}, \citenamefont {Balascuta},
		\citenamefont {Capponi}, \citenamefont {Cernaianu}, \citenamefont {Cuciuc},
		\citenamefont {Cucoanes}, \citenamefont {Dancus}, \citenamefont {Dhal} \emph
		{et~al.}}]{tanaka_mre2020}%
	\BibitemOpen
	\bibfield  {author} {\bibinfo {author} {\bibfnamefont {K.~A.}\ \bibnamefont
			{Tanaka}}, \bibinfo {author} {\bibfnamefont {K.~M.}\ \bibnamefont {Spohr}},
		\bibinfo {author} {\bibfnamefont {D.~L.}\ \bibnamefont {Balabanski}},
		\bibinfo {author} {\bibfnamefont {S.}~\bibnamefont {Balascuta}}, \bibinfo
		{author} {\bibfnamefont {L.}~\bibnamefont {Capponi}}, \bibinfo {author}
		{\bibfnamefont {M.~O.}\ \bibnamefont {Cernaianu}}, \bibinfo {author}
		{\bibfnamefont {M.}~\bibnamefont {Cuciuc}}, \bibinfo {author} {\bibfnamefont
			{A.}~\bibnamefont {Cucoanes}}, \bibinfo {author} {\bibfnamefont
			{I.}~\bibnamefont {Dancus}}, \bibinfo {author} {\bibfnamefont
			{A.}~\bibnamefont {Dhal}}, \emph {et~al.},\ }\bibfield  {title} {\bibinfo
		{title} {Current status and highlights of the {ELI-NP} research program},\
	}\href {https://doi.org/10.1063/1.5093535} {\bibfield  {journal} {\bibinfo
			{journal} {Matter Radiat. at Extremes}\ }\textbf {\bibinfo {volume} {5}},\
		\bibinfo {pages} {024402} (\bibinfo {year} {2020})}\BibitemShut {NoStop}%
	\bibitem [{ast()}]{astragemini}%
	\BibitemOpen
	\href@noop {} {\bibinfo {title} {Astra {G}emini laser}},\ \bibinfo {note}
	{\url{https://www.clf.stfc.ac.uk/Pages/The-Astra-Gemini-Facility.aspx}}\BibitemShut
	{NoStop}%
	\bibitem [{\citenamefont {Gan}\ \emph {et~al.}(2017)\citenamefont {Gan},
		\citenamefont {Yu}, \citenamefont {Li}, \citenamefont {Wang}, \citenamefont
		{Liang}, \citenamefont {Liu}, \citenamefont {Li}, \citenamefont {Guo},
		\citenamefont {Fan}, \citenamefont {Yuan} \emph {et~al.}}]{gan_optexpr2017}%
	\BibitemOpen
	\bibfield  {author} {\bibinfo {author} {\bibfnamefont {Z.}~\bibnamefont
			{Gan}}, \bibinfo {author} {\bibfnamefont {L.}~\bibnamefont {Yu}}, \bibinfo
		{author} {\bibfnamefont {S.}~\bibnamefont {Li}}, \bibinfo {author}
		{\bibfnamefont {C.}~\bibnamefont {Wang}}, \bibinfo {author} {\bibfnamefont
			{X.}~\bibnamefont {Liang}}, \bibinfo {author} {\bibfnamefont
			{Y.}~\bibnamefont {Liu}}, \bibinfo {author} {\bibfnamefont {W.}~\bibnamefont
			{Li}}, \bibinfo {author} {\bibfnamefont {Z.}~\bibnamefont {Guo}}, \bibinfo
		{author} {\bibfnamefont {Z.}~\bibnamefont {Fan}}, \bibinfo {author}
		{\bibfnamefont {X.}~\bibnamefont {Yuan}}, \emph {et~al.},\ }\bibfield
	{title} {\bibinfo {title} {200 {J} high efficiency {T}i:sapphire chirped
			pulse amplifier pumped by temporal dual-pulse},\ }\href
	{https://doi.org/https://doi.org/10.1364/OE.25.005169} {\bibfield  {journal}
		{\bibinfo  {journal} {Opt. Express}\ }\textbf {\bibinfo {volume} {25}},\
		\bibinfo {pages} {5169} (\bibinfo {year} {2017})}\BibitemShut {NoStop}%
	\bibitem [{pea()}]{pearl}%
	\BibitemOpen
	\href@noop {} {\bibinfo {title} {{PEARL} laser complex}},\ \bibinfo {note}
	{\url{https://ipfran.ru/science/laser-physics-and-nonlinear-optics/generation-of-extreme-laser-fields/PEARL-laser-complex}}\BibitemShut
	{NoStop}%
	\bibitem [{\citenamefont {Danson}\ \emph {et~al.}(2019)\citenamefont {Danson},
		\citenamefont {Haefner}, \citenamefont {Bromage}, \citenamefont {Butcher},
		\citenamefont {Chanteloup}, \citenamefont {Chowdhury}, \citenamefont
		{Galvanauskas}, \citenamefont {Gizzi}, \citenamefont {Hein}, \citenamefont
		{Hillier} \emph {et~al.}}]{danson_hplse2019}%
	\BibitemOpen
	\bibfield  {author} {\bibinfo {author} {\bibfnamefont {C.~N.}\ \bibnamefont
			{Danson}}, \bibinfo {author} {\bibfnamefont {C.}~\bibnamefont {Haefner}},
		\bibinfo {author} {\bibfnamefont {J.}~\bibnamefont {Bromage}}, \bibinfo
		{author} {\bibfnamefont {T.}~\bibnamefont {Butcher}}, \bibinfo {author}
		{\bibfnamefont {J.-C.~F.}\ \bibnamefont {Chanteloup}}, \bibinfo {author}
		{\bibfnamefont {E.~A.}\ \bibnamefont {Chowdhury}}, \bibinfo {author}
		{\bibfnamefont {A.}~\bibnamefont {Galvanauskas}}, \bibinfo {author}
		{\bibfnamefont {L.~A.}\ \bibnamefont {Gizzi}}, \bibinfo {author}
		{\bibfnamefont {J.}~\bibnamefont {Hein}}, \bibinfo {author} {\bibfnamefont
			{D.~I.}\ \bibnamefont {Hillier}}, \emph {et~al.},\ }\bibfield  {title}
	{\bibinfo {title} {Petawatt and exawatt class lasers worldwide},\ }\bibfield
	{journal} {\bibinfo  {journal} {High Power Laser Sci. Eng.}\ }\textbf
	{\bibinfo {volume} {7}},\ \href
	{https://doi.org/https://doi.org/10.1017/hpl.2019.36}
	{https://doi.org/10.1017/hpl.2019.36} (\bibinfo {year} {2019})\BibitemShut
	{NoStop}%
	\bibitem [{\citenamefont {Yoon}\ \emph {et~al.}(2021)\citenamefont {Yoon},
		\citenamefont {Kim}, \citenamefont {Choi}, \citenamefont {Sung},
		\citenamefont {Lee}, \citenamefont {Lee},\ and\ \citenamefont
		{Nam}}]{yoon2021realization}%
	\BibitemOpen
	\bibfield  {author} {\bibinfo {author} {\bibfnamefont {J.~W.}\ \bibnamefont
			{Yoon}}, \bibinfo {author} {\bibfnamefont {Y.~G.}\ \bibnamefont {Kim}},
		\bibinfo {author} {\bibfnamefont {I.~W.}\ \bibnamefont {Choi}}, \bibinfo
		{author} {\bibfnamefont {J.~H.}\ \bibnamefont {Sung}}, \bibinfo {author}
		{\bibfnamefont {H.~W.}\ \bibnamefont {Lee}}, \bibinfo {author} {\bibfnamefont
			{S.~K.}\ \bibnamefont {Lee}},\ and\ \bibinfo {author} {\bibfnamefont {C.~H.}\
			\bibnamefont {Nam}},\ }\bibfield  {title} {\bibinfo {title} {Realization of
			laser intensity over $10^{23}$ {W}/cm$^2$},\ }\href
	{https://doi.org/https://doi.org/10.1364/OPTICA.420520} {\bibfield  {journal}
		{\bibinfo  {journal} {Optica}\ }\textbf {\bibinfo {volume} {8}},\ \bibinfo
		{pages} {630} (\bibinfo {year} {2021})}\BibitemShut {NoStop}%
	\bibitem [{\citenamefont {Bromage}\ \emph {et~al.}(2019)\citenamefont
		{Bromage}, \citenamefont {Bahk}, \citenamefont {Begishev}, \citenamefont
		{Dorrer}, \citenamefont {Guardalben}, \citenamefont {Hoffman}, \citenamefont
		{Oliver}, \citenamefont {Roides}, \citenamefont {Schiesser}, \citenamefont
		{Shoup~Iii} \emph {et~al.}}]{bromage2019technology}%
	\BibitemOpen
	\bibfield  {author} {\bibinfo {author} {\bibfnamefont {J.}~\bibnamefont
			{Bromage}}, \bibinfo {author} {\bibfnamefont {S.-W.}\ \bibnamefont {Bahk}},
		\bibinfo {author} {\bibfnamefont {I.~A.}\ \bibnamefont {Begishev}}, \bibinfo
		{author} {\bibfnamefont {C.}~\bibnamefont {Dorrer}}, \bibinfo {author}
		{\bibfnamefont {M.~J.}\ \bibnamefont {Guardalben}}, \bibinfo {author}
		{\bibfnamefont {B.~N.}\ \bibnamefont {Hoffman}}, \bibinfo {author}
		{\bibfnamefont {J.~B.}\ \bibnamefont {Oliver}}, \bibinfo {author}
		{\bibfnamefont {R.~G.}\ \bibnamefont {Roides}}, \bibinfo {author}
		{\bibfnamefont {E.~M.}\ \bibnamefont {Schiesser}}, \bibinfo {author}
		{\bibfnamefont {M.~J.}\ \bibnamefont {Shoup~Iii}}, \emph {et~al.},\
	}\bibfield  {title} {\bibinfo {title} {Technology development for
			ultraintense all-{OPCPA} systems},\ }\href
	{https://doi.org/https://doi.org/10.1017/hpl.2018.64} {\bibfield  {journal}
		{\bibinfo  {journal} {High Power Laser Sci. Eng.}\ }\textbf {\bibinfo
			{volume} {7}},\ \bibinfo {pages} {e4} (\bibinfo {year} {2019})}\BibitemShut
	{NoStop}%
	\bibitem [{nsf()}]{nsf_opal}%
	\BibitemOpen
	\href@noop {} {\bibinfo {title} {{NSF} {OPAL}}},\ \bibinfo {note}
	{\url{https://www.lle.rochester.edu/nsf-opal/}}\BibitemShut {NoStop}%
	\bibitem [{\citenamefont {Khazanov}\ \emph {et~al.}(2023)\citenamefont
		{Khazanov}, \citenamefont {Shaykin}, \citenamefont {Kostyukov}, \citenamefont
		{Ginzburg}, \citenamefont {Mukhin}, \citenamefont {Yakovlev}, \citenamefont
		{Soloviev}, \citenamefont {Kuznetsov}, \citenamefont {Mironov}, \citenamefont
		{Korzhimanov} \emph {et~al.}}]{khazanov2023exawatt}%
	\BibitemOpen
	\bibfield  {author} {\bibinfo {author} {\bibfnamefont {E.}~\bibnamefont
			{Khazanov}}, \bibinfo {author} {\bibfnamefont {A.}~\bibnamefont {Shaykin}},
		\bibinfo {author} {\bibfnamefont {I.}~\bibnamefont {Kostyukov}}, \bibinfo
		{author} {\bibfnamefont {V.}~\bibnamefont {Ginzburg}}, \bibinfo {author}
		{\bibfnamefont {I.}~\bibnamefont {Mukhin}}, \bibinfo {author} {\bibfnamefont
			{I.}~\bibnamefont {Yakovlev}}, \bibinfo {author} {\bibfnamefont
			{A.}~\bibnamefont {Soloviev}}, \bibinfo {author} {\bibfnamefont
			{I.}~\bibnamefont {Kuznetsov}}, \bibinfo {author} {\bibfnamefont
			{S.}~\bibnamefont {Mironov}}, \bibinfo {author} {\bibfnamefont
			{A.}~\bibnamefont {Korzhimanov}}, \emph {et~al.},\ }\bibfield  {title}
	{\bibinfo {title} {e{X}awatt {C}enter for {E}xtreme {L}ight {S}tudies},\
	}\href {https://doi.org/https://doi.org/10.1017/hpl.2023.69} {\bibfield
		{journal} {\bibinfo  {journal} {High Power Laser Sci. Eng.}\ }\textbf
		{\bibinfo {volume} {11}},\ \bibinfo {pages} {e78} (\bibinfo {year}
		{2023})}\BibitemShut {NoStop}%
	\bibitem [{\citenamefont {Shao}\ \emph {et~al.}(2020)\citenamefont {Shao},
		\citenamefont {Li}, \citenamefont {Peng}, \citenamefont {Wang}, \citenamefont
		{Qian}, \citenamefont {Leng},\ and\ \citenamefont {Li}}]{shao2020broad}%
	\BibitemOpen
	\bibfield  {author} {\bibinfo {author} {\bibfnamefont {B.}~\bibnamefont
			{Shao}}, \bibinfo {author} {\bibfnamefont {Y.}~\bibnamefont {Li}}, \bibinfo
		{author} {\bibfnamefont {Y.}~\bibnamefont {Peng}}, \bibinfo {author}
		{\bibfnamefont {P.}~\bibnamefont {Wang}}, \bibinfo {author} {\bibfnamefont
			{J.}~\bibnamefont {Qian}}, \bibinfo {author} {\bibfnamefont {Y.}~\bibnamefont
			{Leng}},\ and\ \bibinfo {author} {\bibfnamefont {R.}~\bibnamefont {Li}},\
	}\bibfield  {title} {\bibinfo {title} {Broad-bandwidth high-temporal-contrast
			carrier-envelope-phase-stabilized laser seed for 100 {PW} lasers},\ }\href
	{https://doi.org/10.1364/OL.390110} {\bibfield  {journal} {\bibinfo
			{journal} {Optics Lett.}\ }\textbf {\bibinfo {volume} {45}},\ \bibinfo
		{pages} {2215} (\bibinfo {year} {2020})}\BibitemShut {NoStop}%
	\bibitem [{\citenamefont {Kostyukov}\ \emph {et~al.}(2018)\citenamefont
		{Kostyukov}, \citenamefont {Artemenko},\ and\ \citenamefont
		{Nerush}}]{kostyukov2018growth}%
	\BibitemOpen
	\bibfield  {author} {\bibinfo {author} {\bibfnamefont {I.~Y.}\ \bibnamefont
			{Kostyukov}}, \bibinfo {author} {\bibfnamefont {I.~I.}\ \bibnamefont
			{Artemenko}},\ and\ \bibinfo {author} {\bibfnamefont {E.~N.}\ \bibnamefont
			{Nerush}},\ }\bibfield  {title} {\bibinfo {title} {Growth rate of {QED}
			cascades in a rotating electric field},\ }\href
	{http://dspace.nbuv.gov.ua/handle/123456789/147446} {\bibfield  {journal}
		{\bibinfo  {journal} {Probl. At. Sci. Technol.}\ ,\ \bibinfo {pages} {259}}
		(\bibinfo {year} {2018})}\BibitemShut {NoStop}%
	\bibitem [{\citenamefont {Esirkepov}\ \emph {et~al.}(2015)\citenamefont
		{Esirkepov}, \citenamefont {Bulanov}, \citenamefont {Koga}, \citenamefont
		{Kando}, \citenamefont {Kondo}, \citenamefont {Rosanov}, \citenamefont
		{Korn},\ and\ \citenamefont {Bulanov}}]{Esirkepov2014}%
	\BibitemOpen
	\bibfield  {author} {\bibinfo {author} {\bibfnamefont {T.~Z.}\ \bibnamefont
			{Esirkepov}}, \bibinfo {author} {\bibfnamefont {S.~S.}\ \bibnamefont
			{Bulanov}}, \bibinfo {author} {\bibfnamefont {J.~K.}\ \bibnamefont {Koga}},
		\bibinfo {author} {\bibfnamefont {M.}~\bibnamefont {Kando}}, \bibinfo
		{author} {\bibfnamefont {K.}~\bibnamefont {Kondo}}, \bibinfo {author}
		{\bibfnamefont {N.~N.}\ \bibnamefont {Rosanov}}, \bibinfo {author}
		{\bibfnamefont {G.}~\bibnamefont {Korn}},\ and\ \bibinfo {author}
		{\bibfnamefont {S.~V.}\ \bibnamefont {Bulanov}},\ }\bibfield  {title}
	{\bibinfo {title} {Attractors and chaos of electron dynamics in
			electromagnetic standing waves},\ }\href
	{https://doi.org/https://doi.org/10.1016/j.physleta.2015.06.017} {\bibfield
		{journal} {\bibinfo  {journal} {Phys. Lett. A}\ }\textbf {\bibinfo {volume}
			{379}},\ \bibinfo {pages} {2044} (\bibinfo {year} {2015})}\BibitemShut
	{NoStop}%
	\bibitem [{\citenamefont {King}\ and\ \citenamefont
		{Hu}(2016)}]{King2016classical}%
	\BibitemOpen
	\bibfield  {author} {\bibinfo {author} {\bibfnamefont {B.}~\bibnamefont
			{King}}\ and\ \bibinfo {author} {\bibfnamefont {H.}~\bibnamefont {Hu}},\
	}\bibfield  {title} {\bibinfo {title} {Classical and quantum dynamics of a
			charged scalar particle in a background of two counterpropagating plane
			waves},\ }\href {https://doi.org/10.1103/PhysRevD.94.125010} {\bibfield
		{journal} {\bibinfo  {journal} {Phys. Rev. D}\ }\textbf {\bibinfo {volume}
			{94}},\ \bibinfo {pages} {125010} (\bibinfo {year} {2016})}\BibitemShut
	{NoStop}%
	\bibitem [{\citenamefont {Niel}\ \emph {et~al.}(2018)\citenamefont {Niel},
		\citenamefont {Riconda}, \citenamefont {Amiranoff}, \citenamefont {Duclous},\
		and\ \citenamefont {Grech}}]{Niel2018}%
	\BibitemOpen
	\bibfield  {author} {\bibinfo {author} {\bibfnamefont {F.}~\bibnamefont
			{Niel}}, \bibinfo {author} {\bibfnamefont {C.}~\bibnamefont {Riconda}},
		\bibinfo {author} {\bibfnamefont {F.}~\bibnamefont {Amiranoff}}, \bibinfo
		{author} {\bibfnamefont {R.}~\bibnamefont {Duclous}},\ and\ \bibinfo {author}
		{\bibfnamefont {M.}~\bibnamefont {Grech}},\ }\bibfield  {title} {\bibinfo
		{title} {From quantum to classical modeling of radiation reaction: A focus on
			stochasticity effects},\ }\href {https://doi.org/10.1103/PhysRevE.97.043209}
	{\bibfield  {journal} {\bibinfo  {journal} {Phys. Rev. E}\ }\textbf {\bibinfo
			{volume} {97}},\ \bibinfo {pages} {043209} (\bibinfo {year}
		{2018})}\BibitemShut {NoStop}%
	\bibitem [{\citenamefont {Fauth}\ \emph {et~al.}(2021)\citenamefont {Fauth},
		\citenamefont {Berges},\ and\ \citenamefont
		{Di~Piazza}}]{fauth2021collisional}%
	\BibitemOpen
	\bibfield  {author} {\bibinfo {author} {\bibfnamefont {G.}~\bibnamefont
			{Fauth}}, \bibinfo {author} {\bibfnamefont {J.}~\bibnamefont {Berges}},\ and\
		\bibinfo {author} {\bibfnamefont {A.}~\bibnamefont {Di~Piazza}},\ }\bibfield
	{title} {\bibinfo {title} {Collisional strong-field {QED} kinetic equations
			from first principles},\ }\href {https://doi.org/10.1103/PhysRevD.104.036007}
	{\bibfield  {journal} {\bibinfo  {journal} {Phys. Rev. D}\ }\textbf {\bibinfo
			{volume} {104}},\ \bibinfo {pages} {036007} (\bibinfo {year}
		{2021})}\BibitemShut {NoStop}%
	\bibitem [{\citenamefont {Gonoskov}\ \emph {et~al.}(2015)\citenamefont
		{Gonoskov}, \citenamefont {Bastrakov}, \citenamefont {Efimenko},
		\citenamefont {Ilderton}, \citenamefont {Marklund}, \citenamefont {Meyerov},
		\citenamefont {Muraviev}, \citenamefont {Sergeev}, \citenamefont {Surmin},\
		and\ \citenamefont {Wallin}}]{gonoskov2015extended}%
	\BibitemOpen
	\bibfield  {author} {\bibinfo {author} {\bibfnamefont {A.}~\bibnamefont
			{Gonoskov}}, \bibinfo {author} {\bibfnamefont {S.}~\bibnamefont {Bastrakov}},
		\bibinfo {author} {\bibfnamefont {E.}~\bibnamefont {Efimenko}}, \bibinfo
		{author} {\bibfnamefont {A.}~\bibnamefont {Ilderton}}, \bibinfo {author}
		{\bibfnamefont {M.}~\bibnamefont {Marklund}}, \bibinfo {author}
		{\bibfnamefont {I.}~\bibnamefont {Meyerov}}, \bibinfo {author} {\bibfnamefont
			{A.}~\bibnamefont {Muraviev}}, \bibinfo {author} {\bibfnamefont
			{A.}~\bibnamefont {Sergeev}}, \bibinfo {author} {\bibfnamefont
			{I.}~\bibnamefont {Surmin}},\ and\ \bibinfo {author} {\bibfnamefont
			{E.}~\bibnamefont {Wallin}},\ }\bibfield  {title} {\bibinfo {title} {Extended
			particle-in-cell schemes for physics in ultrastrong laser fields: Review and
			developments},\ }\href {https://doi.org/10.1103/PhysRevE.92.023305}
	{\bibfield  {journal} {\bibinfo  {journal} {Phys. Rev. E}\ }\textbf {\bibinfo
			{volume} {92}},\ \bibinfo {pages} {023305} (\bibinfo {year}
		{2015})}\BibitemShut {NoStop}%
	\bibitem [{\citenamefont {King}\ \emph {et~al.}(2013)\citenamefont {King},
		\citenamefont {Elkina},\ and\ \citenamefont {Ruhl}}]{king2013photon}%
	\BibitemOpen
	\bibfield  {author} {\bibinfo {author} {\bibfnamefont {B.}~\bibnamefont
			{King}}, \bibinfo {author} {\bibfnamefont {N.}~\bibnamefont {Elkina}},\ and\
		\bibinfo {author} {\bibfnamefont {H.}~\bibnamefont {Ruhl}},\ }\bibfield
	{title} {\bibinfo {title} {Photon polarization in electron-seeded
			pair-creation cascades},\ }\href
	{https://doi.org/https://doi.org/10.1103/PhysRevA.87.042117} {\bibfield
		{journal} {\bibinfo  {journal} {Phys. Rev. A}\ }\textbf {\bibinfo {volume}
			{87}},\ \bibinfo {pages} {042117} (\bibinfo {year} {2013})}\BibitemShut
	{NoStop}%
	\bibitem [{\citenamefont {Seipt}\ and\ \citenamefont
		{Thomas}(2023)}]{seipt2023kinetic}%
	\BibitemOpen
	\bibfield  {author} {\bibinfo {author} {\bibfnamefont {D.}~\bibnamefont
			{Seipt}}\ and\ \bibinfo {author} {\bibfnamefont {A.~G.~R.}\ \bibnamefont
			{Thomas}},\ }\bibfield  {title} {\bibinfo {title} {Kinetic theory for
			spin-polarized relativistic plasmas},\ }\bibfield  {journal} {\bibinfo
		{journal} {Phys. Plasmas}\ }\textbf {\bibinfo {volume} {30}},\ \href
	{https://doi.org/https://doi.org/10.1063/5.0165836}
	{https://doi.org/10.1063/5.0165836} (\bibinfo {year} {2023})\BibitemShut
	{NoStop}%
	\bibitem [{\citenamefont {Zhao}\ \emph {et~al.}(2023)\citenamefont {Zhao},
		\citenamefont {Sun}, \citenamefont {Xue}, \citenamefont {Wan},\ and\
		\citenamefont {Li}}]{zhao2023cascade}%
	\BibitemOpen
	\bibfield  {author} {\bibinfo {author} {\bibfnamefont {Q.}~\bibnamefont
			{Zhao}}, \bibinfo {author} {\bibfnamefont {T.}~\bibnamefont {Sun}}, \bibinfo
		{author} {\bibfnamefont {K.}~\bibnamefont {Xue}}, \bibinfo {author}
		{\bibfnamefont {F.}~\bibnamefont {Wan}},\ and\ \bibinfo {author}
		{\bibfnamefont {J.-X.}\ \bibnamefont {Li}},\ }\bibfield  {title} {\bibinfo
		{title} {Cascade of polarized {C}ompton scattering and {B}reit-{W}heeler pair
			production},\ }\href
	{https://doi.org/https://doi.org/10.1103/PhysRevD.108.116012} {\bibfield
		{journal} {\bibinfo  {journal} {Phys. Rev. D}\ }\textbf {\bibinfo {volume}
			{108}},\ \bibinfo {pages} {116012} (\bibinfo {year} {2023})}\BibitemShut
	{NoStop}%
	\bibitem [{\citenamefont {Sampath}\ and\ \citenamefont
		{Tamburini}(2018)}]{sampath2018towards}%
	\BibitemOpen
	\bibfield  {author} {\bibinfo {author} {\bibfnamefont {A.}~\bibnamefont
			{Sampath}}\ and\ \bibinfo {author} {\bibfnamefont {M.}~\bibnamefont
			{Tamburini}},\ }\bibfield  {title} {\bibinfo {title} {Towards realistic
			simulations of {QED} cascades: {N}on-ideal laser and electron seeding
			effects},\ }\bibfield  {journal} {\bibinfo  {journal} {Phys. Plasmas}\
	}\textbf {\bibinfo {volume} {25}},\ \href
	{https://doi.org/https://doi.org/10.1063/1.5022640}
	{https://doi.org/10.1063/1.5022640} (\bibinfo {year} {2018})\BibitemShut
	{NoStop}%
	\bibitem [{\citenamefont {Jirka}\ \emph {et~al.}(2021)\citenamefont {Jirka},
		\citenamefont {Sasorov}, \citenamefont {Bulanov}, \citenamefont {Korn},
		\citenamefont {Rus},\ and\ \citenamefont {Bulanov}}]{jirka2021reaching}%
	\BibitemOpen
	\bibfield  {author} {\bibinfo {author} {\bibfnamefont {M.}~\bibnamefont
			{Jirka}}, \bibinfo {author} {\bibfnamefont {P.}~\bibnamefont {Sasorov}},
		\bibinfo {author} {\bibfnamefont {S.~S.}\ \bibnamefont {Bulanov}}, \bibinfo
		{author} {\bibfnamefont {G.}~\bibnamefont {Korn}}, \bibinfo {author}
		{\bibfnamefont {B.}~\bibnamefont {Rus}},\ and\ \bibinfo {author}
		{\bibfnamefont {S.~V.}\ \bibnamefont {Bulanov}},\ }\bibfield  {title}
	{\bibinfo {title} {Reaching high laser intensity by a radiating electron},\
	}\href {https://doi.org/https://doi.org/10.1103/PhysRevA.103.053114}
	{\bibfield  {journal} {\bibinfo  {journal} {Phys. Rev. A}\ }\textbf {\bibinfo
			{volume} {103}},\ \bibinfo {pages} {053114} (\bibinfo {year}
		{2021})}\BibitemShut {NoStop}%
	\bibitem [{\citenamefont {Mironov}\ \emph {et~al.}(2016)\citenamefont
		{Mironov}, \citenamefont {Fedotov},\ and\ \citenamefont
		{Narozhny}}]{mironov2016generation}%
	\BibitemOpen
	\bibfield  {author} {\bibinfo {author} {\bibfnamefont {A.~A.}\ \bibnamefont
			{Mironov}}, \bibinfo {author} {\bibfnamefont {A.~M.}\ \bibnamefont
			{Fedotov}},\ and\ \bibinfo {author} {\bibfnamefont {N.~B.}\ \bibnamefont
			{Narozhny}},\ }\bibfield  {title} {\bibinfo {title} {Generation of
			quantum-electrodynamic cascades in oblique collisions of ultrarelativistic
			electrons with an intense laser field},\ }\href
	{https://doi.org/10.1070/QEL16057} {\bibfield  {journal} {\bibinfo  {journal}
			{Quantum Electron.}\ }\textbf {\bibinfo {volume} {46}},\ \bibinfo {pages}
		{305} (\bibinfo {year} {2016})}\BibitemShut {NoStop}%
	\bibitem [{\citenamefont {Bulanov}\ \emph {et~al.}(2010)\citenamefont
		{Bulanov}, \citenamefont {Mur}, \citenamefont {Narozhny}, \citenamefont
		{Nees},\ and\ \citenamefont {Popov}}]{bulanov2010multiple}%
	\BibitemOpen
	\bibfield  {author} {\bibinfo {author} {\bibfnamefont {S.~S.}\ \bibnamefont
			{Bulanov}}, \bibinfo {author} {\bibfnamefont {V.~D.}\ \bibnamefont {Mur}},
		\bibinfo {author} {\bibfnamefont {N.~B.}\ \bibnamefont {Narozhny}}, \bibinfo
		{author} {\bibfnamefont {J.}~\bibnamefont {Nees}},\ and\ \bibinfo {author}
		{\bibfnamefont {V.~S.}\ \bibnamefont {Popov}},\ }\bibfield  {title} {\bibinfo
		{title} {Multiple colliding electromagnetic pulses: a way to lower the
			threshold of e+ e- pair production from vacuum},\ }\href
	{https://doi.org/https://doi.org/10.1103/PhysRevLett.104.220404} {\bibfield
		{journal} {\bibinfo  {journal} {Phys. Rev. Lett.}\ }\textbf {\bibinfo
			{volume} {104}},\ \bibinfo {pages} {220404} (\bibinfo {year}
		{2010})}\BibitemShut {NoStop}%
	\bibitem [{\citenamefont {Derouillat}\ \emph {et~al.}(2018)\citenamefont
		{Derouillat}, \citenamefont {Beck}, \citenamefont {P{\'e}rez}, \citenamefont
		{Vinci}, \citenamefont {Chiaramello}, \citenamefont {Grassi}, \citenamefont
		{Fl{\'e}}, \citenamefont {Bouchard}, \citenamefont {Plotnikov}, \citenamefont
		{Aunai} \emph {et~al.}}]{derouillat2018smilei}%
	\BibitemOpen
	\bibfield  {author} {\bibinfo {author} {\bibfnamefont {J.}~\bibnamefont
			{Derouillat}}, \bibinfo {author} {\bibfnamefont {A.}~\bibnamefont {Beck}},
		\bibinfo {author} {\bibfnamefont {F.}~\bibnamefont {P{\'e}rez}}, \bibinfo
		{author} {\bibfnamefont {T.}~\bibnamefont {Vinci}}, \bibinfo {author}
		{\bibfnamefont {M.}~\bibnamefont {Chiaramello}}, \bibinfo {author}
		{\bibfnamefont {A.}~\bibnamefont {Grassi}}, \bibinfo {author} {\bibfnamefont
			{M.}~\bibnamefont {Fl{\'e}}}, \bibinfo {author} {\bibfnamefont
			{G.}~\bibnamefont {Bouchard}}, \bibinfo {author} {\bibfnamefont
			{I.}~\bibnamefont {Plotnikov}}, \bibinfo {author} {\bibfnamefont
			{N.}~\bibnamefont {Aunai}}, \emph {et~al.},\ }\bibfield  {title} {\bibinfo
		{title} {Smilei: A collaborative, open-source, multi-purpose particle-in-cell
			code for plasma simulation},\ }\href
	{https://doi.org/https://doi.org/10.1016/j.cpc.2017.09.024} {\bibfield
		{journal} {\bibinfo  {journal} {Comp. Phys. Comm.}\ }\textbf {\bibinfo
			{volume} {222}},\ \bibinfo {pages} {351} (\bibinfo {year}
		{2018})}\BibitemShut {NoStop}%
	\bibitem [{\citenamefont {Nikishov}\ and\ \citenamefont
		{Ritus}(1964)}]{nikishov1964quantum}%
	\BibitemOpen
	\bibfield  {author} {\bibinfo {author} {\bibfnamefont {A.~I.}\ \bibnamefont
			{Nikishov}}\ and\ \bibinfo {author} {\bibfnamefont {V.~I.}\ \bibnamefont
			{Ritus}},\ }\bibfield  {title} {\bibinfo {title} {Quantum processes in the
			field of a plane electromagnetic wave and in a constant field {I}},\ }\href
	{http://jetp.ras.ru/cgi-bin/dn/e_019_02_0529.pdf} {\bibfield  {journal}
		{\bibinfo  {journal} {Sov. Phys. JETP}\ }\textbf {\bibinfo {volume} {19}},\
		\bibinfo {pages} {529} (\bibinfo {year} {1964})}\BibitemShut {NoStop}%
	\bibitem [{\citenamefont {Nikishov}\ and\ \citenamefont
		{Ritus}(1967)}]{nikishov1967pair}%
	\BibitemOpen
	\bibfield  {author} {\bibinfo {author} {\bibfnamefont {A.~I.}\ \bibnamefont
			{Nikishov}}\ and\ \bibinfo {author} {\bibfnamefont {V.~I.}\ \bibnamefont
			{Ritus}},\ }\bibfield  {title} {\bibinfo {title} {Pair production by a photon
			and photon emission by an electron in the field of an intense electromagnetic
			wave and in a constant field},\ }\href
	{http://jetp.ras.ru/cgi-bin/dn/e_025_06_1135.pdf} {\bibfield  {journal}
		{\bibinfo  {journal} {Sov. Phys. JETP}\ }\textbf {\bibinfo {volume} {25}},\
		\bibinfo {pages} {1135} (\bibinfo {year} {1967})}\BibitemShut {NoStop}%
	\bibitem [{\citenamefont {Baier}\ and\ \citenamefont
		{Katkov}(1968)}]{Baier1968}%
	\BibitemOpen
	\bibfield  {author} {\bibinfo {author} {\bibfnamefont {V.~N.}\ \bibnamefont
			{Baier}}\ and\ \bibinfo {author} {\bibfnamefont {V.~M.}\ \bibnamefont
			{Katkov}},\ }\bibfield  {title} {\bibinfo {title} {Processes involved in the
			motion of high energy particles in a magnetic field},\ }\href
	{http://jetp.ras.ru/cgi-bin/dn/e_026_04_0854.pdf} {\bibfield  {journal}
		{\bibinfo  {journal} {Sov. Phys. JETP}\ }\textbf {\bibinfo {volume} {26}},\
		\bibinfo {pages} {854} (\bibinfo {year} {1968})}\BibitemShut {NoStop}%
	\bibitem [{\citenamefont {Gelfer}\ \emph {et~al.}(2022)\citenamefont {Gelfer},
		\citenamefont {Fedotov}, \citenamefont {Mironov},\ and\ \citenamefont
		{Weber}}]{gelfer2022nonlinear}%
	\BibitemOpen
	\bibfield  {author} {\bibinfo {author} {\bibfnamefont {E.~G.}\ \bibnamefont
			{Gelfer}}, \bibinfo {author} {\bibfnamefont {A.~M.}\ \bibnamefont {Fedotov}},
		\bibinfo {author} {\bibfnamefont {A.~A.}\ \bibnamefont {Mironov}},\ and\
		\bibinfo {author} {\bibfnamefont {S.}~\bibnamefont {Weber}},\ }\bibfield
	{title} {\bibinfo {title} {Nonlinear {C}ompton scattering in time-dependent
			electric fields beyond the locally constant crossed field approximation},\
	}\href {https://doi.org/10.1103/PhysRevD.106.056013} {\bibfield  {journal}
		{\bibinfo  {journal} {Phys. Rev. D}\ }\textbf {\bibinfo {volume} {106}},\
		\bibinfo {pages} {056013} (\bibinfo {year} {2022})}\BibitemShut {NoStop}%
	\bibitem [{\citenamefont {Di~Piazza}\ \emph {et~al.}(2018)\citenamefont
		{Di~Piazza}, \citenamefont {Tamburini}, \citenamefont {Meuren},\ and\
		\citenamefont {Keitel}}]{DiPiazza2018_LCFA}%
	\BibitemOpen
	\bibfield  {author} {\bibinfo {author} {\bibfnamefont {A.}~\bibnamefont
			{Di~Piazza}}, \bibinfo {author} {\bibfnamefont {M.}~\bibnamefont
			{Tamburini}}, \bibinfo {author} {\bibfnamefont {S.}~\bibnamefont {Meuren}},\
		and\ \bibinfo {author} {\bibfnamefont {C.~H.}\ \bibnamefont {Keitel}},\
	}\bibfield  {title} {\bibinfo {title} {Implementing nonlinear {C}ompton
			scattering beyond the local-constant-field approximation},\ }\href
	{https://doi.org/10.1103/PhysRevA.98.012134} {\bibfield  {journal} {\bibinfo
			{journal} {Phys. Rev. A}\ }\textbf {\bibinfo {volume} {98}},\ \bibinfo
		{pages} {012134} (\bibinfo {year} {2018})}\BibitemShut {NoStop}%
	\bibitem [{\citenamefont {Di~Piazza}\ \emph {et~al.}(2019)\citenamefont
		{Di~Piazza}, \citenamefont {Tamburini}, \citenamefont {Meuren},\ and\
		\citenamefont {Keitel}}]{di2019improved}%
	\BibitemOpen
	\bibfield  {author} {\bibinfo {author} {\bibfnamefont {A.}~\bibnamefont
			{Di~Piazza}}, \bibinfo {author} {\bibfnamefont {M.}~\bibnamefont
			{Tamburini}}, \bibinfo {author} {\bibfnamefont {S.}~\bibnamefont {Meuren}},\
		and\ \bibinfo {author} {\bibfnamefont {C.~H.}\ \bibnamefont {Keitel}},\
	}\bibfield  {title} {\bibinfo {title} {Improved local-constant-field
			approximation for strong-field {QED} codes},\ }\href
	{https://doi.org/https://doi.org/10.1103/PhysRevA.99.022125} {\bibfield
		{journal} {\bibinfo  {journal} {Phys. Rev. A}\ }\textbf {\bibinfo {volume}
			{99}},\ \bibinfo {pages} {022125} (\bibinfo {year} {2019})}\BibitemShut
	{NoStop}%
	\bibitem [{\citenamefont {Ilderton}\ \emph {et~al.}(2019)\citenamefont
		{Ilderton}, \citenamefont {King},\ and\ \citenamefont
		{Seipt}}]{ilderton2019extended}%
	\BibitemOpen
	\bibfield  {author} {\bibinfo {author} {\bibfnamefont {A.}~\bibnamefont
			{Ilderton}}, \bibinfo {author} {\bibfnamefont {B.}~\bibnamefont {King}},\
		and\ \bibinfo {author} {\bibfnamefont {D.}~\bibnamefont {Seipt}},\ }\bibfield
	{title} {\bibinfo {title} {Extended locally constant field approximation for
			nonlinear {C}ompton scattering},\ }\href
	{https://doi.org/https://doi.org/10.1103/PhysRevA.99.042121} {\bibfield
		{journal} {\bibinfo  {journal} {Phys. Rev. A}\ }\textbf {\bibinfo {volume}
			{99}},\ \bibinfo {pages} {042121} (\bibinfo {year} {2019})}\BibitemShut
	{NoStop}%
	\bibitem [{\citenamefont {Raicher}\ \emph {et~al.}(2019)\citenamefont
		{Raicher}, \citenamefont {Eliezer}, \citenamefont {Keitel},\ and\
		\citenamefont {Hatsagortsyan}}]{raicher2019semiclassical}%
	\BibitemOpen
	\bibfield  {author} {\bibinfo {author} {\bibfnamefont {E.}~\bibnamefont
			{Raicher}}, \bibinfo {author} {\bibfnamefont {S.}~\bibnamefont {Eliezer}},
		\bibinfo {author} {\bibfnamefont {C.~H.}\ \bibnamefont {Keitel}},\ and\
		\bibinfo {author} {\bibfnamefont {K.~Z.}\ \bibnamefont {Hatsagortsyan}},\
	}\bibfield  {title} {\bibinfo {title} {Semiclassical limitations for photon
			emission in strong external fields},\ }\href
	{https://doi.org/10.1103/PhysRevA.99.052513} {\bibfield  {journal} {\bibinfo
			{journal} {Phys. Rev. A}\ }\textbf {\bibinfo {volume} {99}},\ \bibinfo
		{pages} {052513} (\bibinfo {year} {2019})}\BibitemShut {NoStop}%
	\bibitem [{\citenamefont {Heinzl}\ \emph {et~al.}(2020)\citenamefont {Heinzl},
		\citenamefont {King},\ and\ \citenamefont {MacLeod}}]{heinzl2020locally}%
	\BibitemOpen
	\bibfield  {author} {\bibinfo {author} {\bibfnamefont {T.}~\bibnamefont
			{Heinzl}}, \bibinfo {author} {\bibfnamefont {B.}~\bibnamefont {King}},\ and\
		\bibinfo {author} {\bibfnamefont {A.~J.}\ \bibnamefont {MacLeod}},\
	}\bibfield  {title} {\bibinfo {title} {Locally monochromatic approximation to
			{QED} in intense laser fields},\ }\href
	{https://doi.org/10.1103/PhysRevA.102.063110} {\bibfield  {journal} {\bibinfo
			{journal} {Phys. Rev. A}\ }\textbf {\bibinfo {volume} {102}},\ \bibinfo
		{pages} {063110} (\bibinfo {year} {2020})}\BibitemShut {NoStop}%
	\bibitem [{\citenamefont {King}(2020)}]{king2020uniform}%
	\BibitemOpen
	\bibfield  {author} {\bibinfo {author} {\bibfnamefont {B.}~\bibnamefont
			{King}},\ }\bibfield  {title} {\bibinfo {title} {Uniform locally constant
			field approximation for photon-seeded pair production},\ }\href
	{https://doi.org/10.1103/PhysRevA.101.042508} {\bibfield  {journal} {\bibinfo
			{journal} {Phys. Rev. A}\ }\textbf {\bibinfo {volume} {101}},\ \bibinfo
		{pages} {042508} (\bibinfo {year} {2020})}\BibitemShut {NoStop}%
	\bibitem [{\citenamefont {Di~Piazza}(2021)}]{di2021wkb}%
	\BibitemOpen
	\bibfield  {author} {\bibinfo {author} {\bibfnamefont {A.}~\bibnamefont
			{Di~Piazza}},\ }\bibfield  {title} {\bibinfo {title} {{WKB} electron wave
			functions in a tightly focused laser beam},\ }\href
	{https://doi.org/https://doi.org/10.1103/PhysRevD.103.076011} {\bibfield
		{journal} {\bibinfo  {journal} {Phys. Rev. D}\ }\textbf {\bibinfo {volume}
			{103}},\ \bibinfo {pages} {076011} (\bibinfo {year} {2021})}\BibitemShut
	{NoStop}%
	\bibitem [{\citenamefont {Nielsen}\ \emph {et~al.}(2022)\citenamefont
		{Nielsen}, \citenamefont {Holtzapple},\ and\ \citenamefont
		{King}}]{nielsen2022high}%
	\BibitemOpen
	\bibfield  {author} {\bibinfo {author} {\bibfnamefont {C.~F.}\ \bibnamefont
			{Nielsen}}, \bibinfo {author} {\bibfnamefont {R.}~\bibnamefont
			{Holtzapple}},\ and\ \bibinfo {author} {\bibfnamefont {B.}~\bibnamefont
			{King}},\ }\bibfield  {title} {\bibinfo {title} {High-resolution modeling of
			nonlinear {C}ompton scattering in focused laser pulses},\ }\href
	{https://doi.org/https://doi.org/10.1103/PhysRevD.106.013010} {\bibfield
		{journal} {\bibinfo  {journal} {Phys. Rev. D}\ }\textbf {\bibinfo {volume}
			{106}},\ \bibinfo {pages} {013010} (\bibinfo {year} {2022})}\BibitemShut
	{NoStop}%
	\bibitem [{\citenamefont {Taub}(1948)}]{taub1948orbits}%
	\BibitemOpen
	\bibfield  {author} {\bibinfo {author} {\bibfnamefont {A.~H.}\ \bibnamefont
			{Taub}},\ }\bibfield  {title} {\bibinfo {title} {Orbits of charged particles
			in constant fields},\ }\href {https://doi.org/10.1103/PhysRev.73.786}
	{\bibfield  {journal} {\bibinfo  {journal} {Phys. Rev.}\ }\textbf {\bibinfo
			{volume} {73}},\ \bibinfo {pages} {786} (\bibinfo {year} {1948})}\BibitemShut
	{NoStop}%
	\bibitem [{\citenamefont {Bezanson}\ \emph {et~al.}(2017)\citenamefont
		{Bezanson}, \citenamefont {Edelman}, \citenamefont {Karpinski},\ and\
		\citenamefont {Shah}}]{bezanson2017julia}%
	\BibitemOpen
	\bibfield  {author} {\bibinfo {author} {\bibfnamefont {J.}~\bibnamefont
			{Bezanson}}, \bibinfo {author} {\bibfnamefont {A.}~\bibnamefont {Edelman}},
		\bibinfo {author} {\bibfnamefont {S.}~\bibnamefont {Karpinski}},\ and\
		\bibinfo {author} {\bibfnamefont {V.~B.}\ \bibnamefont {Shah}},\ }\bibfield
	{title} {\bibinfo {title} {Julia: A fresh approach to numerical computing},\
	}\href {https://doi.org/https://doi.org/10.1137/141000671} {\bibfield
		{journal} {\bibinfo  {journal} {SIAM review}\ }\textbf {\bibinfo {volume}
			{59}},\ \bibinfo {pages} {65} (\bibinfo {year} {2017})}\BibitemShut {NoStop}%
	\bibitem [{Pai()}]{PairAvalanchesQED}%
	\BibitemOpen
	\href@noop {} {\bibinfo {title}
		{\url{https://github.com/amercurib/PairAvalanchesQED.jl}}}\BibitemShut
	{NoStop}%
	\bibitem [{\citenamefont {Lehmann}\ and\ \citenamefont
		{Spatschek}(2012)}]{lehman2012}%
	\BibitemOpen
	\bibfield  {author} {\bibinfo {author} {\bibfnamefont {G.}~\bibnamefont
			{Lehmann}}\ and\ \bibinfo {author} {\bibfnamefont {K.~H.}\ \bibnamefont
			{Spatschek}},\ }\bibfield  {title} {\bibinfo {title} {Phase-space contraction
			and attractors for ultrarelativistic electrons},\ }\href
	{https://doi.org/10.1103/PhysRevE.85.056412} {\bibfield  {journal} {\bibinfo
			{journal} {Phys. Rev. E}\ }\textbf {\bibinfo {volume} {85}},\ \bibinfo
		{pages} {056412} (\bibinfo {year} {2012})}\BibitemShut {NoStop}%
	\bibitem [{\citenamefont {Gong}\ \emph {et~al.}(2016)\citenamefont {Gong},
		\citenamefont {Hu}, \citenamefont {Shou}, \citenamefont {Qiao}, \citenamefont
		{Chen}, \citenamefont {Xu}, \citenamefont {He},\ and\ \citenamefont
		{Yan}}]{Gong2016_attractors}%
	\BibitemOpen
	\bibfield  {author} {\bibinfo {author} {\bibfnamefont {Z.}~\bibnamefont
			{Gong}}, \bibinfo {author} {\bibfnamefont {R.}~\bibnamefont {Hu}}, \bibinfo
		{author} {\bibfnamefont {Y.}~\bibnamefont {Shou}}, \bibinfo {author}
		{\bibfnamefont {B.}~\bibnamefont {Qiao}}, \bibinfo {author} {\bibfnamefont
			{C.}~\bibnamefont {Chen}}, \bibinfo {author} {\bibfnamefont {F.}~\bibnamefont
			{Xu}}, \bibinfo {author} {\bibfnamefont {X.}~\bibnamefont {He}},\ and\
		\bibinfo {author} {\bibfnamefont {X.}~\bibnamefont {Yan}},\ }\bibfield
	{title} {\bibinfo {title} {Radiation reaction induced spiral attractors in
			ultra-intense colliding laser beams},\ }\href
	{https://doi.org/https://doi.org/10.1016/j.mre.2016.10.005} {\bibfield
		{journal} {\bibinfo  {journal} {Matter Radiat. Extremes}\ }\textbf {\bibinfo
			{volume} {1}},\ \bibinfo {pages} {308} (\bibinfo {year} {2016})}\BibitemShut
	{NoStop}%
	\bibitem [{\citenamefont {Kirk}(2016)}]{Kirk2016}%
	\BibitemOpen
	\bibfield  {author} {\bibinfo {author} {\bibfnamefont {J.~G.}\ \bibnamefont
			{Kirk}},\ }\bibfield  {title} {\bibinfo {title} {Radiative trapping in
			intense laser beams},\ }\href {https://doi.org/10.1088/0741-3335/58/8/085005}
	{\bibfield  {journal} {\bibinfo  {journal} {Plasma Phys. Control. Fusion}\
		}\textbf {\bibinfo {volume} {58}},\ \bibinfo {pages} {085005} (\bibinfo
		{year} {2016})}\BibitemShut {NoStop}%
	\bibitem [{\citenamefont {Gonoskov}\ \emph {et~al.}(2014)\citenamefont
		{Gonoskov}, \citenamefont {Bashinov}, \citenamefont {Gonoskov}, \citenamefont
		{Harvey}, \citenamefont {Ilderton}, \citenamefont {Kim}, \citenamefont
		{Marklund}, \citenamefont {Mourou},\ and\ \citenamefont
		{Sergeev}}]{gonoskov2014anomalous}%
	\BibitemOpen
	\bibfield  {author} {\bibinfo {author} {\bibfnamefont {A.}~\bibnamefont
			{Gonoskov}}, \bibinfo {author} {\bibfnamefont {A.}~\bibnamefont {Bashinov}},
		\bibinfo {author} {\bibfnamefont {I.}~\bibnamefont {Gonoskov}}, \bibinfo
		{author} {\bibfnamefont {C.}~\bibnamefont {Harvey}}, \bibinfo {author}
		{\bibfnamefont {A.}~\bibnamefont {Ilderton}}, \bibinfo {author}
		{\bibfnamefont {A.}~\bibnamefont {Kim}}, \bibinfo {author} {\bibfnamefont
			{M.}~\bibnamefont {Marklund}}, \bibinfo {author} {\bibfnamefont
			{G.}~\bibnamefont {Mourou}},\ and\ \bibinfo {author} {\bibfnamefont
			{A.}~\bibnamefont {Sergeev}},\ }\bibfield  {title} {\bibinfo {title}
		{Anomalous radiative trapping in laser fields of extreme intensity},\ }\href
	{https://doi.org/https://doi.org/10.1103/PhysRevLett.113.014801} {\bibfield
		{journal} {\bibinfo  {journal} {Phys. Rev. Lett.}\ }\textbf {\bibinfo
			{volume} {113}},\ \bibinfo {pages} {014801} (\bibinfo {year}
		{2014})}\BibitemShut {NoStop}%
	\bibitem [{\citenamefont {Bashinov}\ \emph {et~al.}(2022)\citenamefont
		{Bashinov}, \citenamefont {Efimenko}, \citenamefont {Muraviev}, \citenamefont
		{Volokitin}, \citenamefont {Meyerov}, \citenamefont {Leuchs}, \citenamefont
		{Sergeev},\ and\ \citenamefont {Kim}}]{bashinov2022particle}%
	\BibitemOpen
	\bibfield  {author} {\bibinfo {author} {\bibfnamefont {A.~V.}\ \bibnamefont
			{Bashinov}}, \bibinfo {author} {\bibfnamefont {E.~S.}\ \bibnamefont
			{Efimenko}}, \bibinfo {author} {\bibfnamefont {A.~A.}\ \bibnamefont
			{Muraviev}}, \bibinfo {author} {\bibfnamefont {V.~D.}\ \bibnamefont
			{Volokitin}}, \bibinfo {author} {\bibfnamefont {I.~B.}\ \bibnamefont
			{Meyerov}}, \bibinfo {author} {\bibfnamefont {G.}~\bibnamefont {Leuchs}},
		\bibinfo {author} {\bibfnamefont {A.~M.}\ \bibnamefont {Sergeev}},\ and\
		\bibinfo {author} {\bibfnamefont {A.~V.}\ \bibnamefont {Kim}},\ }\bibfield
	{title} {\bibinfo {title} {Particle trajectories, gamma-ray emission, and
			anomalous radiative trapping effects in magnetic dipole wave},\ }\href
	{https://doi.org/https://doi.org/10.1103/PhysRevE.105.065202} {\bibfield
		{journal} {\bibinfo  {journal} {Phys. Rev. E}\ }\textbf {\bibinfo {volume}
			{105}},\ \bibinfo {pages} {065202} (\bibinfo {year} {2022})}\BibitemShut
	{NoStop}%
	\bibitem [{\citenamefont {Narozhny}\ and\ \citenamefont
		{Fofanov}(2000)}]{narozhny2000scattering}%
	\BibitemOpen
	\bibfield  {author} {\bibinfo {author} {\bibfnamefont {N.~B.}\ \bibnamefont
			{Narozhny}}\ and\ \bibinfo {author} {\bibfnamefont {M.~S.}\ \bibnamefont
			{Fofanov}},\ }\bibfield  {title} {\bibinfo {title} {Scattering of
			relativistic electrons by a focused laser pulse},\ }\href
	{https://doi.org/10.1134/1.559160} {\bibfield  {journal} {\bibinfo  {journal}
			{JETP}\ }\textbf {\bibinfo {volume} {90}},\ \bibinfo {pages} {753} (\bibinfo
		{year} {2000})}\BibitemShut {NoStop}%
	\bibitem [{\citenamefont {Landau}\ and\ \citenamefont
		{Lifshitz}()}]{landau2013classical}%
	\BibitemOpen
	\bibfield  {author} {\bibinfo {author} {\bibfnamefont {L.~D.}\ \bibnamefont
			{Landau}}\ and\ \bibinfo {author} {\bibfnamefont {I.~M.}\ \bibnamefont
			{Lifshitz}},\ }\href@noop {} {\emph {\bibinfo {title} {Theoretical Physics:
				The Classical Theory of Fields}}},\ Vol.~\bibinfo {volume} {2}\ (\bibinfo
	{publisher} {Course of Theoretical Physics Series, Pergamon Press, London,
		1988})\BibitemShut {NoStop}%
\end{thebibliography}
%\bibliographystyle{unsrt}

%apsrev4-2.bst 2019-01-14 (MD) hand-edited version of apsrev4-1.bst
%Control: key (0)
%Control: author (8) initials jnrlst
%Control: editor formatted (1) identically to author
%Control: production of article title (0) allowed
%Control: page (0) single
%Control: year (1) truncated
%Control: production of eprint (0) enabled
%

\end{document}